\documentclass[conference]{IEEEtran}
\IEEEoverridecommandlockouts
\usepackage{cite}
\usepackage{amsmath,amssymb,amsfonts}
\usepackage{algorithmic}
\usepackage{graphicx}
\usepackage{textcomp}
\usepackage{xcolor}
\usepackage{verbatim}
\usepackage{subfigure}
\def\BibTeX{{\rm B\kern-.05em{\sc i\kern-.025em b}\kern-.08em
    T\kern-.1667em\lower.7ex\hbox{E}\kern-.125emX}}

\begin{document}

\title{Feature Extraction, Modulation and Recognition of Mixed Signal Based on SVM
}

\author{Rong Han, Zihuai Lin\\
	School of Electrical and Information Engineering, The University of 
	Sydney, Australia\\
	Emails:	zihuai.lin@sydney.edu.au. 
}

\maketitle

\begin{abstract}
This paper introduces likelihood-based and feature-based modulation recognition methods. In the feature-based modulation simulation part, instantaneous feature, cyclic spectrum, high-order cumulants, and wavelet transform features are used as the entry point, and six digital signals including $2\mathrm{ASK} , 4\mathrm{ASK}, \mathrm{BPSK}, \mathrm{QPSK}, 2\mathrm{FSK}$ and $4\mathrm{FSK}$ are simulated, showing the difference of signals in multiple dimensions.
\end{abstract}


\section{Introduction}

Modulation recognition is to determine the modulation format of the signal based on the signal information received by the receiver. Modulation recognition technology has both practical value and theoretical significance \cite{dobre2007survey,Jiahao2022}.

In terms of commercial applications, modulation recognition technology has the following application values\cite{DorieL.2006Aslm}: first, in adaptive modulation communication systems, if the receiver uses modulation recognition technology, the additional protocol header overhead in the above method can be saved, and the protocol complexity can be reduced. Second, in the software-defined radio system, the modulation recognition technology can support the digital signal processing module at the receiving antenna to dynamically configure and recognize the modulation format of the received signal and improve the flexibility of the module to meet the design requirements. Third, for multimedia transmission applications, the receiver adopts modulation recognition technology to adapt to the modulation change requirements of the adaptive modulation communication system, thereby providing underlying technical support for upper-level applications\cite{alma991031525652105106}.

In military applications, modulation recognition technology has the following application value: In information countermeasures, usually the receiver does not know the modulation format of the received signal. At this time, modulation recognition technology can determine the modulation format of the received signal, which is a key step in signal demodulation\cite{alma991031525652105106}. 

At the same time, due to the rapid development and widespread popularity of radio communication technology, the signal environment has become increasingly complex, and modulation recognition technology has encountered many new challenges. The modulation recognition of mixed-signals in the same channel is one of the main technical problems. The modulation recognition of mixed-signals in the same channel is very different from the conventional single-signal modulation recognition. However, most of the existing modulation recognition methods are aimed at the single-signal and cannot be directly used for multi-signal identification. Therefore, it is urgent to study new methods to solve the problem of mixed-signal modulation recognition.

Modulation recognition technology can be divided into two categories from a theoretical perspective: the first category is likelihood-based modulation recognition; the second category is feature-based modulation recognition. This section first introduces the current research status of likelihood-based modulation recognition technology; second, introduces the current research status of feature-based modulation recognition technology; third, introduces the related research of modulation recognition based on machine learning that has become popular in recent years. Finally, introduces non-orthogonal multiple access (NOMA) which offers the research direction of mixed signal modulation recognition. 

\subsection{Likelihood-based Modulation Recognition}
For the likelihood-based modulation recognition technology, from the perspective of the recognition process, it can be divided into the following steps: first, process the signal received by the receiving end antenna to obtain the sampled signal sequence; then, calculate the sampled signal sequence to obtain likelihood value; Finally, the likelihood value is compared to determine the modulation to which the signal belongs. According to the calculation method of the likelihood value, the likelihood-based modulation recognition can be divided into: average likelihood ratio test (ALRT), generalized likelihood ratio test (GLRT), and Hybrid Likelihood Ratio Test (HLRT).

ALRT assumes that the unknown random variable obeys a certain probability distribution, calculates the average of the likelihood values through integration, and compares the results to determine the modulation pattern\cite{SapianoP.C1996MlPc,WenWei2000Mcfd,SillsJ.A1999Mmcf}. Because the computational complexity of ALRT is too high, some quasi-ALRT methods are proposed to reduce computational complexity\cite{HuangCy1995LMFM}. GLRT assumes that unknown parameters are unknown certain values, and on this basis, calculates the likelihood value to determine the modulation pattern\cite{PolydorosA1990Otda}. HLRT assumes that some unknown parameters obey a certain probability distribution, and some unknown parameters are unknown certain values, and based on this, the likelihood value is calculated to determine the modulation pattern\cite{LiangHong2002Aalm,ChuggK.M1995Clpe,DobreO.A2006LAfL}. 

In recent years, the research of modulation recognition technology based on likelihood ratio has been further studied mainly from more complex channel models, more complex physical scenes and more complex signal models\cite{ZhuZ.2015MciM,JingwenZhang2017CMCf}.

\subsection{Feature-based Modulation Recognition}
For the feature-based modulation recognition technology, from the perspective of the recognition process, it can be divided into the following steps: First, process the signal received by the receiving end antenna to obtain a sampled signal sequence. Then, process the sampled signal sequence to obtain signal characteristics. Finally, combine various classification tools to classify signal features and determine the modulation format to which the signal belongs.

The feature-based modulation part is the feature extraction operation by analyzing the signal time domain or transform domain information in many aspects. These feature parameters that characterize the signal are the core content of the statistical pattern recognition method, and the obviousness of the feature discrimination is directly related to the classification accuracy. In recent years, feature-based modulation methods have become more and more completed. Following sections would do a brief introduction to some common-used features.

\subsubsection*{(1) Higher-order cumulants (HOC) features }
Higher-order cumulants are essentially the characteristics of higher-order statistics, and the computational complexity increases significantly as the order of the cumulants increases. Because the cumulative amount of Gaussian white noise above the second order is zero, it can effectively reduce the interference of additive noise, and the demand of amount of observation data is not high, so it often appears in modulation recognition algorithms.

\cite{FengXiang2012HMCA} proposes a support vector machines based hierarchical algorithm for the automatic classification of QAM modulation signals which is less complex computationally and has faster classifier training speed compared with other algorithms. \cite{ZhangQ2009Otsc} considers the identification of single carrier signal with cyclic prefix, and good experimental performance is achieved. However, these methods did not consider the influence of channel fading, and the problems are studied under the premise that the observation sample contains only one modulation type. \cite{YanPeng-Zhan2010ARoD} proposes an algorithm for automatic recognition of digital communication signals by applying high order cumulants and support vector machines. The paper identifies eight types of signals by binary tree-based SVM as the classifier and rate of accurate classification is over 90\% when SNR is above 5dB. 

\subsubsection*{(2) Instantaneous characteristics features }
Different types of modulation signals have different performances in the statistical characteristics of instantaneous information, and there are large differences in instantaneous amplitude, phase, and frequency. It is feasible to modulate and identify signals by constructing instantaneous features. This type of feature parameters is less constrained by prior information and simple to extract, but the anti-noise performance is not ideal.

In 1995, Azzouz and Nandi extracted five kinds of information features such as signal instantaneous amplitude, instantaneous frequency, and nonlinear phase. When the signal-to-noise ratio is 10dB, the recognition rate of 2ASK, 4ASK, 2PSK, 4PSK, 2FSK, and 4FSK signals is larger than 90\%\cite{AzzouzE.E1995Aiod}. In the next few years, the two have successively proposed 9 classical instantaneous information features, which can realize the modulation recognition of 13 analog and digital signals\cite{AzzouzE1996Pfar,alma991031688625105106}. 

\subsubsection*{(3) Cyclic spectrum features}
Communication signals often modulate certain parameters of periodic signals which are to be transmitted, so its first-order or second-order statistical characteristics (average value, correlation function, etc.) often show periodicity in time. This periodicity is not reflected in the power spectrum, but is shown as the correlation characteristics between different frequency bands, namely the cyclostationary characteristics. Since the cyclic spectrum correlation theory was proposed, studies have been done about applying the application of this method to signal modulation recognition. Modulation recognition based on cyclic spectrum mainly includes two parts: constructing recognition features and designing classifiers. The signal recognition features based on the cyclic spectrum have the following categories. One is the intuitive characteristics of the cyclic spectrum, such as the number, position, and intensity of the peaks of the cyclic spectrum axis section and the mean and variance of the characteristic surface; the other is the envelope characteristics of the cyclic spectrum, such as the envelope of the cyclic spectrum. The design of the classifier mainly includes decision trees, neural networks, support vector machines, hidden Markov processes and so on.

Since the 1950s, cyclostationary signal processing technology has entered an era of rapid development. In the late 1980s, the cyclic spectrum theory proposed by W.A. Gardner has been applied to the field of signal modulation recognition\cite{GardnerWilliamA1986Tsct}.
In 2017, Yan.X et al. proposed a graphical feature based on the periodicity and symmetry of the cyclic spectrum. Simulations show that this scheme has higher classification accuracy\cite{XiaoYan2017IRMC}.

\subsubsection*{(4) Wavelet transform features }
Wavelet transform is an important characteristic parameter that can simultaneously reflect the characteristics of time domain and frequency domain. It has different resolutions in different frequency bands, so as to achieve the adaptive analysis effect of time subdivision at high frequency and frequency subdivision at low frequency. Moreover, the existence of multiple types of wavelet functions can be applied to a variety of application scenarios, but the analysis effects of different wavelet functions are very different, and how to choose the best wavelet is a difficult point in current research.

\cite{LiangYe2018RoDS} designs and establishes an automatic recognition method of three typical digital communication signal modulation types based on wavelet transform and neural network. \cite{ChenJian2006Miod} Uses a new method of digital modulation identification with wavelet transform. The paper use two ways to get the characteristics. One is to get the local maximum with the continuous wavelet transform and the other is the multi resolution analysis. Such method the speed of modulation and the classification accuracy.

\subsection{Modulation recognition based on support vector machine (SVM)}
In recent years, the support vector machine (SVM) developed on the basis of Vapnip's statistical learning theory, is becoming a new research hotspot in the field of machine learning. SVM is a structured machine learning method which has solved many problems like the model selection problem in neural networks, over-learning and under-learning problems, nonlinear and dimensional disaster problems, and local minimum points. Because of its excellent learning performance, it has been successfully applied in many fields, such as speech recognition, remote sensing image analysis, face recognition, automatic text classification and so on. In the field of pattern recognition, in general, statistical learning theory and support vector machine determine the classification surface based on the limited learning sample so that the expected risk is minimized when the unknown sample is estimated\cite{SunJian-Cheng2006Mamr}.

\subsection{Non-orthogonal multiple access (NOMA)}
In the past 20 years, with the rapid development of mobile communication technology and the continuous evolution of technical standards, the fourth-generation mobile communication technology (4G) based on orthogonal frequency division multiple access technology (OFDMA) is proposed \cite{cellular1,cellular2,cellular3,cellular4,cellular5,cellular6,cellular7,cellular8,cellular9,cellular10,cellular11,cellular12,cellular13,cellular14,MIMO_capacity}, and its data service transmission rate has reached hundreds of megabits or even gigabits per second which can meet the needs of broadband mobile communications applications to a greater extent in the future. However, as the demand for popularization of smart terminals and mobile new services continues to increase, the demand for wireless transmission rates is increasing exponentially, and the transmission rate of wireless communications will still be difficult to meet the application requirements of future mobile communications. 5G is positioned in a wireless network with higher spectrum efficiency, faster speed, and larger capacity, where the spectrum efficiency needs to be increased by 5 to 15 times compared with 4G.

While achieving good system throughput, in order to keep the low cost of reception, orthogonal multiple access technology is adopted in 4G. However, in response to the need for 5G spectrum efficiency to increase by 5 to 15 times, the industry proposes to adopt a new multiple access multiplexing method, namely NOMA. NOMA is different from previous multiple access technologies. NOMA uses non-orthogonal power domains to distinguish users. The so-called non-orthogonal means that data between users can be transmitted at the same time and frequency, and users can be distinguished only by the difference in power. NOMA uses non-orthogonal transmission at the transmitter to actively import interference information, and at the receiver, correct demodulation is achieved through serial interference cancellation technology. Compared with orthogonal transmission, the receiver complexity is improved, but higher spectrum efficiency can be obtained. The basic idea of non-orthogonal transmission is to use complex receiver design in exchange for higher spectrum efficiency.

In this paper, we first introduce likelihood-based and feature-based modulation recognition methods. In the feature-based modulation simulation part, instantaneous feature, cyclic spectrum, high-order cumulants, and wavelet transform features are used as the entry point, and six digital signals including $2\mathrm{ASK} , 4\mathrm{ASK}, \mathrm{BPSK}, \mathrm{QPSK}, 2\mathrm{FSK}$ and $4\mathrm{FSK}$ are simulated, showing the difference of signals in multiple dimensions.

Then, in the recognition of mixed signal modulation based on likelihood value part, this paper divides the three kinds of digital signals $\{\mathrm{BPSK}, \mathrm{QPSK}, 8\mathrm{PSK}\}$ into six mixed signals according to the signal power ratio, which is equal to 2:1, namely $\{2\mathrm{BPSK}+\mathrm{QPSK}, 2\mathrm{BPSK}+8\mathrm{PSK}, 2\mathrm{QPSK}+\mathrm{BPSK}, 2\mathrm{QPSK}+8\mathrm{PSK}, 2\mathrm{PSK}8+\mathrm{BPSK}, 2\mathrm{PSK}8+\mathrm{QPSK}\}$, and uses traditional manually-designed classification rules to do the mixed-signal modulation recognition to obtain the corresponding accuracy curve; in the recognition of mixed signal based on feature extraction part, this paper divides three digital signals $\{4\mathrm{ASK}, \mathrm{QPSK}, 4\mathrm{FSK}\}$ into six mixed signals according to the signal power ratio, which is equal to 2:1, namely $\{2\mathrm{ASK}4+\mathrm{PSK}4,2\mathrm{ASK}4+\mathrm{FSK}4,2\mathrm{PSK}4+\mathrm{ASK}4,2\mathrm{PSK}4+\mathrm{FSK}4,2\mathrm{FSK}4+\mathrm{ASK}4 , 2\mathrm{FSK}4+\mathrm{PSK}4\}$, and uses the decision tree classifier to perform the mixed signal modulation classification according to the above signal features to obtain the corresponding accuracy.

Besides, this paper studies the machine learning modulation recognition technology based on support vector machines (SVM), and uses $\{2\mathrm{ASK}, 4\mathrm{ASK}, \mathrm{BPSK}, \mathrm{QPSK}, 2\mathrm{FSK}, 4\mathrm{FSK}\}$ six single signals and $\{2\mathrm{ASK}4+\mathrm{PSK}4,2\mathrm{ASK}4+\mathrm{FSK}4,2\mathrm{PSK}4+\mathrm{ASK}4,2\mathrm{PSK}4 +\mathrm{FSK}4,2\mathrm{FSK}4+\mathrm{ASK}4,2\mathrm{FSK}4+\mathrm{PSK4}\}$ six mixed signals to do the modulation recognition and classification. Theoretical analysis and simulation results show that a well-designed and trained SVM can perform feature extraction and do the signal recognition at the same time, and SVM classification recognition has better classification accuracy compared to the manually designed classification rules.

Finally, this paper select features from above mentioned methods to do the feature combination and puts the combined features as new features into SVM for signal modulation recognition and finds the feature combination with the highest recognition rate.


In the following, we will give a detailed introduction to the method of signal modulation recognition. Section \ref{sec:LMR} will introduce the single and mixed signal models based on the ALRT method and its basic theories. Section \ref{sec:feature} will introduce the single and mixed signal models based on the feature extraction method and its basic theories, and do the simulation based on six kinds of single signals \{2ASK, 4ASK, BPSK, QPSK, 2FSK, 4FSK\} and six kinds of mixed signals $\{2\mathrm{ASK}4+\mathrm{PSK}4,2\mathrm{ASK}4+\mathrm{FSK}4,2\mathrm{PSK}4+\mathrm{ASK}4,2\mathrm{PSK}4+\mathrm{FSK}4,2\mathrm{FSK}4+ASK4,2\mathrm{FSK}4+\mathrm{PSK4}\}$. In Section \ref{sec:SVM}, the basic theory of SVM classification method will be introduced.

\section{Likelihood-based Modulation Recognition}
\label{sec:LMR}
\subsection{Signal model}
\subsubsection{Single signal}
Consider that the signal received by the receiver is down-converted to obtain a baseband signal, where the complex envelope of the baseband signal is:
\begin{equation}
r(t)=s\left(t ; u_{k}\right)+g(t)
\end{equation}
where $g(t)$ is the mean value which is zero. The double-sideband power spectral density of complex Gaussian white noise is $N_{0}$. $s(t;u_{k})$ is the comprehensive representation of the modulated digital signal without noise interference. Its specific representation is as follows:
\begin{equation}
s\left(t ; u_{k}\right)=a e^{j\left(2 \pi \Delta f+\theta_{c}\right)} \sum_{n=0}^{L-1} s_{n}^{k, i} e^{j \theta_{n}} v(t-n T-\phi T)
\end{equation}
where $u_{k}$ represents a multi-dimensional variable set, namely the statistical values of unknown signal-related variables and channel-related variables under the k-th modulation system,
\begin{equation}
u_{k}=\left\{a, \theta_{c}, \Delta f,\left\{s^{k, i}\right\}_{i=1}^{M_{k}}, h(t), \phi\right\}
\end{equation}
The physical meanings of the above variables are:
\begin{itemize}
  \item [1)] $a$ represents the amplitude of the signal.
  \item [2)] $\theta_{c} $represents the fixed phase offset caused by the initial phase of the carrier and the propagation delay.
  \item [3)] $N$ represents the number of symbols in the observation interval.
  \item [4)] $s^{k, i}_{n}$ represents the i-th constellation point under the k-th modulation system, where $i \in\left\{1, \ldots, M_{k}\right\}, \quad k \in\{1, \ldots, C\}$. The probability distribution function of each constellation point is the same. $M_{k}$ represents the number of symbol types of the system the k-th modulation. $C$ represents the number of types of modulation systems concerned. Generally, the power of modulated symbols needs to be normalized. The normalization method is as follows:
\begin{equation}
\frac{1}{M_{k}} \sum_{i=1}^{M_{k}}\left|s^{k, i}\right|^{2}=1
\end{equation}
  \item [5)] $\Delta f$ indicates the frequency offset caused by the down conversion process.
  \item [6)] $\theta_{n}$ indicates phase jitter.
  \item [7)] $T$ indicates the symbol period.
  \item [8)] $\phi$ indicates symbol timing offset, $0 \leq \phi<1$
  \item [9)] $v(t)$ represents the common influence of the channel influence $h(t)$ and the pulse forming function $p(t)$, $v(t)=h(t)*p(t)$ , where * represents the convolution operation. Usually the pulse shaping function $p(t)$ is the root mean square raised cosine roll-off filter whose period is $T$.
\end{itemize}

$r(t)$ obtained above is a continuous signal in the time domain.  Discretizing $r(t)$ for subsequent modulation recognition processing, using traditional signal detection theory, $r(t)$ can be expressed as a discrete sampling sequence $r$ , $\boldsymbol{r}=[r(1), \ldots, r(L)]$. The signal sequence after passing the matched filter is as follows:
\begin{equation}
r(n)=\frac{1}{T} \int_{n T}^{(n+1) T} r(t) p(t-n T) d t
\end{equation}
For the discrete sampling sequence $r$ obtained in the above process , it is usually called a symbol sequence. After further transformation, we get the transmitted signal model, which is:
\begin{equation}
r(n)=e^{j 2 \pi f_{0} T n+j \theta_{n}} \sum_{\ell=0}^{L-1} s(\ell) h(n T-\ell T-\epsilon T)+g(n).
\end{equation}
Here $s(\ell)$ is the transmitting symbol sequence, $h(\cdot)$ is the channel response function, $T$ is the symbol interval, $\epsilon(0 \leq \epsilon<1)$ is the synchronization error, $f_{0}$ is the frequency offset, $\theta_{n}$ is the phase jitter, $g(t)$ is the noise, and $\sum_{\ell=0}^{L-1} s(\ell) h(n T-\ell T)$ is the inter-symbol interference.

In cooperative communications, it can be assumed that the symbol interval $T$ is known, the signal is synchronized, and the phase jitter $\theta_{n}$ is eliminated by preprocessing. Finally, the frequency offset, channel response, and noise interference are retained, and the symbol sequence used can be simplified as follows:
\begin{equation}
r(n)=e^{j 2 \pi f_{0} T n} s(n) h(n)+g(n)
\end{equation}
\subsubsection{ Mixed signal}
In the mixed-signal modulation recognition part, the recognition is conducted two cochannel signals received by a single receiver. Each signal can be denoted as:
\begin{equation}
r_{i}(n)=e^{j 2 \pi f_{0} T_{i}} S_{i}(n) h_{i}(n), \quad i=1,2
\end{equation}
Under the condition of two cochannel signals, received signal can be modeled as:
\begin{equation}
r(t)=\Sigma_{i=1}^{2} S_{i}(n) h_{i}(n)+g(n)
\end{equation}
\subsection{Average likelihood ratio test algorithm}
The likelihood-based method models the modulation recognition problem as a complex hypothesis testing problem. It often depends on the probability distribution of unknowns in the model. Unknowns are usually amplitude, phase, noise, etc. in modulation recognition. The number of assumptions for this problem is usually equal to the number of signal modulation types in the signal set. Under the assumption $H_{i}$ , $i$ represents that the received signal belongs to the i-th modulation type, and the probability distribution of the received signal can be determined by the probability distribution of the unknown quantity or can be approximated by the estimate of the unknown quantity. 

The ALRT method is one of the detection methods based on LB. ALRT treats the unknown as a random variable. The probability distribution function of the received signal under each assumption is calculated by their average, and each unknown requires a known distribution. If the assumed distribution of the unknown is consistent with the actual distribution, the ALRT method will obtain the maximum accuracy probability of the classification problem. Usually based on the assumption that the distribution of unknown quantities is idealized based on the ALRT method, the upper bound of the modulation recognition classification problem is obtained. The problem is that, with the increase of the number of unknowns, computational complexity of ALRT will be high, even mathematically impossible in some cases \cite{WenWei2000Mcfd}.

Consider the signal model as the baseband symbol sequence output by the matched filter:
\begin{equation}
r(n)=e^{j 2 \pi f_{0} T n} s(n) h(n)+g(n)
\end{equation}
where $s(n)$ is the transmitted symbol sequence, $h(n)$ is the channel response function, $ f_{0}$ is the frequency offset, and $g(n)$ is the noise.
The transmitted signal sequence $s=[s(1), s(2), \ldots, s(N)]$ is generated by the $i-th$ modulation type $M_{i}$ , that is assumption $H_{i}$.
The probability distribution of the received signal sequence in this section is:
\begin{equation}
f\left(\boldsymbol{r} \mid H_{i}\right)=\prod_{n=1}^{N} f\left(r(n) \mid H_{i}\right)
\end{equation}
where $f\left(\boldsymbol{r} \mid H_{i}\right)$ is the PDF of the signal $r(n)$ , and further suppose that the total number of categories in the classified signal set is $L_{n}$ and the channel state is known, then the likelihood function is:
\begin{equation}
f\left(r(n) \mid H_{i}\right)=\frac{1}{L_{m}} \sum_{l=1}^{L_{m}} \frac{1}{\pi \sigma_{\omega}^{2}} \exp \left\{-\frac{\left|r(n)-A e^{j 2 \pi f_{0} T n+j \theta_{n}} s_{i l}\right|}{\sigma_{\omega}^{2}}\right\}
\end{equation}
where $s_{i l}$ is the l-th transmitted signal under the hypothesis $H_{i}$ of the i-th modulation category $M_{i}$ and the prior probability of each hypothesis $H_{i}$ is $\kappa_{i}$ , which satisfies $\sum_{i=1}^{M} \kappa_{i}=1$. According to Bayes' criterion, the posterior probability of $H_{i}$ is:
\begin{equation}
P\left(H_{i} \mid \boldsymbol{r}\right)=\frac{f\left(\boldsymbol{r}, H_{i}\right)}{f(\boldsymbol{r})}=\frac{\kappa_{i} f\left(\boldsymbol{r} \mid H_{i}\right)}{f(\boldsymbol{r})}
\end{equation}
Finally, according to the maximum likelihood decision criterion, the decision process is as follows:

\centerline{Accept $H_{k}$ if $\kappa_{k} f\left(r \mid H_{k}\right) \geq \kappa_{i} f\left(r \mid H_{i}\right), \forall k !=i$}

\section{Feature-based Modulation Recognition} \label{sec:feature}
\subsection{Signal model}
\subsubsection{Single signal}
Digital baseband signals cannot propagate in channels with bandpass characteristics. In order to match the two, the carrier must be modulated with digital baseband signals, and the carrier characteristics can be changed by changing one or more of the carrier parameters. Different types of modulation signals can be obtained by changing different carrier descriptions. This section will introduce the digital models of MASK, MPSK and MFSK
\paragraph{Amplitude shift keying modulation (MASK)}\mbox{}\\
Amplitude shift keying modulation adopts the method of changing the amplitude of the carrier to transmit digital information, and the phase and frequency of the carrier are not changed. In M-ary modulation, the carrier amplitude has M values. The signal expression of $MASK$ is:
\begin{equation}
\begin{aligned} x_{A S K} &=a(t) \cos \left(2 \pi f_{c} t+\phi_{0}\right) \\ &=\sum_{n=-\infty}^{\infty} a_{n} q\left(t-n T_{d}\right) \cos \left(2 \pi f_{c} t+\phi_{0}\right) \end{aligned}
\end{equation}
where $ f_{c}$ is the carrier frequency, $\phi_{0}$ is the initial phase, $T_{d}$ is the symbol interval, and $\boldsymbol{a_{n}}={0,1,2, \ldots, M-1}$ is the level of the $n-th$ symbol.
\paragraph{Phase shift keying modulation (MPSK)}\mbox{}\\
Phase shift keying adopts the method of changing the phase state of the carrier signal to transmit digital information without changing the amplitude and frequency of the carrier. The $M$ values of the baseband symbols correspond to the $M$ phases of the carrier, and the signal expression of $MPSK$ is:
\begin{eqnarray}
x_{P S K}(t)&=&\cos \left(2 \pi f_{c} t+\phi(t)+\phi_{0}\right)\nonumber \\
\phi(t) 
&=&\sum_{n=-\infty}^{\infty} \theta_{n} q\left(t-n T_{d}\right)
\end{eqnarray}
where $ f_{c}$ is the carrier frequency, $\phi_{0}$ is the initial phase, $T_{d}$ is the symbol interval, $\theta_{n}$ is $M$ phases uniformly distributed between $[0,2\pi]$ , and the phase difference is $2\pi/M$.
\paragraph{Frequency shift keying modulation (MFSK)}\mbox{}\\
Frequency shift keying adopts the method of changing the carrier frequency to transmit digital information, without changing the amplitude and phase of the carrier. The expression of $MFSK$ is:
\begin{eqnarray}
x_{FMSK}(t)&=&
p_{k} \cos \frac{\pi t}{2 T_{d}} \cos \left(2 \pi f_{c} t+\phi_{0}\right)
\nonumber \\
&& -q_{k} \sin \frac{\pi t}{2 T_{d}} \sin \left(2 \pi f_{c} t+\phi_{0}\right)
\end{eqnarray}
In the formula: $(k-1) T_{d} \leq t \leq k T_{d}, \quad p_{k}=\pm 1, \quad q_{k}=a_{k} p_{k}=\pm 1$. $f_{c}$is the carrier frequency, $\phi_{0}$ is the initial phase. $\cos \left(\pi t / 2 T_{d}\right)$ and $\sin \left(\pi t / 2 T_{d}\right)$ is referred to as a weighting function.
\subsubsection{Nonlinear phase extraction of the digital signal}
The general expression of the digital signal $u(t)$ is:
\begin{equation}
u(t)=a(t) \cos \left[2 \pi f_{c}(t) t+\varphi_{N L}(t)\right]
\end{equation}
Where $ a(t)$ is the instantaneous amplitude of the signal, $ f_{c}(t)$ is the carrier frequency, $\varphi_{N L}(t)$ is the nonlinear phase of the signal.
Let $v(t)$ denotes the Hilbert transform of $u(t)$, then:
$v(t)=u(t) * \frac{1}{\pi t}=\int_{-\infty}^{\infty} u(\tau) \frac{1}{\pi(t-\tau)} \mathrm{d} \tau$
Then the analytical expression $z(t)$ of the digital signal is:
\begin{equation}
z(t)=u(t)+jv(t)
\end{equation}
According to the signal analysis expression, the instantaneous information statistics can be obtained. The specific calculation formula is as follows:
\begin{enumerate}
\item  instantaneous amplitude
\begin{equation}
a(t)=\sqrt{u^{2}(t)+v^{2}(t)}
\end{equation}
\item  instantaneous phase 
\begin{equation}
\varphi(t)=\arctan \left(\frac{v(t)}{u(t)}\right)
\end{equation}
\item  instantaneous frequency 
\begin{equation}
f(t)=\frac{1}{2 \pi} \frac{\mathrm{d}}{\mathrm{d} t}[\varphi(t)]
\end{equation}
\end{enumerate}
According to the formula $u(t)$, the instantaneous phase expression $\varphi(t)$ can be obtained:
\begin{equation}
\varphi(t)=2 \pi f_{c} t+\phi_{N L}(t)
\end{equation}
In the formula, $f_{c}$ and $\varphi(t)$ have the same meaning in the above formula , and the phase sequence after A/D sampling is:
\begin{equation}
\varphi(i)=2 \pi f_{c} / f_{s} i+\phi_{N L}(i)
\end{equation}
In the above formula, the instantaneous phase is divided into two parts: the linear part $2 \pi f_{c} / f_{s} i$ that changes with time and the nonlinear part $\phi_{N L}(i)$, but the actual situation is that the instantaneous phase $\phi_{i}$ extracted by the Hilbert transform is limited, because the actual instantaneous phase is mod by $2pi$, the value range of $\phi_{i}$ is limited to $[-pi,pi]$. When $f_{z}$ and $f_{z}$ satisfy relation $ f_{z}/ f_{z}=n_{1}$, once the instantaneous phase difference of two adjacent samples is much larger than $2pi/n_{1}$, it is necessary to add a modified phase sequence to restore the original phase information.
First, calculate the corrected phase sequence $C(i)$:
\begin{equation}
C(i)=\left\{\begin{array}{l}C(i-1)+2 \pi, \text { if } \hat{\varphi}(i-1)-\hat{\varphi}(i)>\pi \\ C(i-1)-2 \pi, \text { if } \hat{\varphi}(i)-\hat{\varphi}(i-1)>\pi \\ C(i-1), \text { others }\end{array}\right.
\end{equation}
The corrected phase sequence is expressed as:
\begin{equation}
\phi(i)=\varphi(i)+C(i)
\end{equation}
Finally, the nonlinear phase expression result can be obtained:
\begin{equation}
\phi_{N L}(i)=\phi(i)-2 \pi f_{c} / f_{s} i
\end{equation}
\subsubsection{Mixed signal}
In the mix-signal modulation recognition part, the recognition is conducted two cochannel signals received by a single receiver. Each signal can be denoted as:
\begin{equation}
S(t)=\sqrt{E} \sum_{n} a_{n} g\left(t-n T_{s}\right) \mathrm{e}^{\mathrm{j}\left(\omega_{c} t+\theta_{c}\right)}+n(t)
\end{equation}
Where $E$ is the energy of the transmitted symbol waveform, $a$ is the symbol sequence, $g$ is the transmitted symbol waveform, $T$ is the symbol duration, $\omega_{c}$ and $\theta_{c}$ are respectively frequency and phase of the carrier, and $n(t)$ represents the additive white gaussian noise with zero mean. 
Under the condition of two cochannel signals, received signal can be modeled as:
\begin{equation}
s(t)=\sum_{i=1}^{2} s_{i}(t)+n(t)
\end{equation}
\subsection{Higher order cumulants}
\subsubsection{Definitions of higher-order cumulants and higher-order moments}
Suppose the probability density of a random variable $x$ is $f(x)$ , and its characteristic function $\phi(\omega)$ is defined as:
\begin{equation}
\phi(\omega)=\int_{-\infty}^{+\infty} f(x) \mathrm{e}^{\mathrm{j} \omega x} \mathrm{d} x
\end{equation}
In the formula: the characteristic function $\phi(\omega)$ is also called the Fourier transform of the probability density $f(x)$.
More generally, the common form of the characteristic function is:
\begin{equation}
\phi(s)=E\left[\mathrm{e}^{s x}\right]=\int_{-\infty}^{+\infty} f(x) \mathrm{e}^{s x} \mathrm{d} x
\end{equation}
Find the $k-th$ derivative:
\begin{equation}
\phi^{k}(s)=E\left[x^{k} \mathrm{e}^{s x}\right]
\end{equation}
Then the $k-th$ derivative of $\phi(s)$ at the origin is:
\begin{equation}
\phi^{k}(0)=E[x^{k}]=m_{k}
\end{equation}
Where $m_{k}$ is the $k-th$ moment of $x$.
Define $\phi(s)$ as the the first characteristic function of random variable $x$, which is also called moment generating function. The second characteristic function of $x$ can be obtained by taking the logarithm of $\phi(s)$, which is also called the cumulant generating function $\psi(s)$ , defined as
\begin{equation}
\psi(s)=\ln \phi(s)
\end{equation}
Where $c_{k}$ is the $k-th$ order cumulant of $x$ , which is defined as:
\begin{equation}
c_{k}=\frac{1}{j^{k}}\left[\frac{\mathrm{d}^{k} \psi(s)}{\mathrm{d} s^{k}}\right]_{s=0}
\end{equation}

The above discussion is the case of a single random variable. More generally, let $x=\left[x_{1}, x_{2}, \cdots, x_{n}\right]$ be a stationary random process, and its characteristic function is defined as:
\begin{equation}
\phi\left(\omega_{1}, \omega_{2}, \cdots, \omega_{n}\right)=E\left\{\exp \left[j\left(\omega_{1} x_{1}+\omega_{2} x_{2}+\cdots+\omega_{n} x_{n}\right)\right]\right\}
\end{equation}

Then the joint $r=k_{1}+k_{2}+\cdots+k_{n}$ moment of $x=\left[x_{1}, x_{2}, \cdots, x_{n}\right]$ is:
\begin{eqnarray}
&\operatorname{mom}\left[x_{1}^{k_{1}}, x_{2}^{k_{2}}, \cdots, x_{n}^{k_{n}}\right]=E\left\{x_{1}^{k_{1}}, x_{2}^{k_{2}}, \cdots, x_{n}^{k_{n}}\right\} \nonumber\\
&\quad=(-j)^{r}\left[\frac{\partial \phi\left(\omega_{1}, \omega_{2}, \cdots, \omega_{n}\right)}{\partial \omega_{1}^{k_{1}}, \partial \omega_{2}^{k_{2}}, \cdots, \partial \omega_{n}^{k_{n}}}\right]_{\omega_{1}=\omega_{2}=\cdots=\omega_{n}=0}
\end{eqnarray}

Then the joint $r=k_{1}+k_{2}+\cdots+k_{n}$ order cumulant of $x=\left[x_{1}, x_{2}, \cdots, x_{n}\right]$ is:
\begin{eqnarray}
&\operatorname{cum}\left[x_{1}^{k_{1}}, x_{2}^{k_{2}}, \cdots, x_{n}^{k_{n}}\right] \nonumber\\
&=(-j)^{r}\left[\frac{\partial\left[\ln \phi\left(\omega_{1}, \omega_{2}, \cdots, \omega_{n}\right)\right]}{\partial \omega_{1}^{k_{1}}, \partial \omega_{2}^{k_{2}}, \cdots, \partial \omega_{n}^{k_{n}}}\right]_{\omega_{1}=\omega_{2}=\cdots=\omega_{n}=0}
\end{eqnarray}
where $\ln \phi\left(\omega_{1}, \omega_{2}, \cdots, \omega_{n}\right)$ is the cumulative generating function of $x=\left[x_{1}, x_{2}, \cdots, x_{n}\right]$.

\subsubsection{The relationship between higher-order cumulants and higher-order moments}
Let $x(k)$ be a complex stationary random process with an average value of 0 , then the second-order cumulant of $x(k)$ is:
\begin{align}
 C_{20}\left(l_{1}\right)=E\left[x(k) x\left(k+l_{1}\right)\right]
\\ C_{21}\left(l_{1}\right)=E\left[x(k) x^{*}\left(k+l_{1}\right)\right]
\end{align}
The fourth-order cumulant of $x(k)$ is:
\begin{align}
C_{40}\left(l_{1}, l_{2}, l_{3}\right)=\operatorname{cum}\left[x(k), x\left(k+l_{1}\right), x\left(k+l_{2}\right), x\left(k+l_{3}\right)\right]
\\C_{41}\left(l_{1}, l_{2}, l_{3}\right)=\operatorname{cum}\left[x(k), x\left(k+l_{1}\right), x\left(k+l_{2}\right), x^{*}\left(k+l_{3}\right)\right]
\\C_{42}\left(l_{1}, l_{2}, l_{3}\right)=\operatorname{cum}\left[x^{*}(k), x\left(k+l_{1}\right), x\left(k+l_{2}\right), x^{*}\left(k+l_{3}\right)\right]
\end{align}
The sixth-order cumulant of  $x(k)$ is:
\begin{eqnarray}
&C_{60}\left(l_{1}, l_{2}, l_{3}, l_{4}, l_{5}\right)
= \nonumber\\
&\operatorname{cum}\left[x(k), x\left(k+l_{1}\right), x\left(k+l_{2}\right), x\left(k+l_{3}\right), x\left(k+l_{4}\right), x\left(k+l_{5}\right)\right] \nonumber 
\end{eqnarray}
where $E[\cdot]$ is the mean value calculation, $x(k+l_{i})$ is the time delay of $x(k)$, and the time delay is $l_{i}$ , and $x^{*}\left(k+l_{i}\right)$ is the conjugate of $x(k+l_{i})$ .
If the stationary random process $x(k)$ is independent and identically distributed, when $l_{1}=l_{2}=l_{3}=l_{4}=l_{5}=l_{6}=l_{7}=0$, the higher-order cumulant can be expressed in the form of higher-order moments as:
\begin{align}
C_{20}&=M_{20}  &&  \\
C_{21}&=M_{21}  &&  \\
C_{40}&=M_{40}-3 M_{20}^{2}  &&  \\
C_{41}&=M_{41}-3 M_{21} M_{20} &&  \\
C_{42}&=M_{42}-\left|M_{20}\right|^{2}-2 M_{21}^{2} &&  \\
C_{60}&=M_{60}-15 M_{40} M_{20}+30\left(M_{20}\right)^{3} &&  \\
C_{63}&=M_{63}-9 C_{42} C_{21}-6 C_{21}^{3} &&  \\
C_{80}&=M_{80}-28 M_{20} M_{60}-35 M_{40}^{2}+420 M_{40} M_{20}^{2}-630 M_{20}^{4}
\end{align}
In the formula, $M_{p q}=E\left[x(k)^{p-q} x^{*}(k)^{q}\right]$ is the $p-th$ order mixing moment of $x(k)$ .
\subsubsection{Simulation}
\paragraph{Single signal}\mbox{}\\
The theoretical values of each order cumulant of $\mathrm{MASK}$, $\mathrm{MPSK}$ and $\mathrm{MFSK}$ are shown in the table:
\begin{center}
\begin{tabular}{cccccc}
\hline
Signal Type& $|C_{40}|$& $|C_{41}|$& $|C_{42}|$& $|C_{60}|$& $|C_{63}|$ \\
\hline
2ASK& $2 E^{2}$& $2 E^{2}$& $2 E^{2}$& $16 E^{3}$& $16 E^{3}$\\
4ASK& $1.36 E^{2}$& $1.36 E^{2}$& $1.36 E^{2}$& $8.32 E^{2}$& $8.32 E^{2}$\\
BPSK& $2 E^{2}$& $2 E^{2}$& $2 E^{2}$& $16 E^{3}$& $16 E^{3}$\\
QPSK& $E^{2}$& $0$& $E^{2}$& $0$& $4 E^{3}$\\
2FSK& $0$& $0$& $E^{2}$& $0$& $4 E^{3}$\\
4FSK& $0$& $0$& $E^{2}$& $0$& $4 E^{3}$\\
\hline
\end{tabular}
\end{center}

\begin{figure}[htb]

\label{Simulation values of $C_{63}$}
\includegraphics[width=0.48\textwidth]{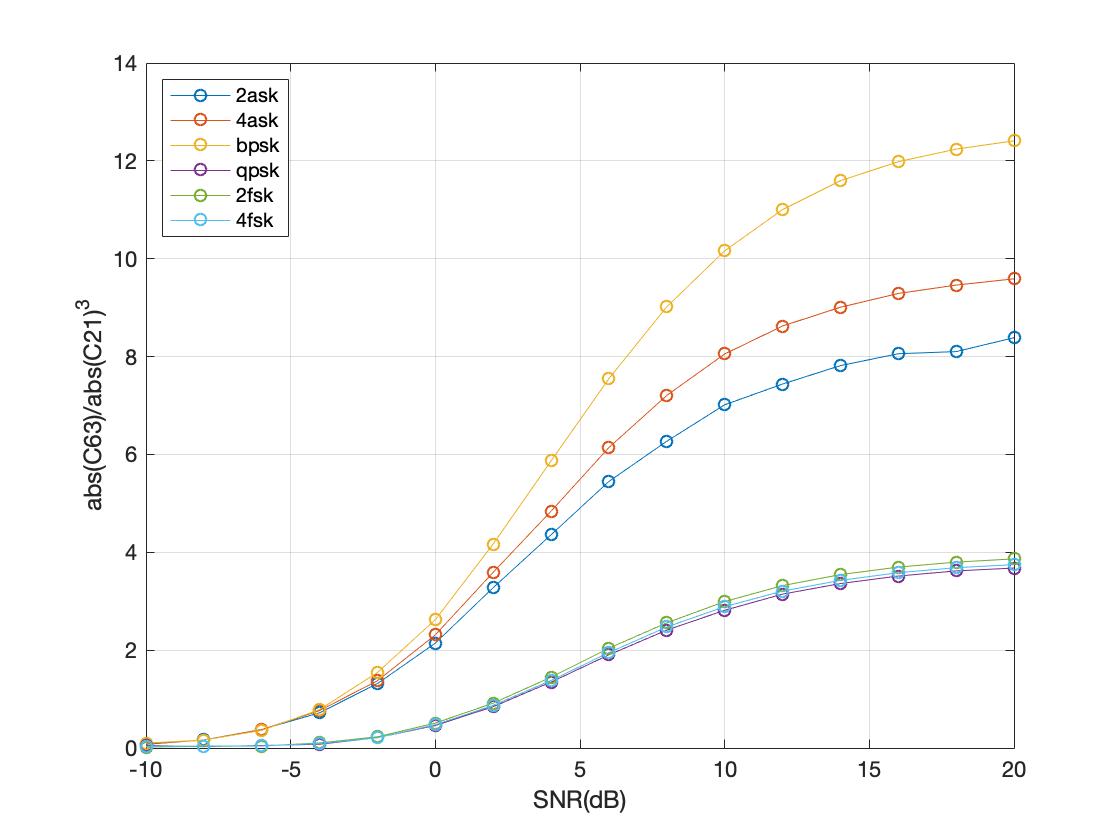}
\caption{Simulation values of higher order cumulants of single signal, $C_{63}$}
\label{Simulation values of higher order cumulants of single signal}
\end{figure}

In this section, $\{2\mathrm{ASK}, 4\mathrm{ASK}, \mathrm{BPSK}, \mathrm{QPSK}, 2\mathrm{FSK}, 4\mathrm{FSK}\}$ would be used to do the simulation and show the value of each features. Computer simulation is used to simulate the feature parameters.
The simulation uses MATLAB software and uses random sequence as signal model. The baseband signal is obtained after down-conversion, and then modulation recognition is performed. With carrier frequency of symbol modulation equal to 70Hz, sampling rate equal to 400Hz, symbol rate equal to 2 bps, number of symbols equal to 1000, noise chosen as Gaussian white noise and signal-to-noise ratio from -10 to 20dB, take the average of 500 simulations for each signal. The simulation results of the five feature parameters of $C_{40}$, $C_{41}$, $C_{42}$, $C_{60}$, $C_{63}$ are shown in Figs. \ref{Simulation values of C_{40}} - \ref{Simulation values of higher order cumulants of single signal}, respectively.

\begin{figure}[htb]
\includegraphics[width=0.48\textwidth]{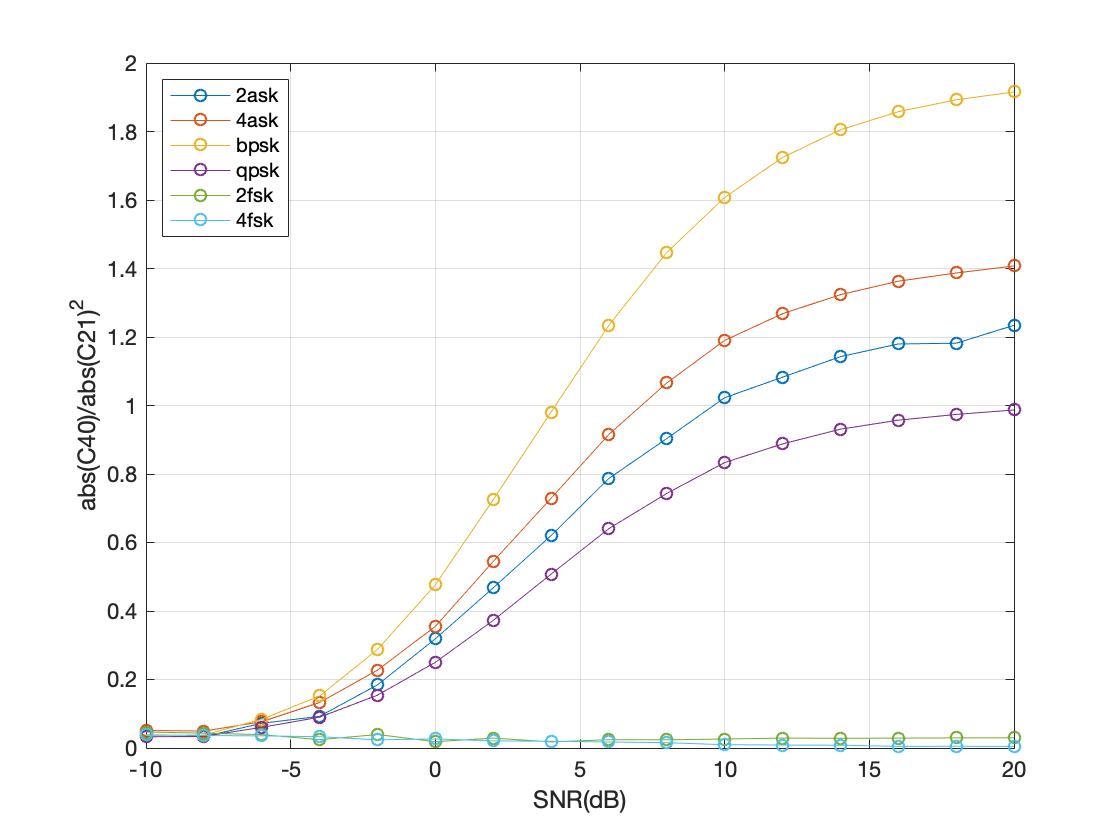}
\caption{Simulation values of higher order cumulants of single signal, $C_{40}$}
\label{Simulation values of C_{40}}
\end{figure}

\begin{figure}[htb]
\label{Simulation values of $C_{41}$}
\includegraphics[width=0.48\textwidth]{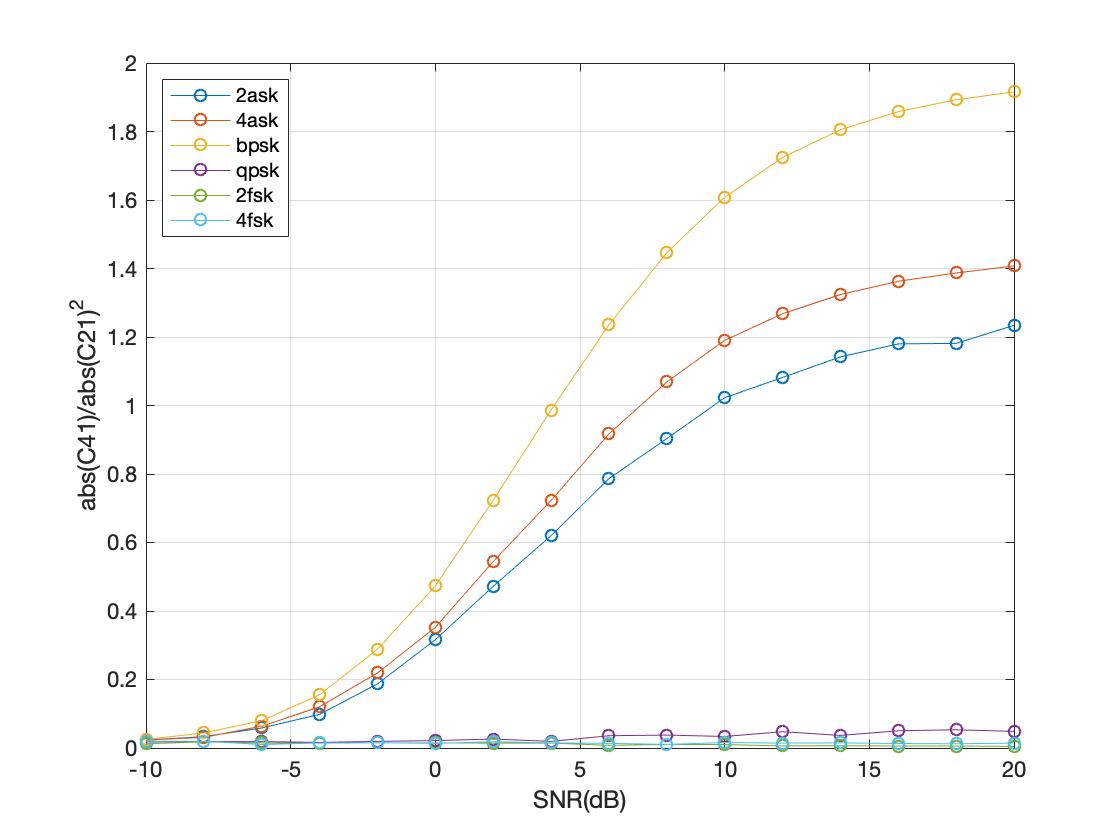}
\caption{Simulation values of higher order cumulants of single signal, $C_{41}$}
\end{figure}

\begin{figure}[htb]
\label{Simulation values of $C_{42}$}
\includegraphics[width=0.48\textwidth]{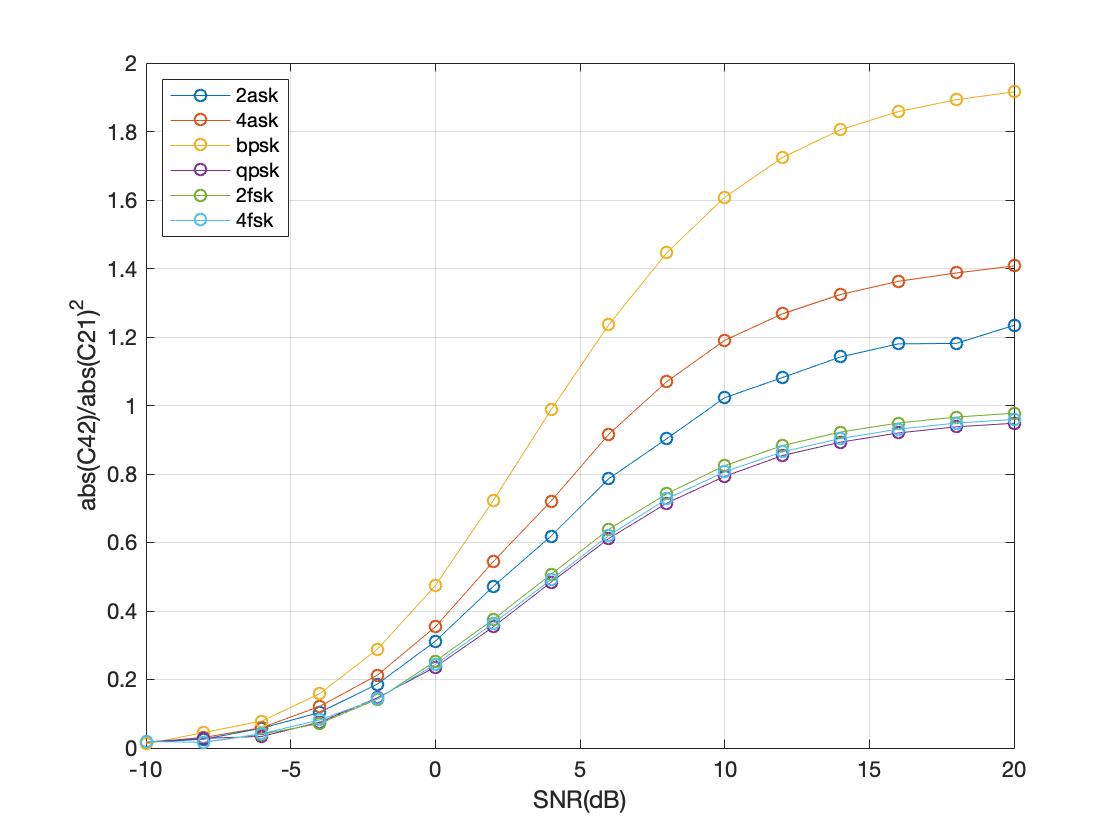}
\caption{Simulation values of higher order cumulants of single signal, $C_{42}$}
\end{figure}

\begin{figure}[htb]

\label{Simulation values of $C_{60}$}
\includegraphics[width=0.48\textwidth]{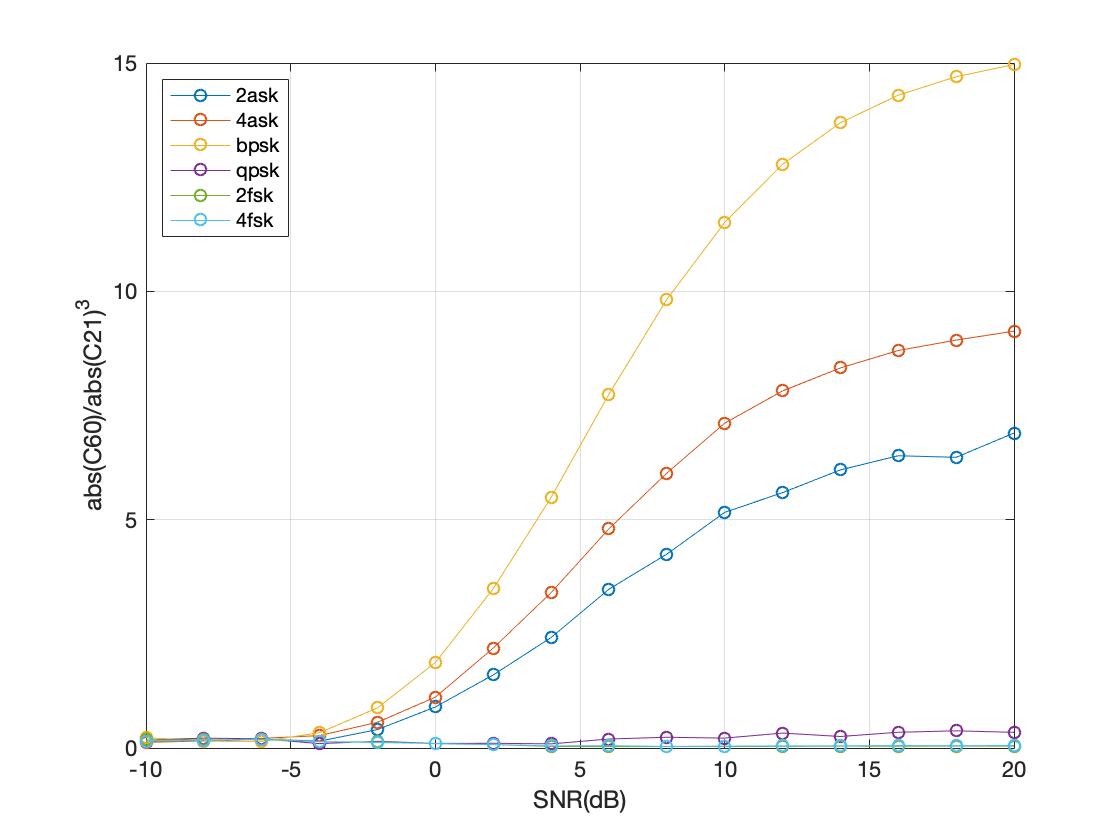}
\caption{Simulation values of higher order cumulants of single signal, $C_{60}$}
\end{figure}

\begin{figure}[htb]
\includegraphics[width=0.48\textwidth]{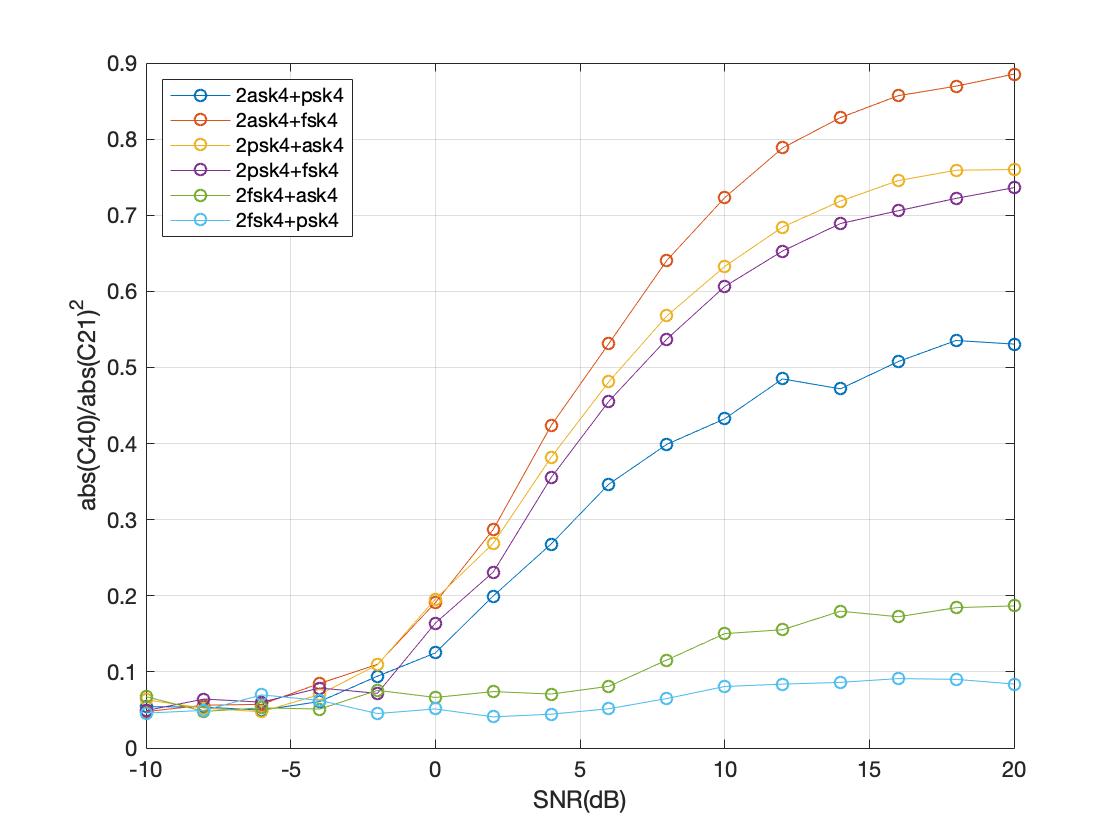}
\caption{Simulation values of higher order cumulants of mixed signal, $C_{40}$}
\label{MixedSimulationC40}
\end{figure}

\begin{figure}[htbp]

\includegraphics[width=0.48\textwidth]{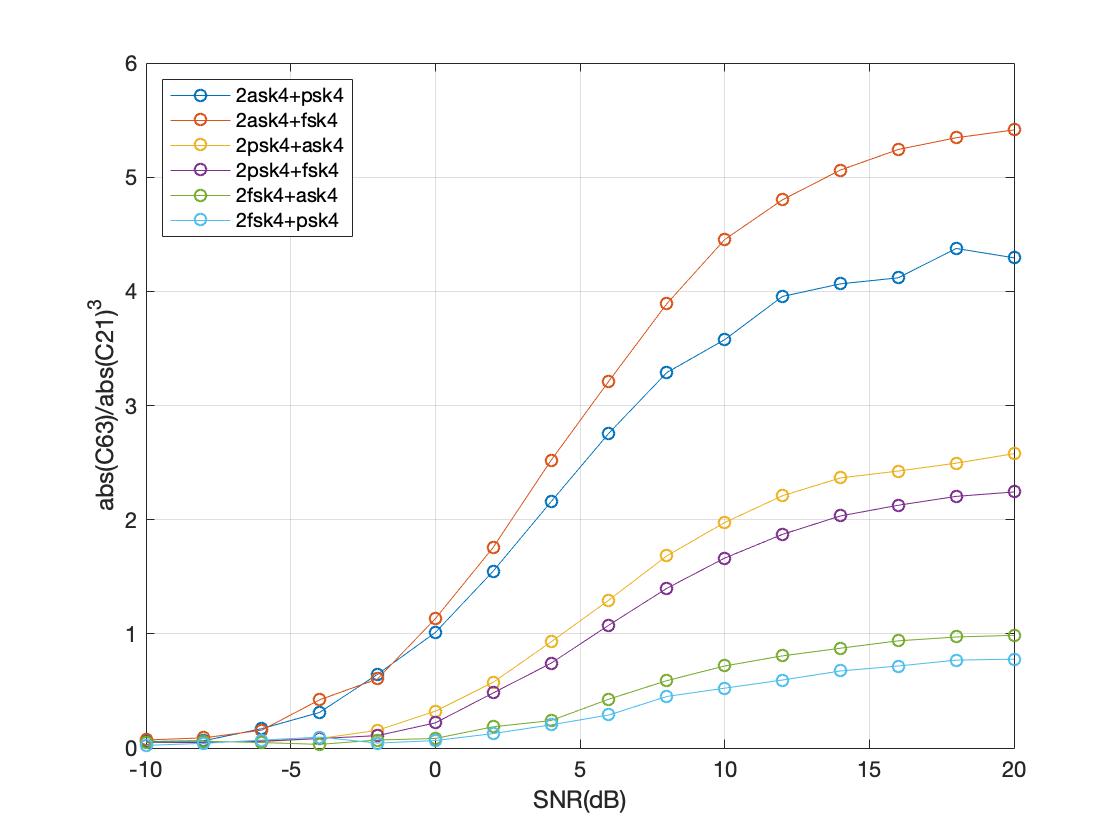}
\caption{Simulation values of higher order cumulants of mixed signal, $C_{63}$}
\label{Simulation values of higher order cumulants of mixed signal}
\end{figure}

\paragraph{Mixed signal}\mbox{}\\
In this section, three digital signals $\{4\mathrm{ASK}, \mathrm{QPSK}, 4\mathrm{FSK}\}$ are divided into six mixed signals according to the signal power ratio, which is equal to 2:1, namely $\{2\mathrm{ASK}4+\mathrm{PSK}4,2\mathrm{ASK}4+\mathrm{FSK}4,2\mathrm{PSK}4+\mathrm{ASK}4,2\mathrm{PSK}4+\mathrm{FSK}4,2\mathrm{FSK}4+\mathrm{ASK}4 , 2\mathrm{FSK}4+\mathrm{PSK}4\}$. The signals would be used to do the simulation and get the results of each value of different features. The parameters sets are the same as those in the single signal section.The simulation results of the five feature parameters of $C_{40}$, $C_{41}$, $C_{42}$, $C_{60}$, $C_{63}$ are shown in Figs. \ref{MixedSimulationC40} - \ref{Simulation values of higher order cumulants of mixed signal}, respectively.

\begin{figure}[htb]
\label{Simulation values of $C_{41}$}
\includegraphics[width=0.48\textwidth]{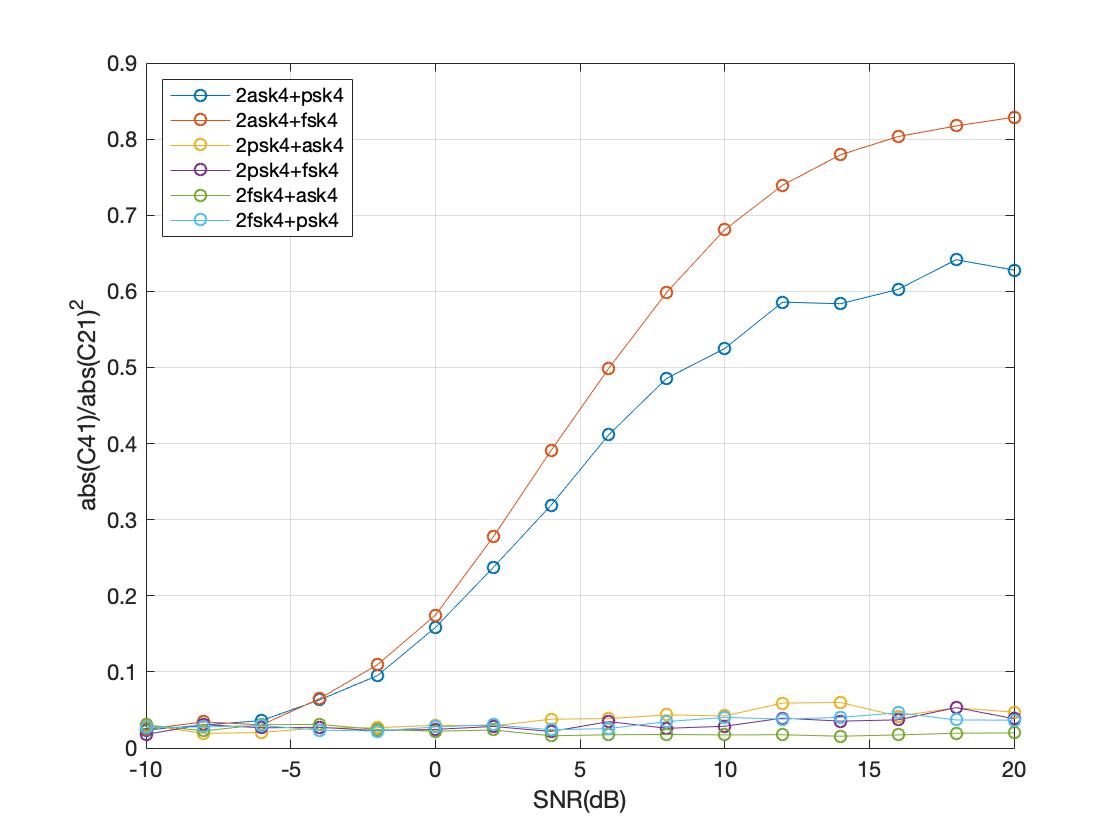}
\caption{Simulation values of higher order cumulants of mixed signal, $C_{41}$}
\end{figure}

\begin{figure}[htb]
\label{Simulation values of $C_{42}$}
\includegraphics[width=0.48\textwidth]{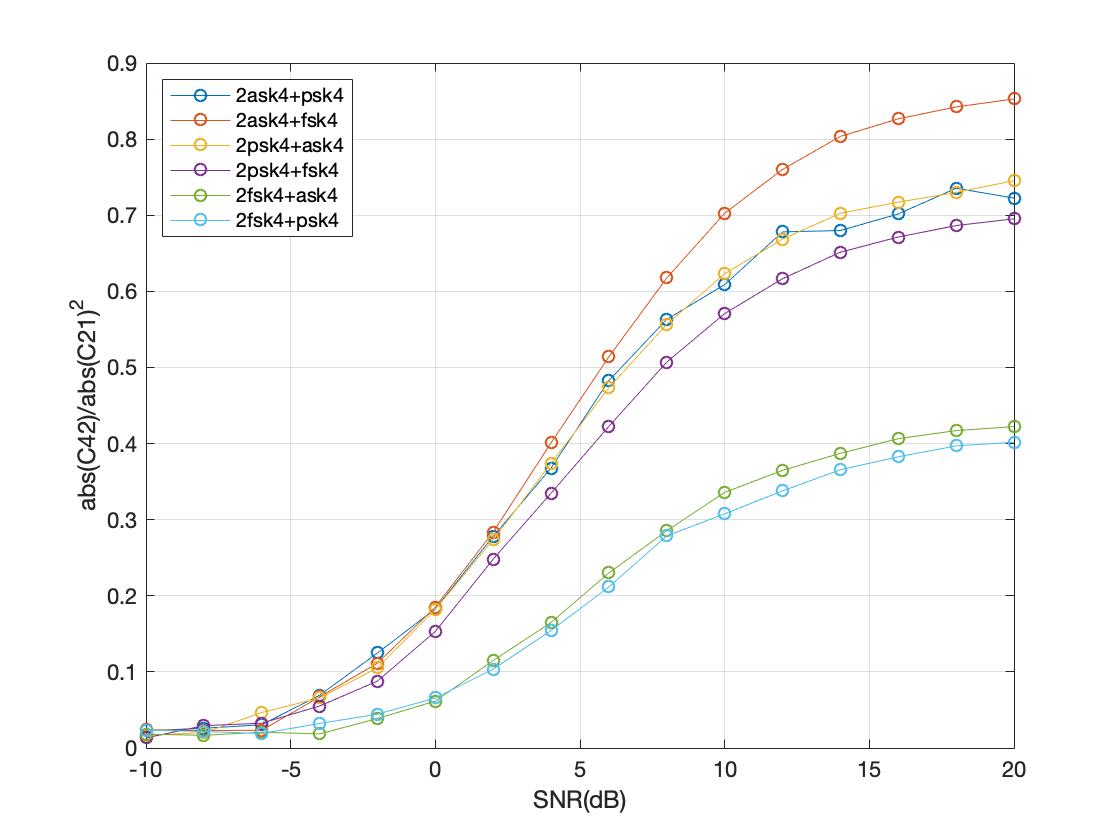}
\caption{Simulation values of higher order cumulants of mixed signal, $C_{42}$}
\end{figure}

\begin{figure}[htb]
\label{Simulation values of $C_{60}$}
\includegraphics[width=0.48\textwidth]{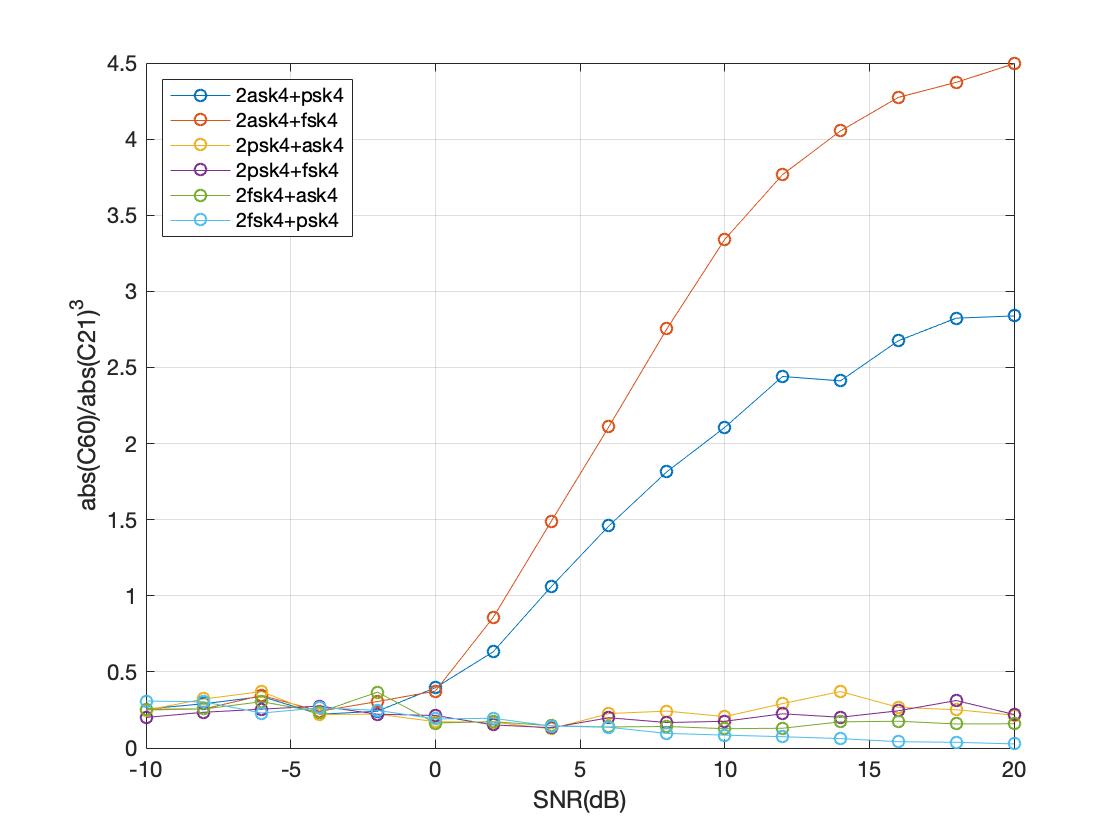}
\caption{Simulation values of higher order cumulants of mixed signal, $C_{60}$}
\end{figure}




\subsection{Instantaneous feature}
\subsubsection{Basic theory}
Based on the instantaneous amplitude of the signal, the signal instantaneous phase, and the instantaneous frequency, related instantaneous features could be extracted to recognize different types of modulation signals.
Several instantaneous features are introduced respectively in the following paragraphs.
The first feature is the maximum value of the power spectral density of the instantaneous normalized amplitude at the center. The expression is as follows:
\begin{equation}
gamma_{\max }=\frac{\max \left|D F T\left(A_{c n}\right)\right|^{2}}{N}
\end{equation}
In the formula, $N$ represents the number of sampling points of the received signal,  $DFT$ is the discrete Fourier transform, $ A_{c n}$ represents the instantaneous normalized amplitude of the center of the signal, $A_{n}[n]=A_{n}[n] / \mu_{A}$, $\mu_{A}$ represents the average value of the instantaneous amplitude, and the specific calculation formulas of $ A_{c n}$ and $\mu_{A}$ are as follows:
\begin{align}
A_{c n}[n]=A_{n}[n]-1\\
\mu_{A}=\frac{1}{N} \sum_{n=1}^{N} A[n]
\end{align}
The second feature is the standard deviation of the absolute value of the instantaneous normalized amplitude of the center, which is expressed as:
\begin{equation}
\sigma_{\text {aa}}=\sqrt{\frac{1}{N}\left(\sum_{n=1}^{N} A_{c n}^{2}[n]\right)-\left(\frac{1}{N} \sum_{n=1}^{N} A_{c n}[n]\right)^{2}}
\end{equation}
The third feature is the standard deviation of absolute value of the instantaneous phase:
\begin{equation}
\sigma_{a p}=\sqrt{\frac{1}{N_{c}}\left(\sum_{A_{n}[n]>A_{t}} \phi_{N L}^{2}[n]\right)- \left(\frac{1}{N_{c}} \sum_{A_{n}[n]>A_{t}}\left|\phi_{N L}[n]\right|\right)^{2}}
\end{equation}
Among them, $N_{c}$ represents the number of sampling points of the received signal subject to $A_{n}[n]>A_{t}$ , and the function of threshold $A_{t}$ is to filter out the sampling points with low amplitude, because the sampling points with low amplitude are too much affected by noise. $\phi_{N L}$ represents the nonlinearity component in the instantaneous phase. Generally $\phi_{N L}[n]$ is calculated by subtracting linear weight from unfolded phase $\phi[n]$. The specific formula is as follows:
\begin{equation}
\phi_{N L}[n]=\phi[n]-\frac{2 \pi f_{c} n}{f_{s}}
\end{equation}
In the formula above, the unfolded phase $\phi[n]$ is obtained by adding the correction sequence $C[n]$ to the signal instantaneous phase sequence $\hat{\phi}[n]$ , namely:
\begin{equation}
\phi[n]= \hat{\phi}[n]+C[n]
\end{equation}
Among them, the calculation formula of the modified sequence $C[n]$ is:
\begin{equation}
C[n]=\left\{\begin{array}{c}C[n-1]-2 \pi, \text { if } \hat{\phi}[n+1]-\hat{\phi}[n]>\pi \\ C[n-1]+2 \pi, \text { if } \hat{\phi}[n]-\hat{\phi}[n+1]>\pi \\ C[n-1], \text { others }\end{array}\right.
\end{equation}
The fourth feature is the standard deviation of the direct instantaneous phase, which is expressed as follows:
\begin{equation}
\sigma_{d p}=\sqrt{\frac{1}{N_{c}}\left(\sum_{A_{n}[n]>A_{t}} \phi_{N L}^{2}[n]\right)-\left(\frac{1}{N_{c}} \sum_{A_{n}[n]>A_{t}} \phi_{N L}[n]\right)^{2}}
\end{equation}
The fifth feature is the standard deviation of the absolute value of the instantaneous normalized frequency at the center, which is expressed as follows:
\begin{equation}
\sigma_{a f}=\sqrt{\frac{1}{N_{c}}\left(\sum_{A_{n}[n]>A_{t}} f_{N}^{2}[n]\right)-\left(\frac{1}{N_{c}} \sum_{A_{n}[n]>A_{t}} f_{N}[n]\right)^{2}}
\end{equation}
In the formula above, $f_{N}[n]$ represents the center instantaneous normalized frequency, which is calculated by the center instantaneous frequency  $f_{m}[n]$ and the symbol rate $R_{s}$ together, namely:
\begin{equation}
f_{N}[n]=\frac{f_{m}[n]}{R_{s}}
\end{equation}
Among them, the center instantaneous frequency $f_{m}[n]$ is calculated from the frequency average $\mu_{f}$ and instantaneous frequency $f[n]$ , namely:
\begin{align}
f_{m}[n]=f[n]-\mu_{f}\\
\mu_{f}=\frac{1}{N} \sum_{n=1}^{N} f[n]
\end{align}
Among them, the instantaneous frequency $f[n]$ is generally obtained by calculating the difference of the non-folding instantaneous phase, the formula is as follows:
\begin{equation}
f[n]=\frac{f_{s}}{2 \pi}(\phi[n+1]-\phi[n])
\end{equation}
The sixth feature is the standard deviation of the amplitude envelope, the formula is as follows:
\begin{equation}
E=\sqrt{\frac{\left(\sum_{n=1}^{N} A[n]-\mu_{A}\right)^{2}}{N-1}}
\end{equation}
The seventh feature is the maximum value of the normalized power spectrum, the formula is as follows:
\begin{equation}
p=\max \left|\frac{\left|D F T\left(S_{n}\right)\right|}{\sum_{n=1}^{N}\left|D F T\left(S_{n}\right)\right|}\right|
\end{equation}

\subsubsection{Simulation}
\begin{figure}[htb]
\centering  

\includegraphics[width=0.48\textwidth]{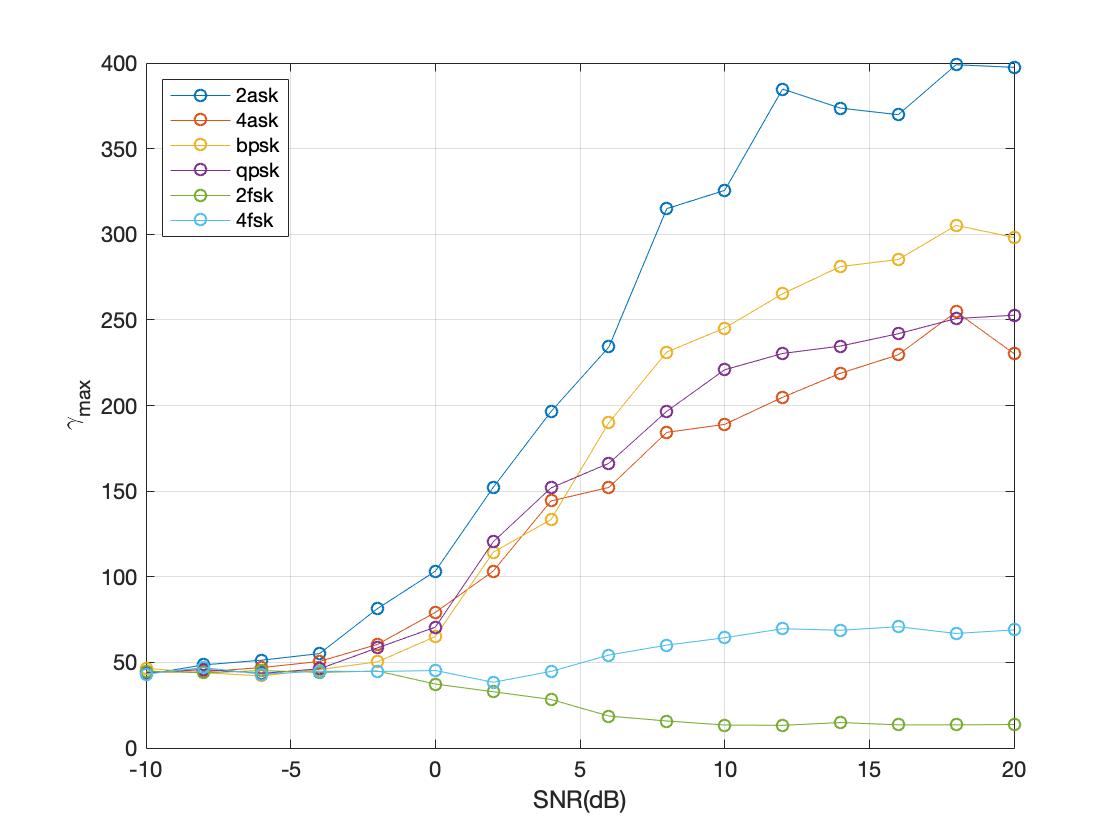}
\caption{Simulation values of instantaneous features of single signal, feature1}
\label{Simulationvaluesfeature1}
\end{figure}

\begin{figure}[htb]

\includegraphics[width=0.48\textwidth]{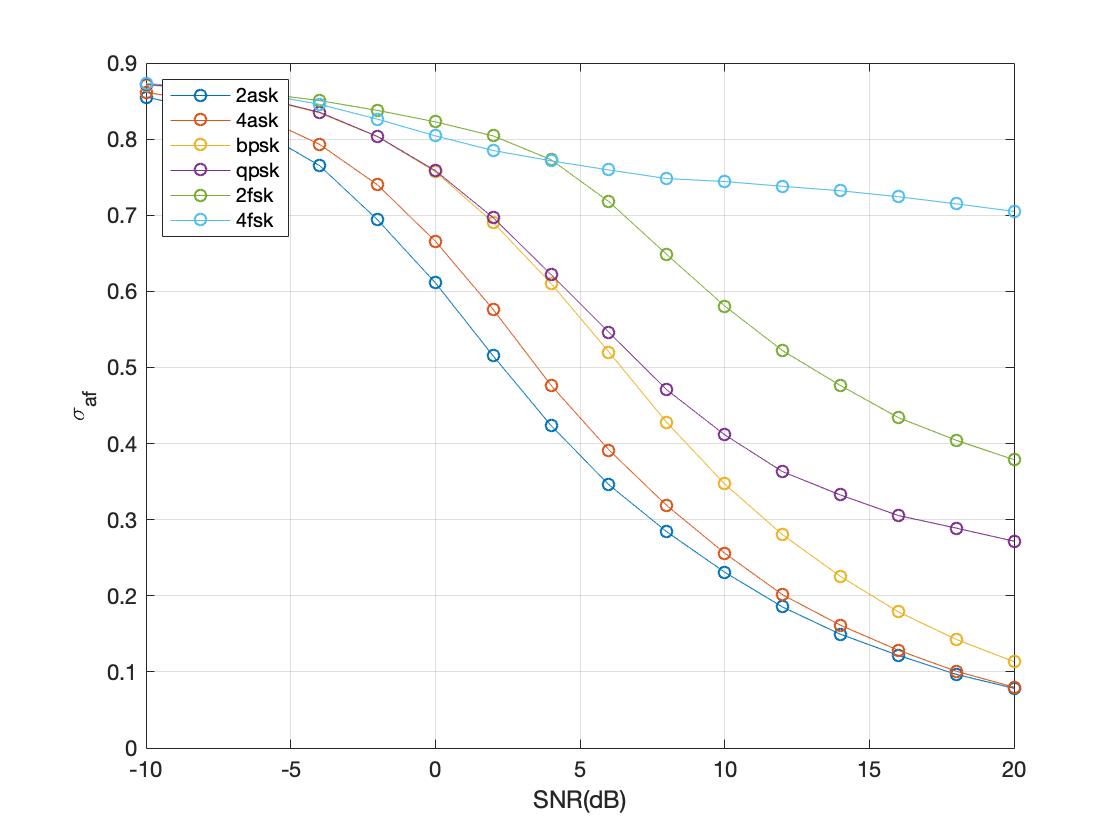}
\caption{Simulation values of instantaneous features of single signal, feature5}
\label{Simulationvaluesfeature5}
\end{figure}

\paragraph{Single signal}\mbox{}\\
In this section, $\{2\mathrm{ASK}, 4\mathrm{ASK}, \mathrm{BPSK}, \mathrm{QPSK}, 2\mathrm{FSK}, 4\mathrm{FSK}\}$ would be used to do the simulation and show the value of each features. Computer simulation is used to simulate the feature parameters. The simulation uses MATLAB software and uses random sequence as signal model. The baseband signal is obtained after down-conversion, and then modulation recognition is performed. With carrier frequency of symbol modulation equal to 70Hz, sampling rate equal to 400Hz, symbol rate equal to 2 bps, number of symbols equal to 1000, noise chosen as Gaussian white noise and signal-to-noise ratio from -10 to 20 dB, take the average of 500 simulations for each signal. The simulation results of the seven feature parameters are shown in Figs. \ref{Simulationvaluesfeature1} - \ref{Simulationvaluesfeature5}.

\begin{figure}[htb]

\includegraphics[width=0.48\textwidth]{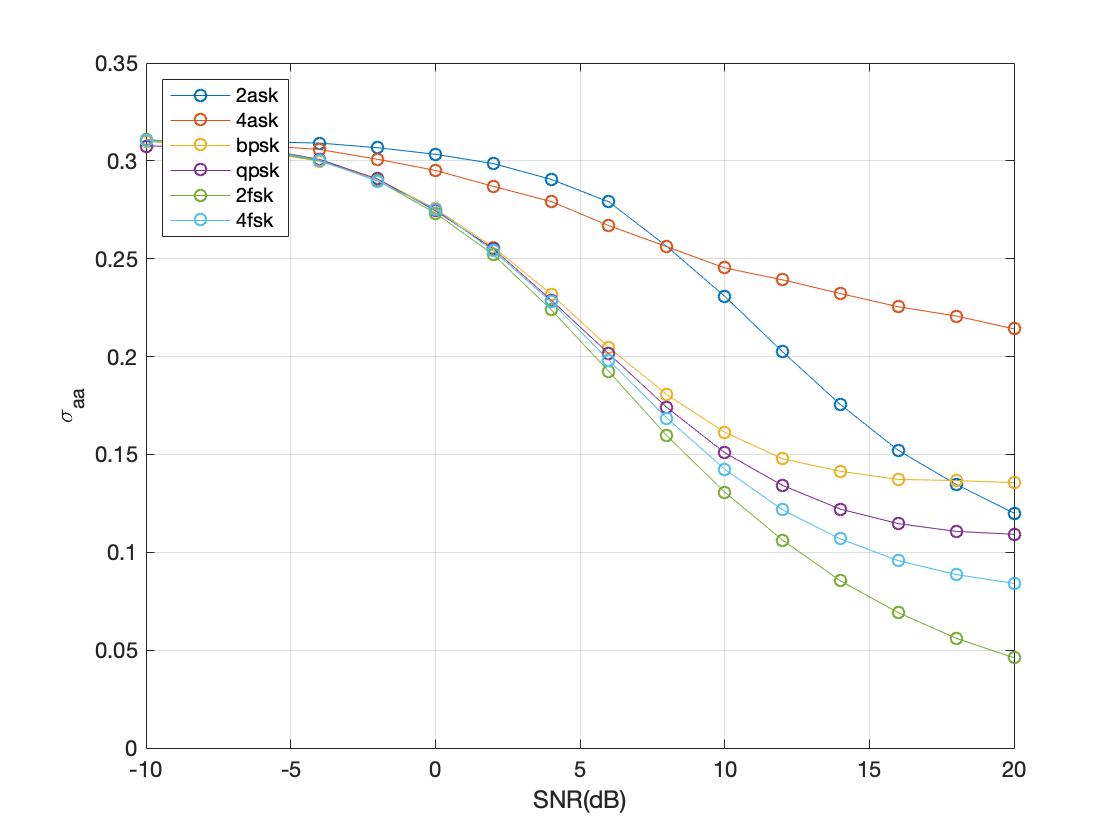}
\label{Simulation values of feature2}
\caption{Simulation values of instantaneous features of single signal, feature2}
\end{figure}

\begin{figure}[htb]
\label{Simulation values of feature3}
\includegraphics[width=0.48\textwidth]{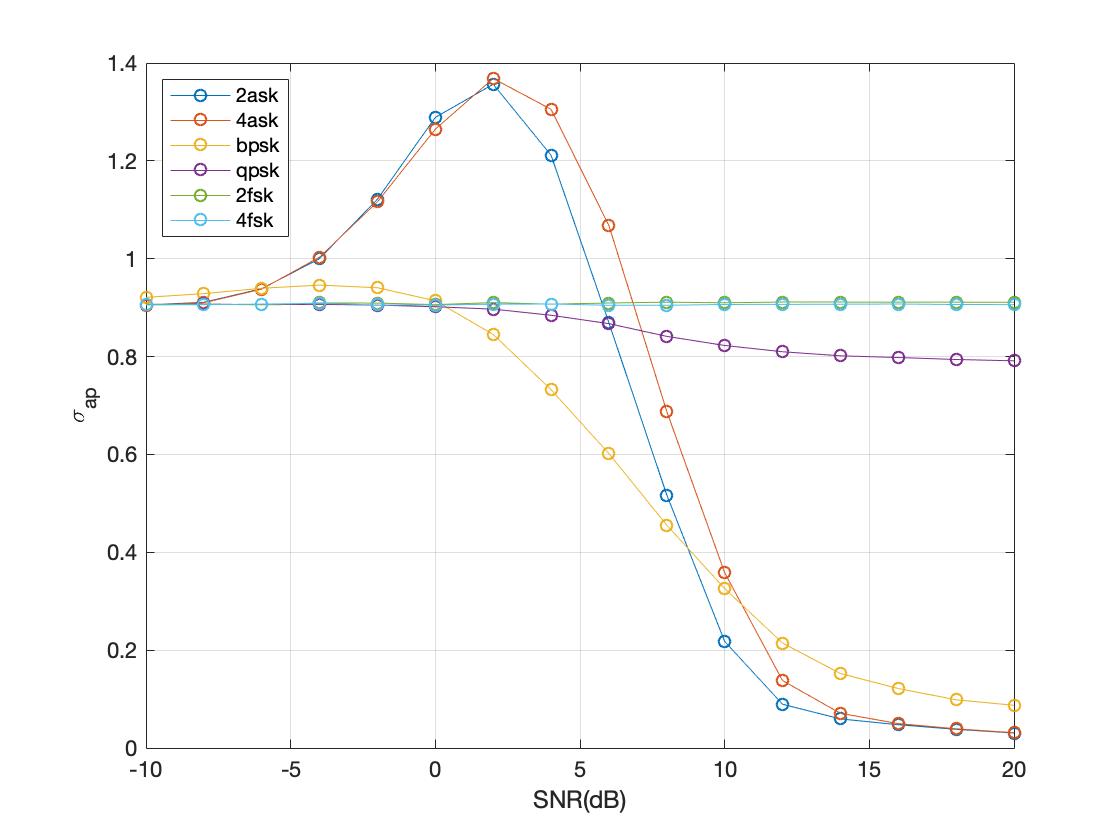}
\caption{Simulation values of instantaneous features of single signal, feature3}
\label{Simulation values of feature3}
\end{figure}

\begin{figure}[htb]

\includegraphics[width=0.48\textwidth]{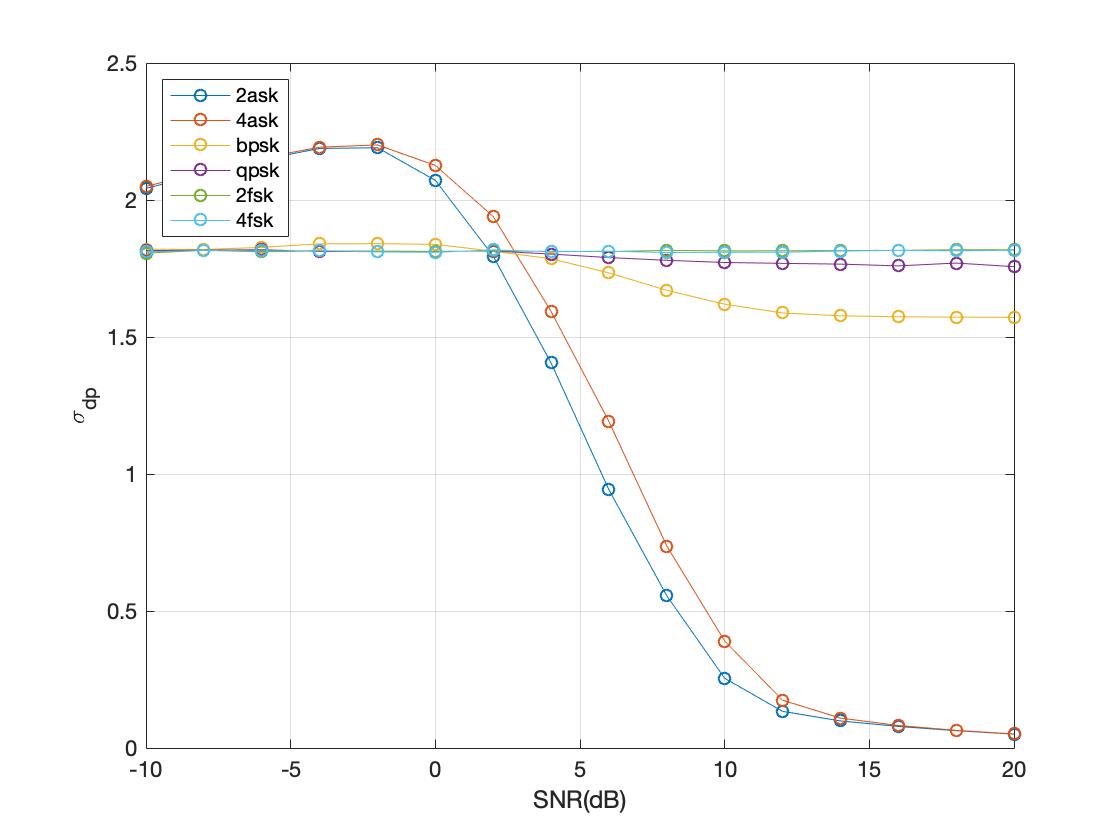}
\caption{Simulation values of instantaneous features of single signal, feature4}
\label{Simulation values of feature4}
\end{figure}

\begin{figure}[htb]
\centering  
\includegraphics[width=0.48\textwidth]{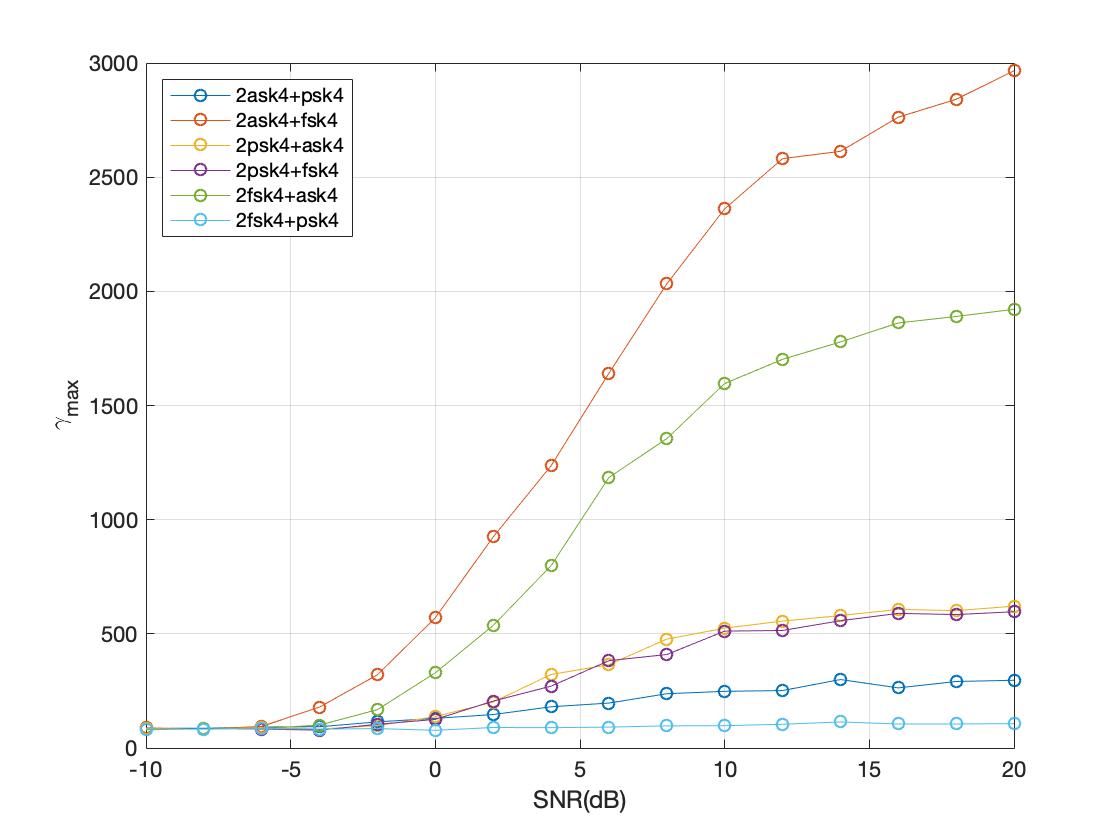}
\caption{Simulation values of instantaneous features of mixed signal, feature1}
\label{MixedSimulation values of feature1}
\end{figure}

\begin{figure}[htb]
\includegraphics[width=0.48\textwidth]{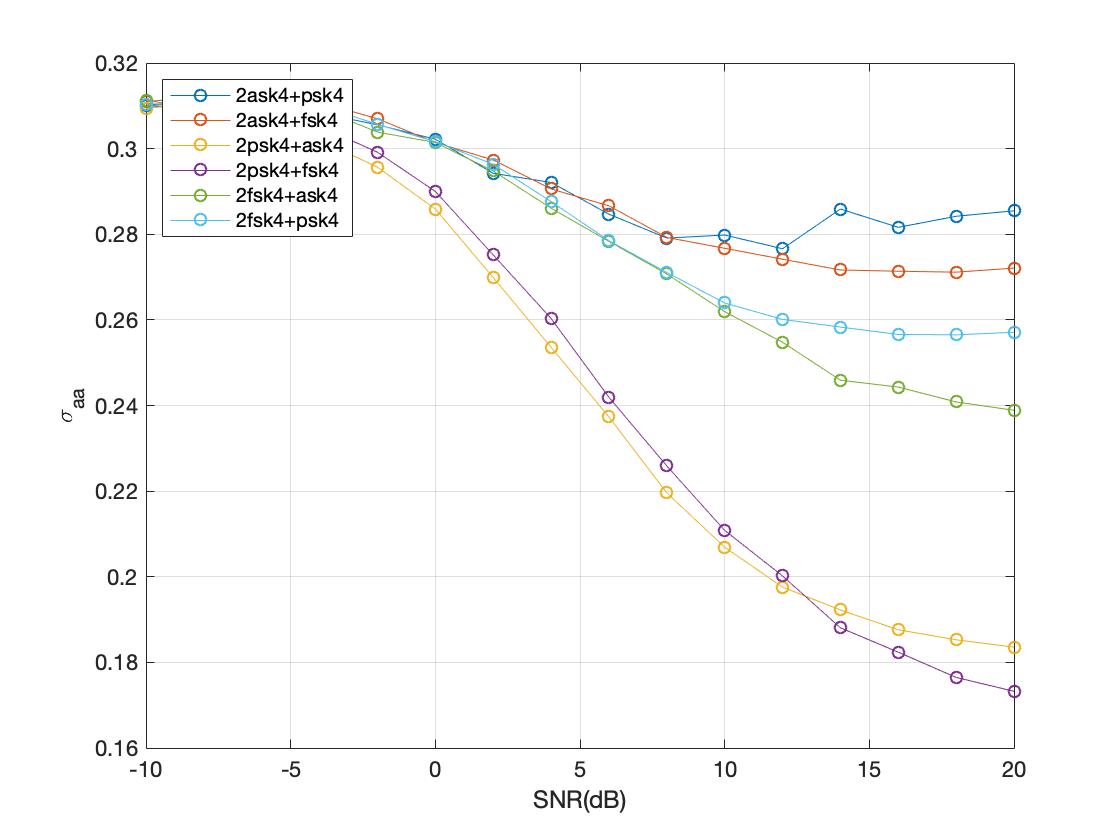}
\caption{Simulation values of instantaneous features of mixed signal, feature2}
\label{MixedSimulation values of feature2}
\end{figure}

\begin{figure}[htb]


\includegraphics[width=0.48\textwidth]{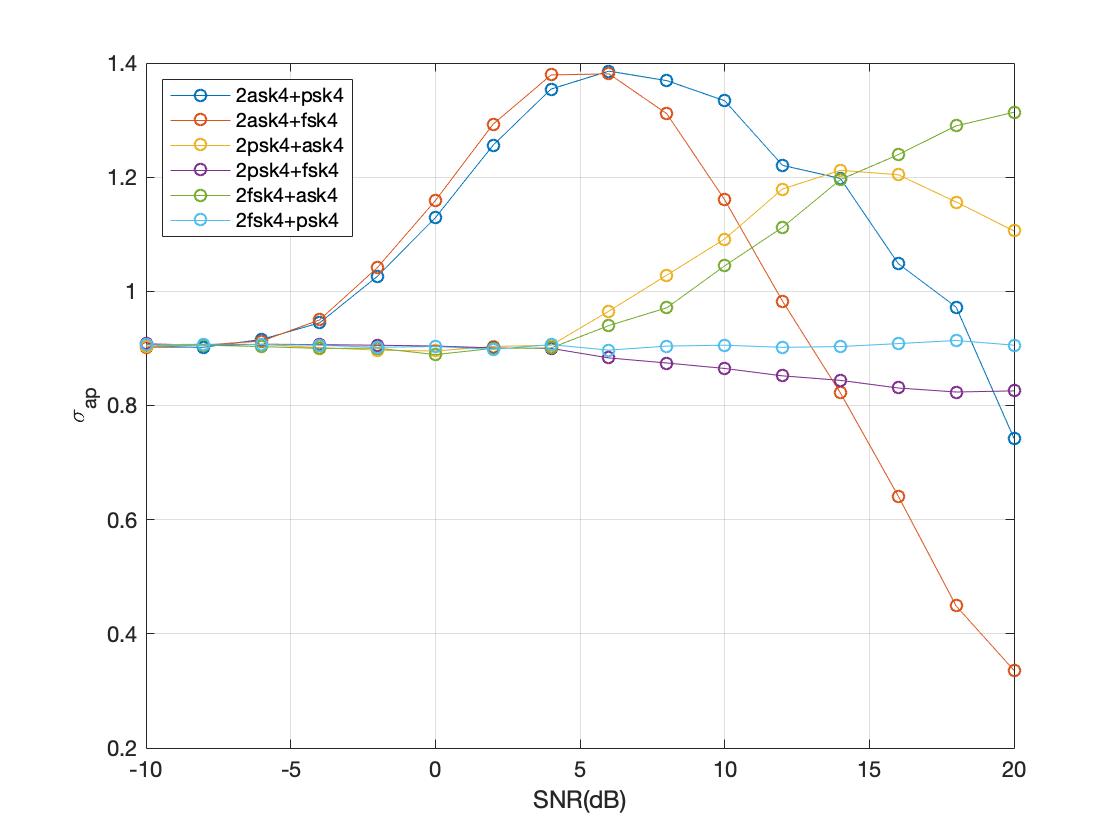}
\caption{Simulation values of instantaneous features of mixed signal, feature3}
\label{MixedSimulation values of feature3}
\end{figure}

\begin{figure}[htb]
\includegraphics[width=0.48\textwidth]{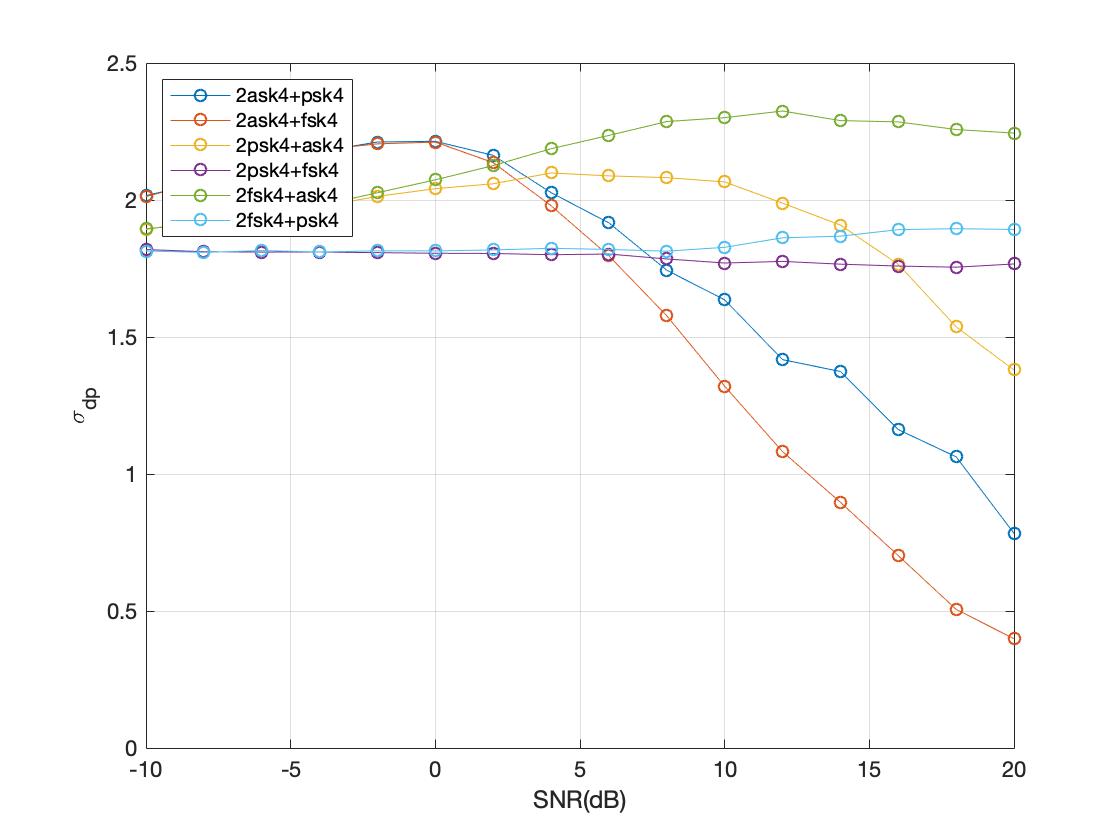}
\caption{Simulation values of instantaneous features of mixed signal, feature4}
\label{MixedSimulation values of feature4}
\end{figure}

\begin{figure}[htb]
\includegraphics[width=0.48\textwidth]{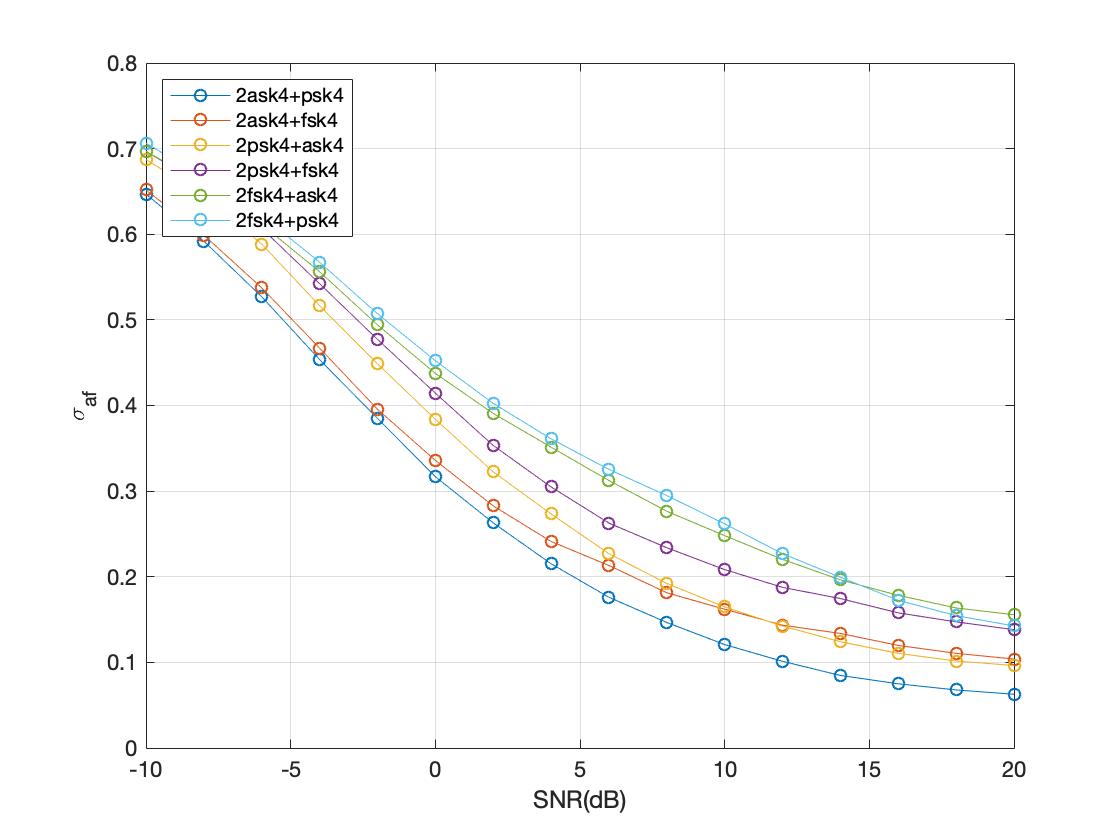}
\caption{Simulation values of instantaneous features of mixed signal, feature5}
\label{MixedSimulation values of feature5}
\end{figure}

\paragraph{Mixed signal}\mbox{}\\
In this section, three digital signals $\{4\mathrm{ASK}, \mathrm{QPSK}, 4\mathrm{FSK}\}$ are divided into six mixed signals according to the signal power ratio, which is equal to 2:1, namely $\{2\mathrm{ASK}4+\mathrm{PSK}4,2\mathrm{ASK}4+\mathrm{FSK}4,2\mathrm{PSK}4+\mathrm{ASK}4,2\mathrm{PSK}4+\mathrm{FSK}4,2\mathrm{FSK}4+\mathrm{ASK}4 , 2\mathrm{FSK}4+\mathrm{PSK}4\}$. The signals would be used to do the simulation and get the results of each value of different features. The parameters sets are the same as those in the single signal section. The simulation results of the seven feature parameters are shown in Figs. \ref{MixedSimulation values of feature1} - \ref{MixedSimulation values of feature5}.

\subsection{Cyclic spectrum}
\subsubsection{Basic theory}
Let signal $x(t)$ be a generalized cyclostationary random process, and its autocorrelation function is defined as:
\begin{equation}
\begin{aligned} R_{x}(t+\tau / 2, t-\tau / 2)&= R_{x}(t+\tau / 2+T, t-\tau / 2+T)\\&=E[x(t+\tau / 2+T) \cdot x(t-\tau / 2+T)] \end{aligned}
\end{equation}
In the formula, $T$ is the signal symbol period and $\tau$ is the delay time. Then the cyclic autocorrelation function of the signal $x(t)$ is defined as:
\begin{equation}
\begin{aligned} R_{x}^{\alpha}(\tau)=& \lim _{T \rightarrow \infty} \frac{1}{T} \int_{-T / 2}^{T / 2} x(t+\tau / 2) \cdot x^{*}(t-\tau / 2) \mathrm{e}^{-\mathrm{j} 2 \pi \alpha t} \mathrm{d} t \end{aligned}
\end{equation}
Let the frequency $\alpha$ which makes $R_{x}^{\alpha}(\tau) \neq 0$ be the cycle frequency of the signal, where $\alpha=k/T$ and $k$ is an integer. There may be multiple cyclic frequencies of a cyclostationary signal, including zero frequency and non-zero frequency. When $\alpha=0$, $R_{x}^{\alpha}(\tau)$ is a stationary signal autocorrelation function; when $\alpha \neq 0$, $R_{x}^{\alpha}(\tau)$ is the period weighted form of $R_{x}(\tau)$, which is referred to as a periodic autocorrelation function, i.e. cyclic autocorrelation function. The Fourier transform $S_{x}^{\alpha}(f)$ of the cyclic autocorrelation function $R_{x}^{\alpha}(\tau)$ is the cyclic spectrum of the signal $x(t)$:
\begin{equation}
S_{x}^{\alpha}(f)=\int_{-\infty}^{+\infty} R_{x}^{\alpha}(\tau) \mathrm{e}^{-\mathrm{j} 2 \pi f \tau} \mathrm{d} \tau
\end{equation}

Envelope characteristics of ASK, PSK, FSK signals are quite different. Several features are selected according to the characteristics of signals. The first feature $\sigma_{S 0}$ is the amplitude envelope variance of the section of normalized cyclic spectrum $S_{x}^{\alpha}(0)$ of signal $x(t)$. The second feature $\sigma_{S f_{c}}$ is the amplitude envelope variance of the section of normalized cyclic spectrum $S_{x}^{\alpha}(f_{c})$ of signal $x(t)$. The third feature is the mean value of the amplitude envelope of the section of normalized cyclic spectrum $S_{x}^{\alpha}(0)$. The fourth feature $\beta$ is the maximum value of the normalized cyclic spectrum $S_{x}^{\alpha}(0)$ (the ratio of the maximum value of the spectrum correlation on the $\alpha$ axis to the maximum value of the spectrum correlation on the $f$ axis. The fifth feature $P$ is the average energy of the value point of $S_{x}^{2}(f_{c})$ on the $f$ axis.

\subsubsection{simulation}
\begin{figure}[htb]
\centering  
\includegraphics[width=0.48\textwidth]{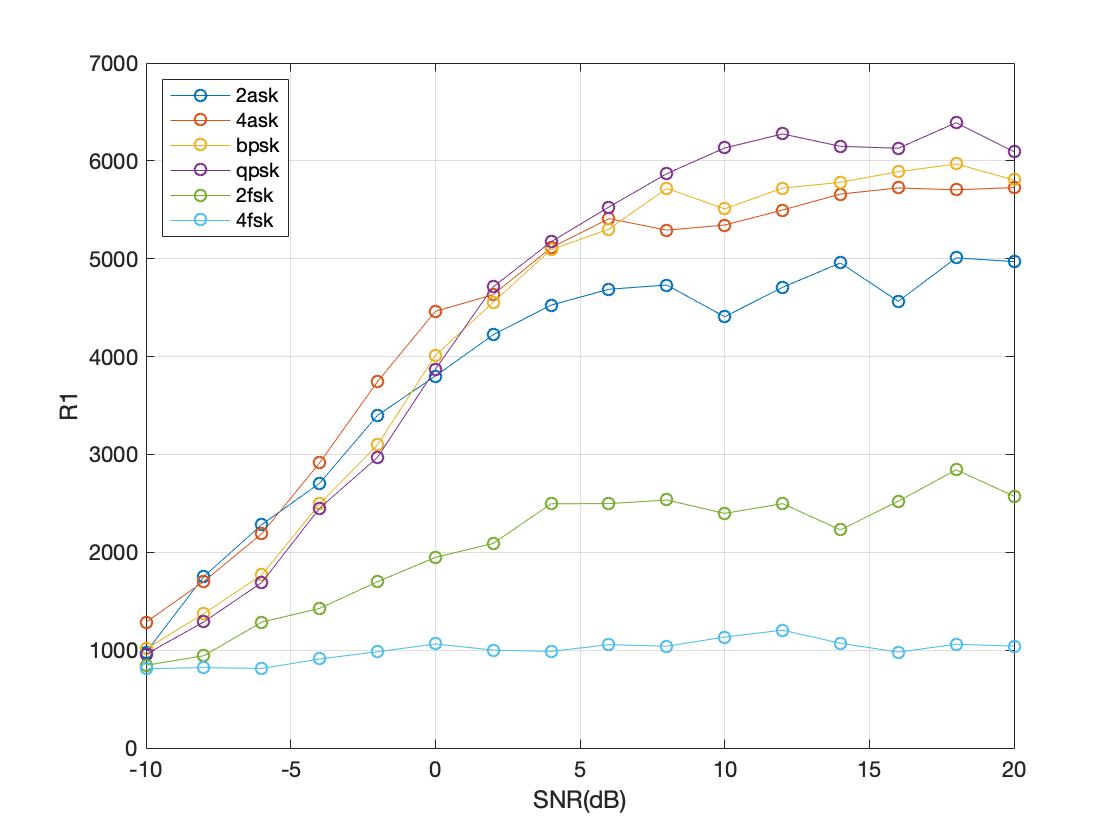}
\caption{Simulation values of cyclic spectrum of single signal, feature1}
\label{Simulationvalues of feature1}
\end{figure}

\begin{figure}[htb]

\includegraphics[width=0.48\textwidth]{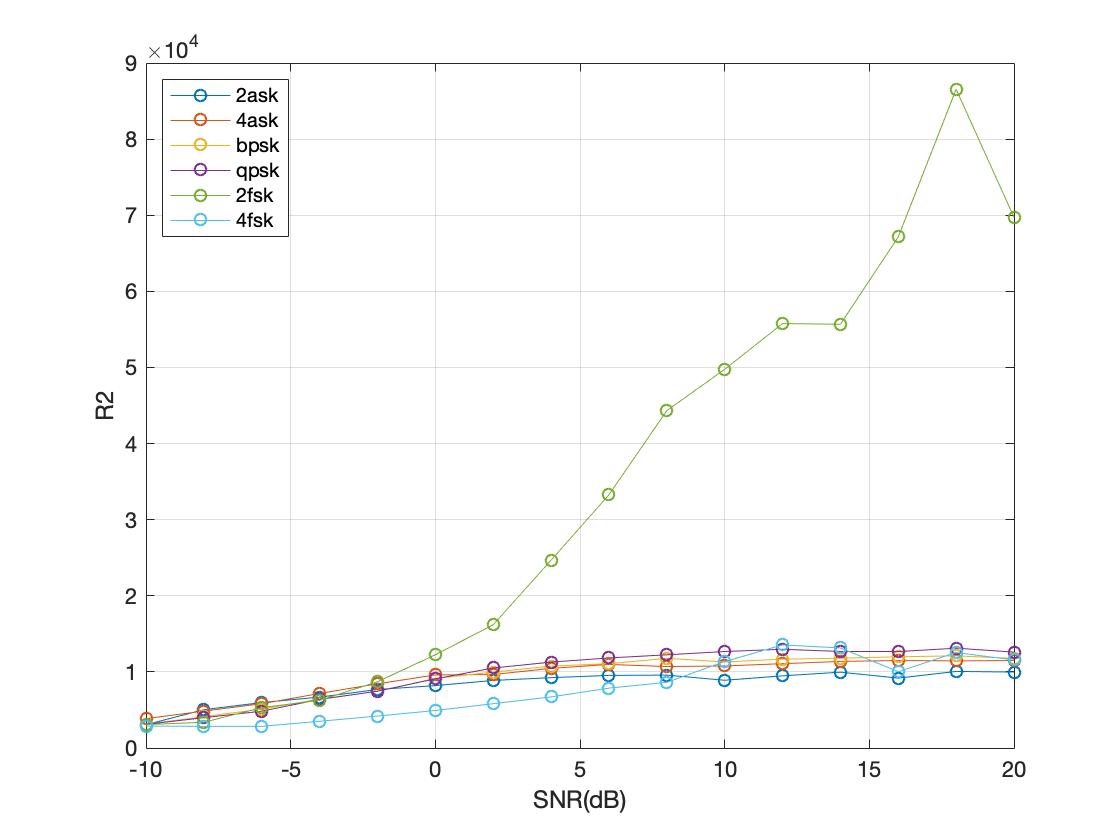}
\caption{Simulation values of cyclic spectrum of single signal, feature2}
\label{Simulationvalues of feature2}
\end{figure}

\begin{figure}[htb]

\includegraphics[width=0.48\textwidth]{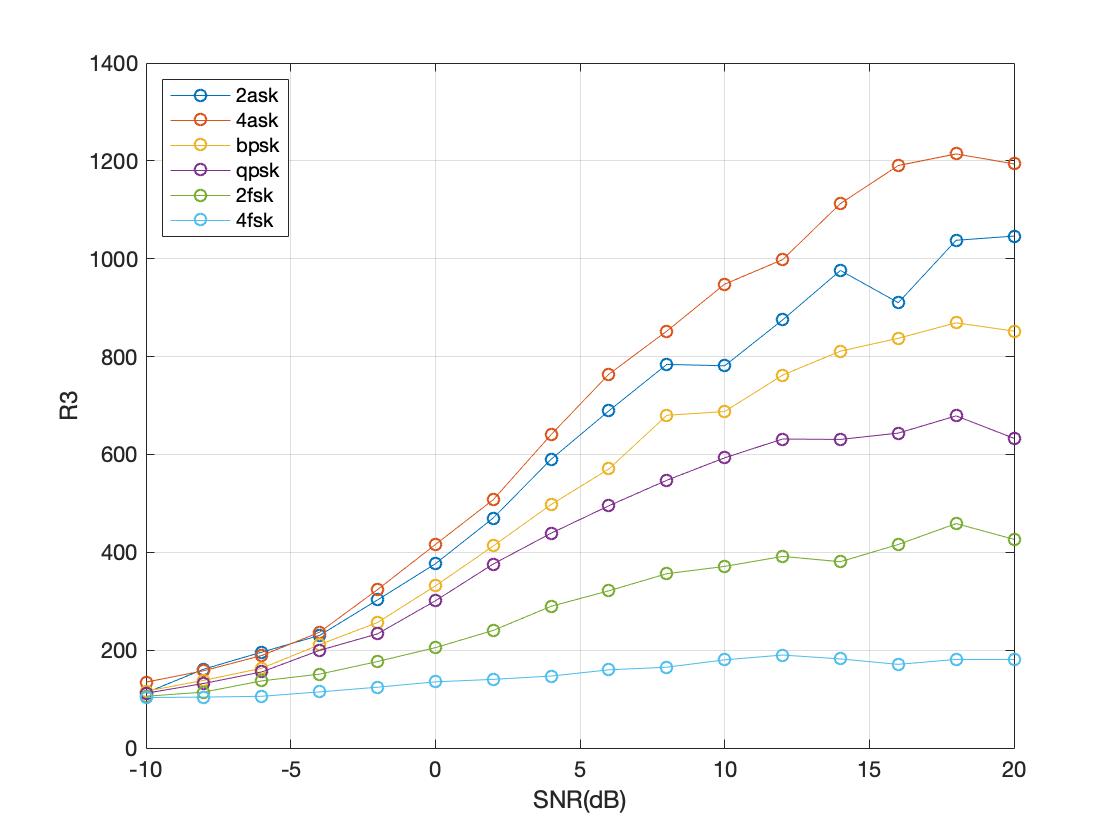}
\caption{Simulation values of cyclic spectrum of single signal, feature3}
\label{Simulationvalues of feature3}
\end{figure}

\begin{figure}[htb]

\includegraphics[width=0.48\textwidth]{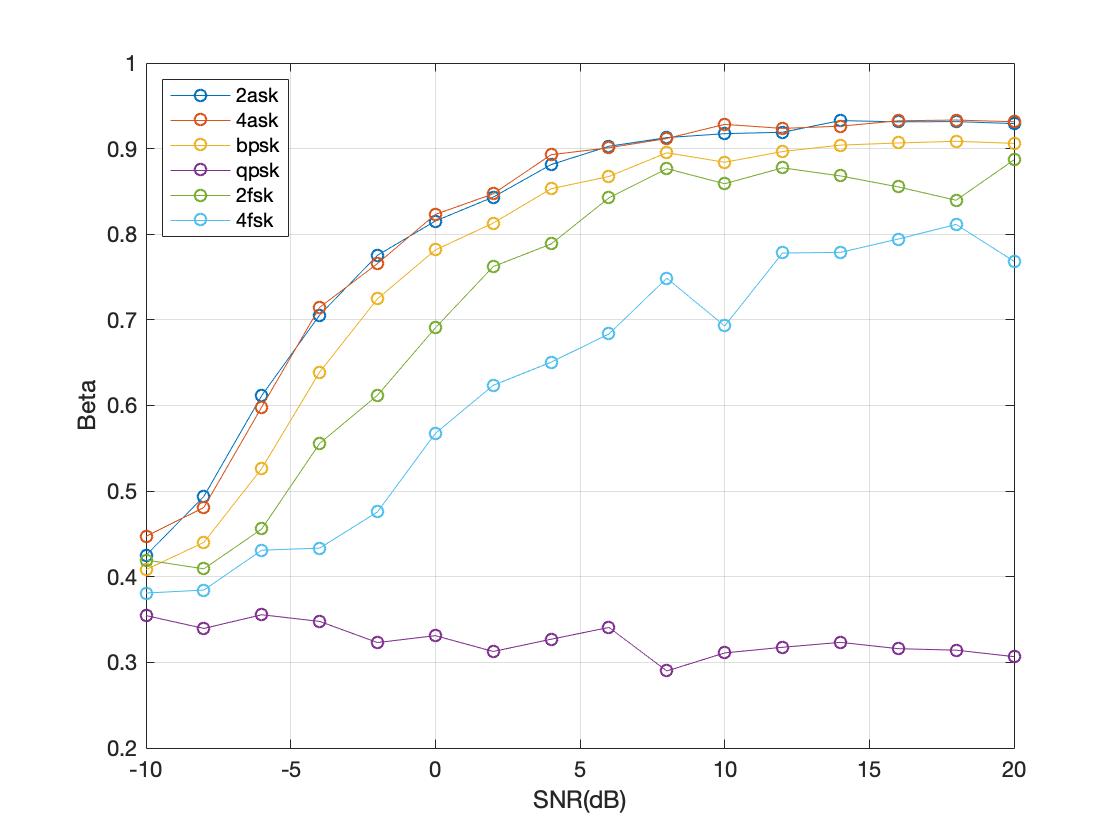}
\caption{Simulation values of cyclic spectrum of single signal, feature4}
\label{Simulationvalues of feature4}
\end{figure}

\begin{figure}[htb]

\includegraphics[width=0.48\textwidth]{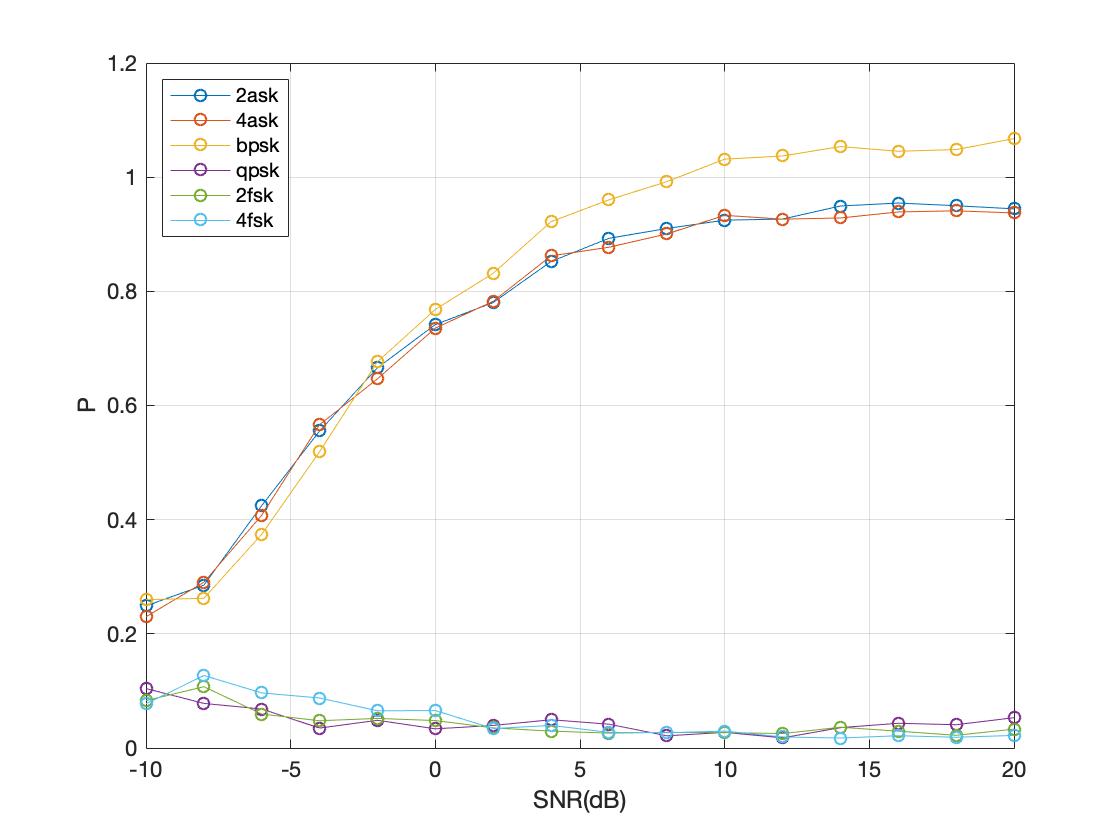}
\caption{Simulation values of cyclic spectrum of single signal, feature5}
\label{Simulationvalues of feature5}
\end{figure}

\paragraph{Single signal}\mbox{}\\
In this section, $\{2\mathrm{ASK}, 4\mathrm{ASK}, \mathrm{BPSK}, \mathrm{QPSK}, 2\mathrm{FSK}, 4\mathrm{FSK}\}$ are used for the simulation and show the value of each features. Computer simulation is used to simulate the feature parameters.
The simulation is based on MATLAB. The baseband signal is obtained after down-conversion, and then modulation recognition is performed. With carrier frequency of symbol modulation equal to 70Hz,  we set the sampling rate equal to 400Hz, symbol rate equal to 2bps, number of symbols equal to 1000, noise chosen as the Gaussian white noise and signal-to-noise ratio ranging from -10 to 20dB, and take the average of 500 simulations for each signal. The simulation results of five feature parameters are shown in Figs. \ref{Simulationvalues of feature1} - \ref{Simulationvalues of feature5}

\begin{figure}[htb]
\centering  
\includegraphics[width=0.48\textwidth]{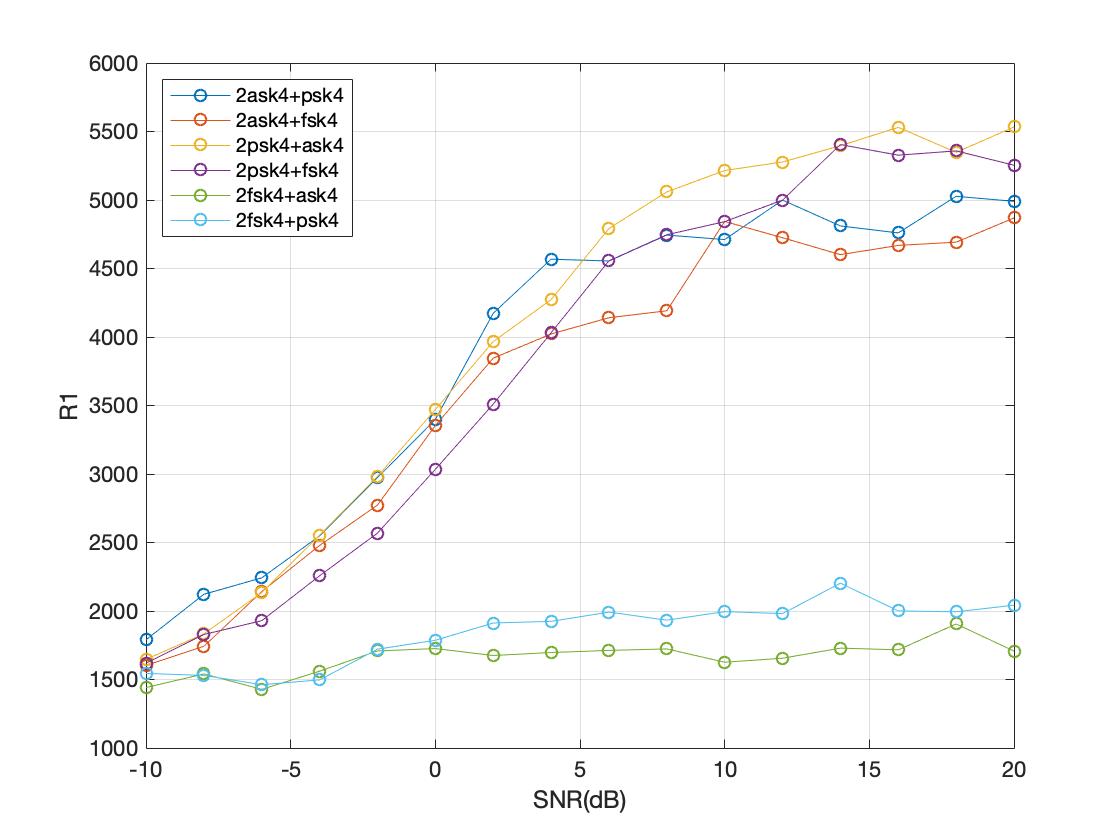}
\caption{Simulation values of cyclic spectrum of mixed signal, feature1}
\label{Simulationfeature1}
\end{figure}

\begin{figure}[htb]

\includegraphics[width=0.48\textwidth]{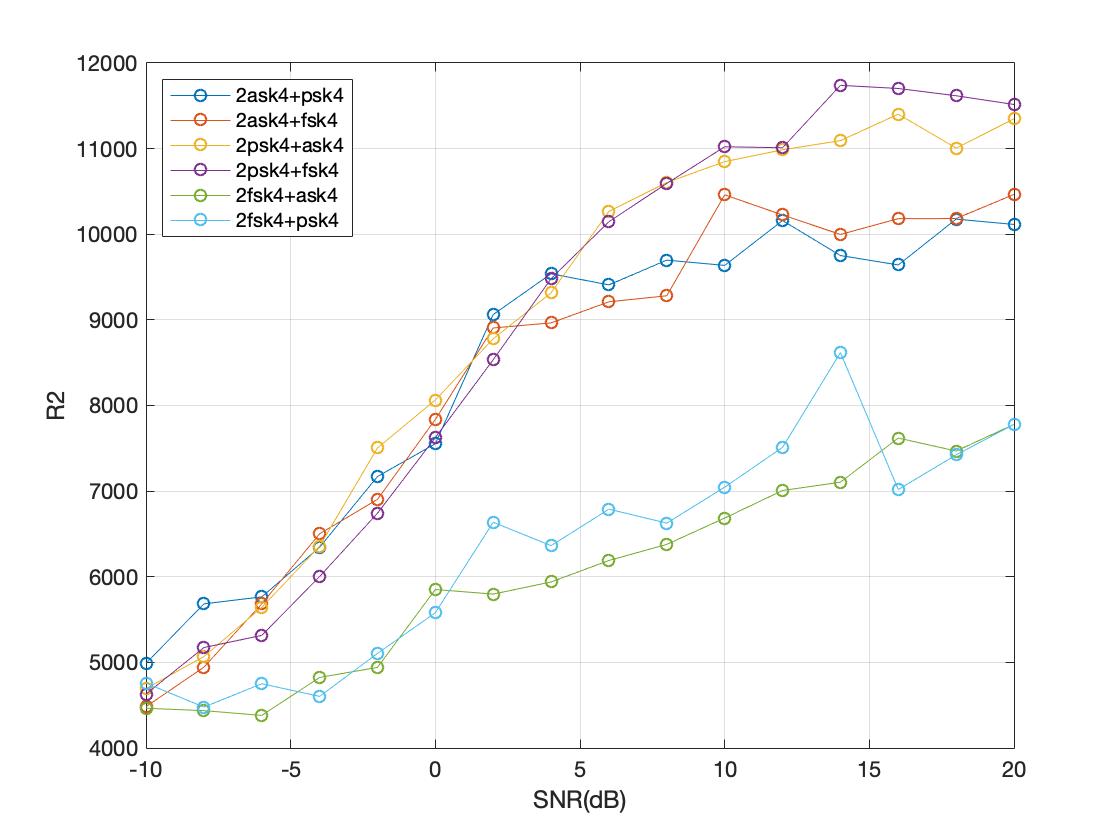}
\caption{Simulation values of cyclic spectrum of mixed signal, feature2}
\label{Simulationfeature2}
\end{figure}

\begin{figure}[htb]

\includegraphics[width=0.48\textwidth]{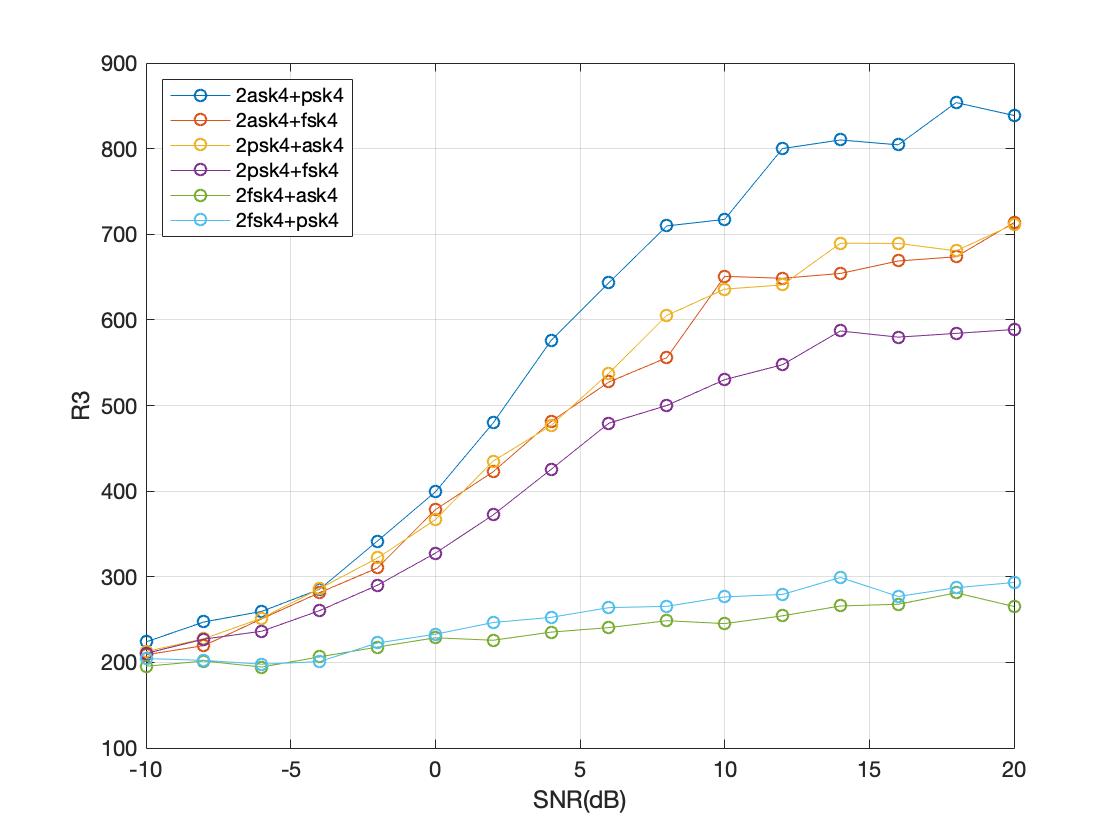}
\caption{Simulation values of cyclic spectrum of mixed signal, feature3}
\label{Simulationfeature3}
\end{figure}

\begin{figure}[htb]
\label{Simulation values of feature4}
\includegraphics[width=0.48\textwidth]{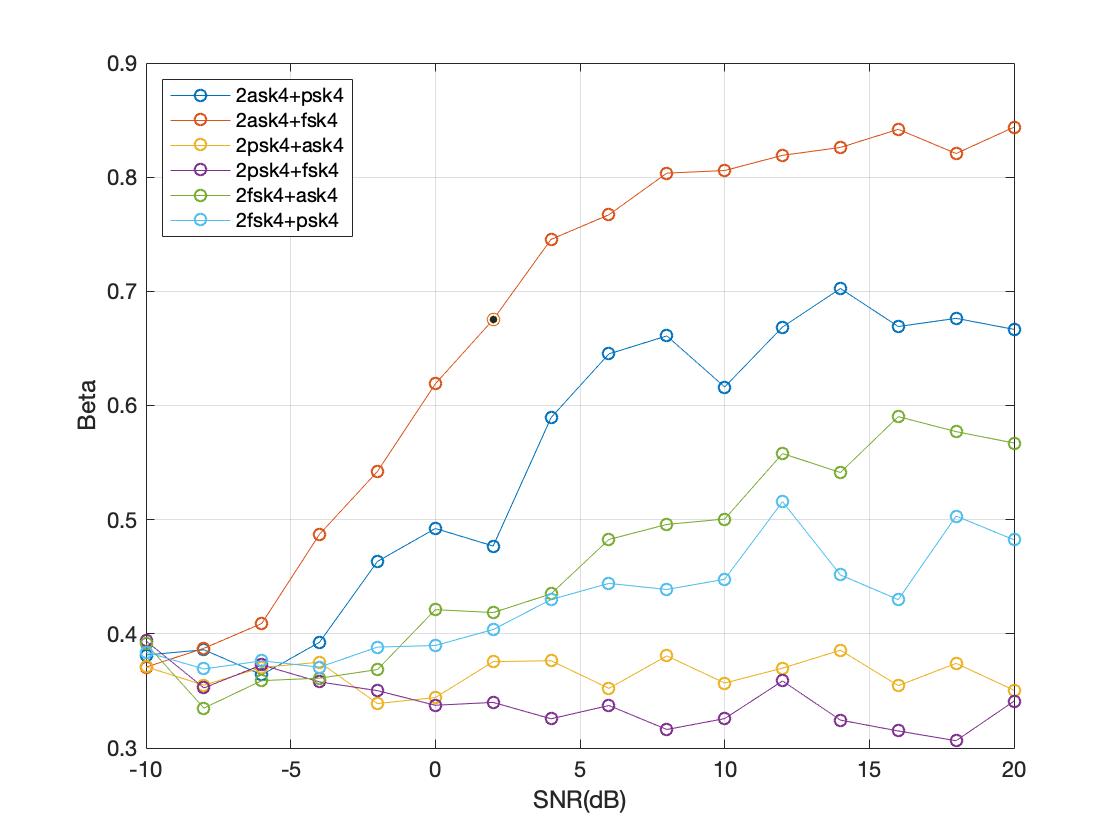}
\caption{Simulation values of cyclic spectrum of mixed signal, feature4}
\label{Simulationfeature4}
\end{figure}

\begin{figure}[htb]

\label{Simulation values of feature5}
\includegraphics[width=0.48\textwidth]{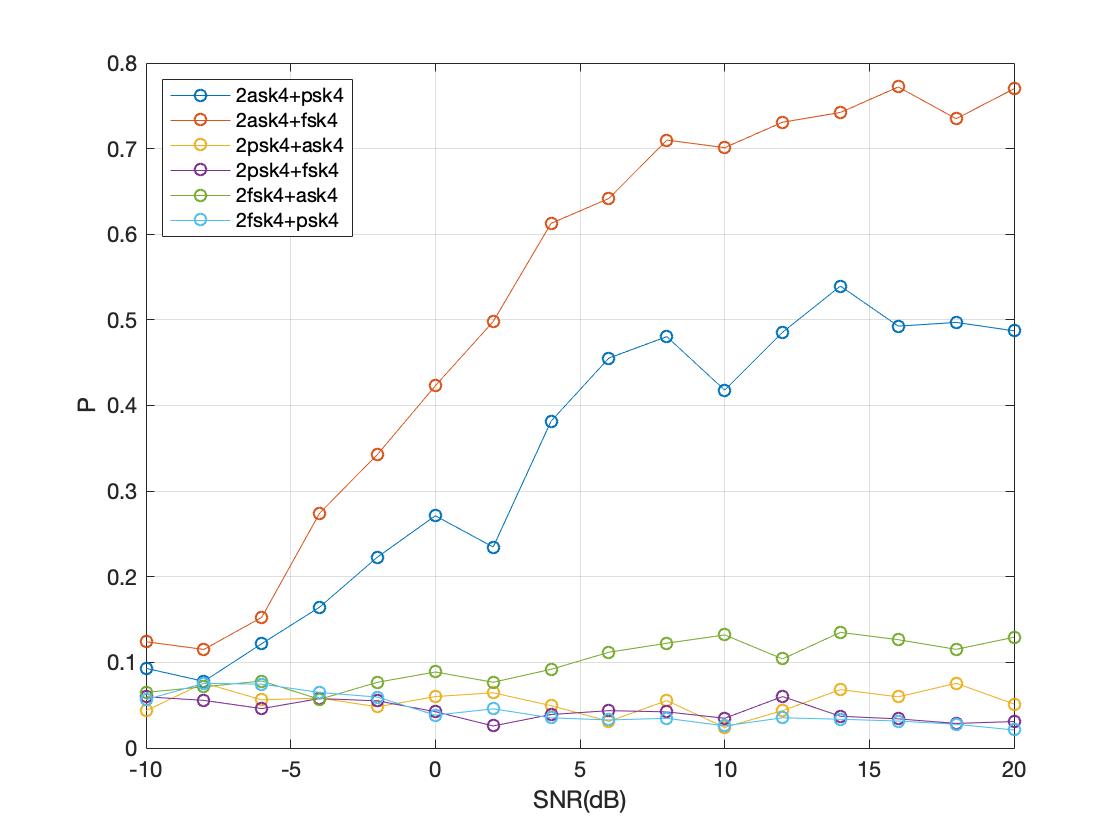}
\caption{Simulation values of cyclic spectrum of mixed signal, feature5}
\label{Simulationfeature5}
\end{figure}

\paragraph{Mixed signal}\mbox{}\\
In this section, three digital signals $\{4\mathrm{ASK}, \mathrm{QPSK}, 4\mathrm{FSK}\}$ are divided into six mixed signals according to the signal power ratio, which is equal to 2:1, namely $\{2\mathrm{ASK}4+\mathrm{PSK}4,2\mathrm{ASK}4+\mathrm{FSK}4,2\mathrm{PSK}4+\mathrm{ASK}4,2\mathrm{PSK}4+\mathrm{FSK}4,2\mathrm{FSK}4+\mathrm{ASK}4 , 2\mathrm{FSK}4+\mathrm{PSK}4\}$. The signals would be used to do the simulation and get the results of each value of different features. The parameters sets are the same as those in the single signal section.The simulation results of five feature parameters are shown in Figs. \ref{Simulationfeature1} - \ref{Simulationfeature5}.


\subsection{Wave transform}
\subsubsection{Basic Theory}
\begin{figure}[htbp]
\centering
\includegraphics[width=8cm]{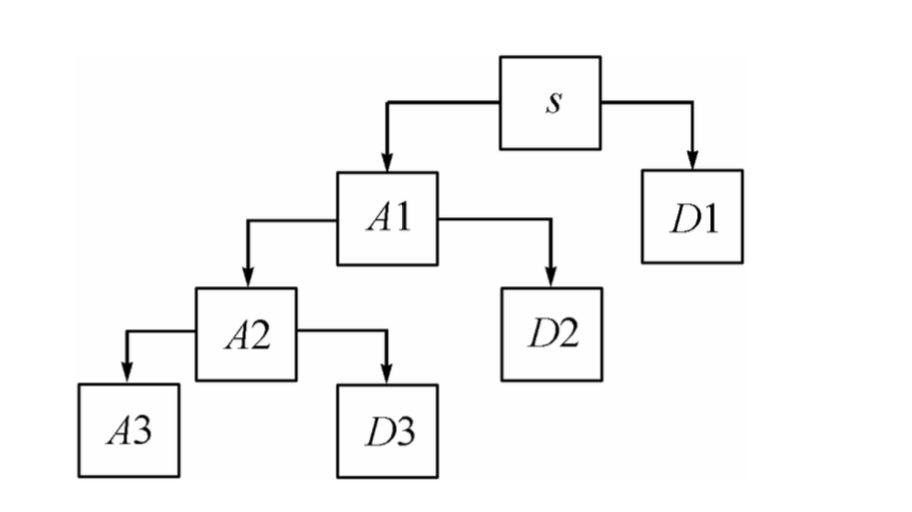}
\caption{The structure of multiresolution analysis at level 3}
\label{The structure of multiresolution analysis at level 3}
\end{figure}

Multiresolution analysis of Mallet algorithm\cite{MallatS.G1989Atfm} is to do the further decomposition of the low-frequency part of the signal. According to the number of decomposition layers, the discrete approximation and discrete details of the signal under different frequency channels can be obtained. Fig. \ref{The structure of multiresolution analysis at level 3} 
shows a three-layer decomposition structure diagram, and the signal $s$ can be expressed as.
\begin{equation}
s=A_{3}+D_{3}+D_{2}+D_{1}
\end{equation}

Multi-layer decomposition only further decomposes the low-frequency space. As the number of decomposition layers increases, the frequency resolution will become higher and higher. Therefore, when the signal has different frequency components, the signal features can be extracted through multi-layer wavelet decomposition.
In this paper, the five-scale wavelet decomposition method will be used, and the detailed energy values obtained after the five-scale wavelet decomposition will be used as five features.

\subsubsection{Simulation}
\paragraph{Single signal}\mbox{}

\begin{figure}[htbp]
\centering  
\label{Simulation values of feature1}
\includegraphics[width=0.48\textwidth]{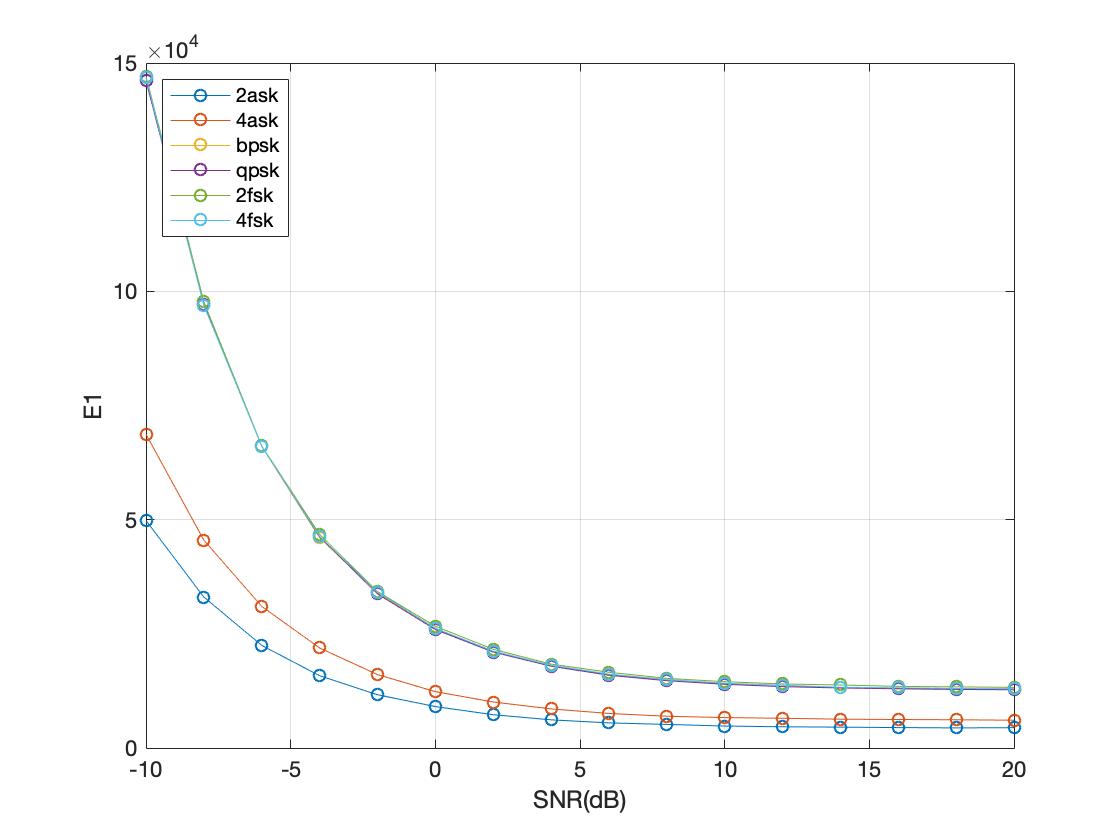}
\caption{Simulation values of wavelet transform of single signal, feature1}
\label{feature1}
\end{figure}

\begin{figure}[htb]
\label{Simulation values of feature2}
\includegraphics[width=0.48\textwidth]{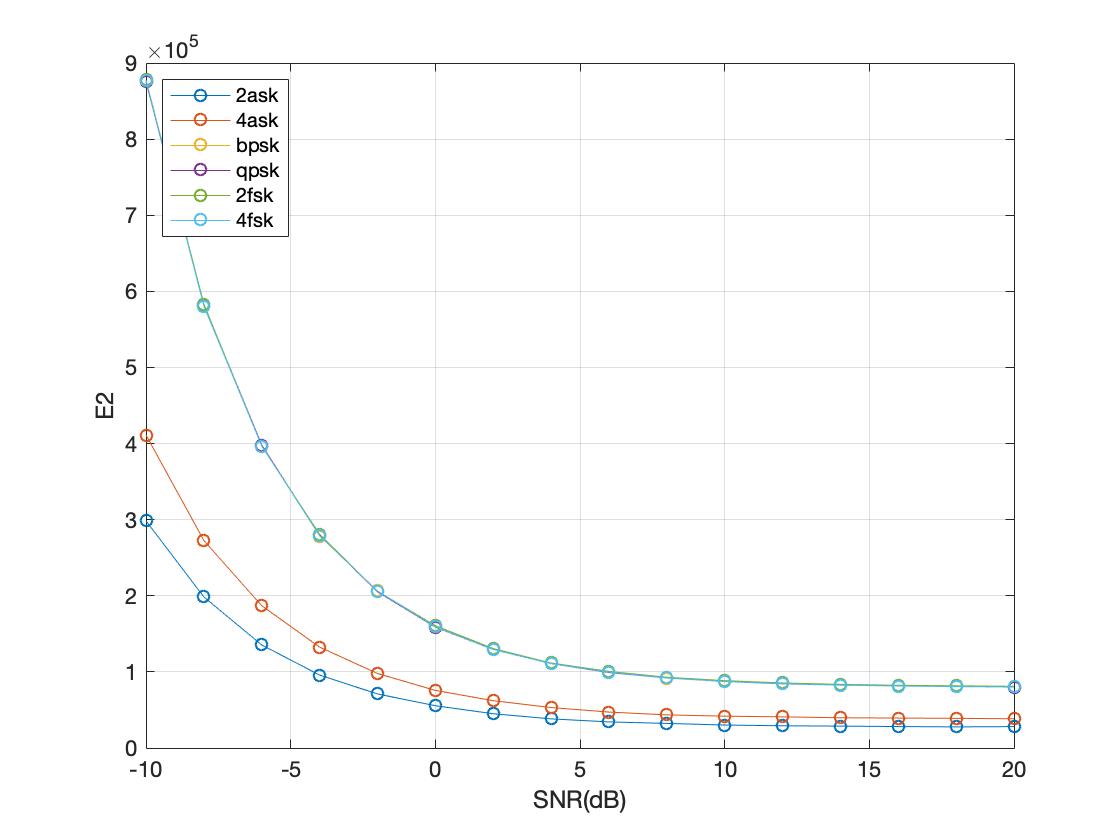}
\caption{Simulation values of wavelet transform of single signal, feature2}
\label{feature2}
\end{figure}

\begin{figure}[htb]

\label{Simulation values of feature3}
\includegraphics[width=0.48\textwidth]{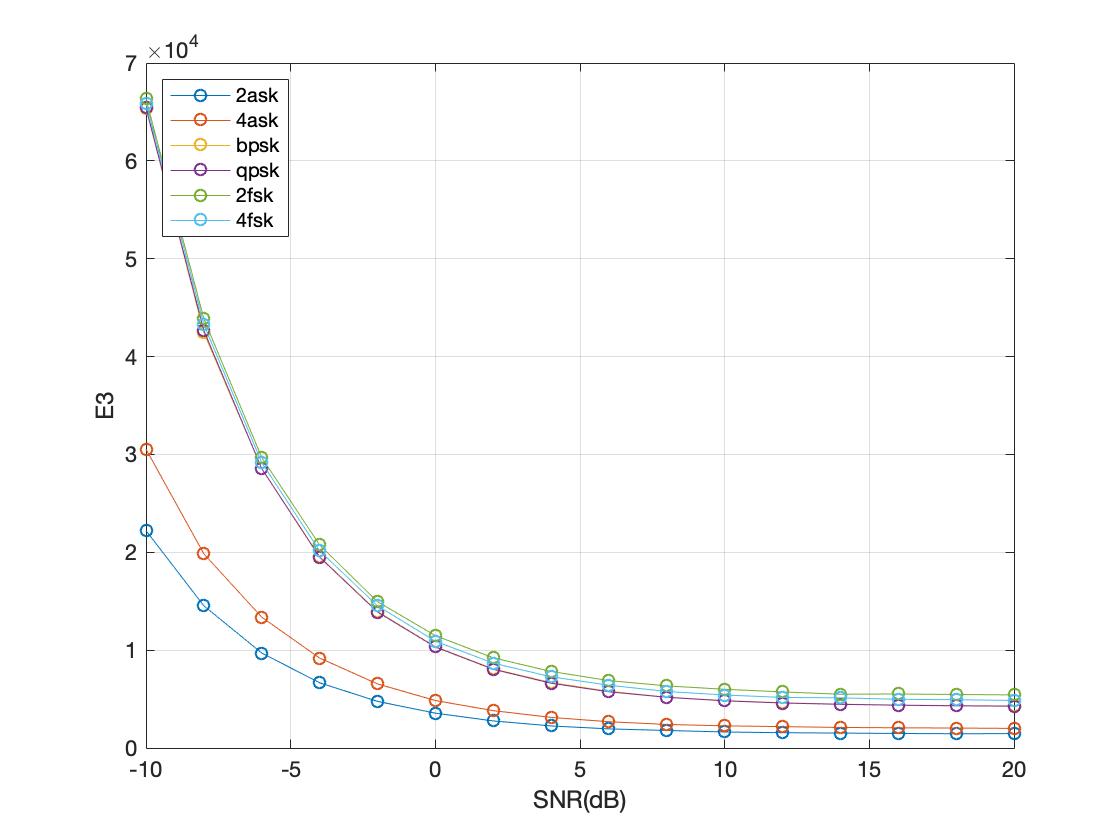}
\caption{Simulation values of wavelet transform of single signal, feature3}
\label{feature3}
\end{figure}

\begin{figure}[htb]
\label{Simulation values of feature4}
\includegraphics[width=0.48\textwidth]{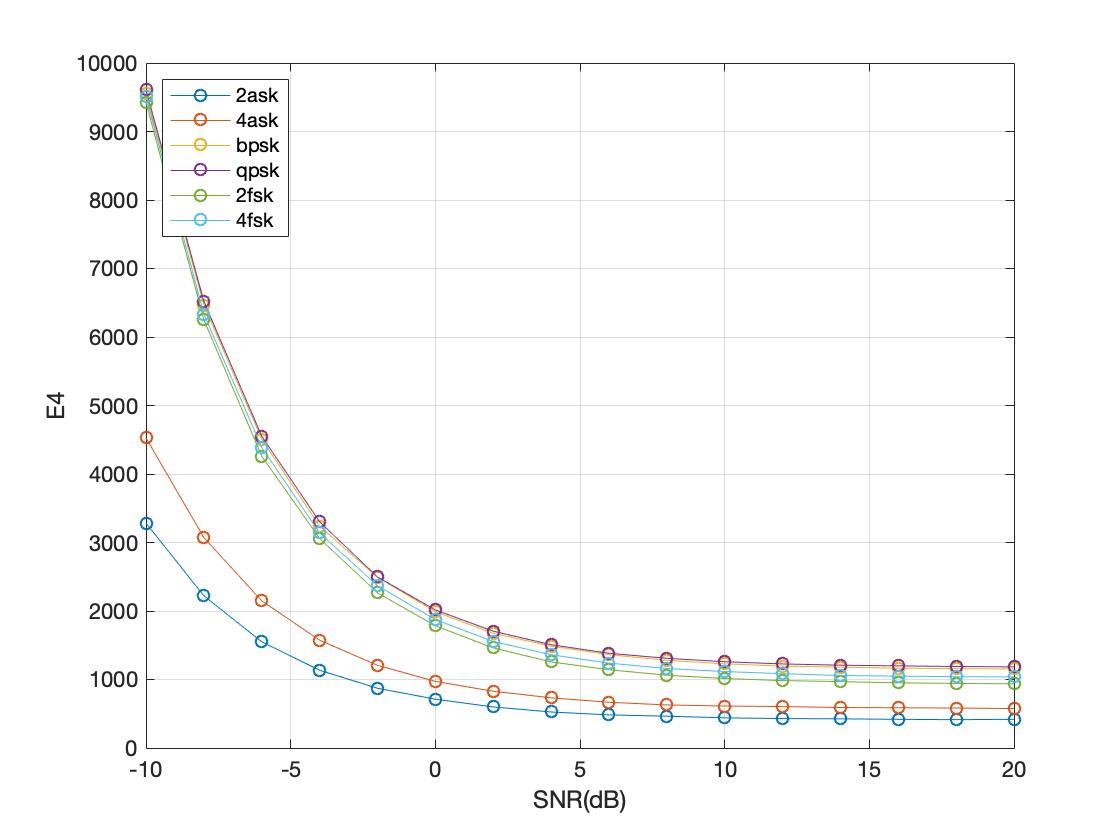}
\caption{Simulation values of wavelet transform of single signal, feature4}
\label{feature4}
\end{figure}

\begin{figure}[htb]
\label{Simulation values of feature5}
\includegraphics[width=0.48\textwidth]{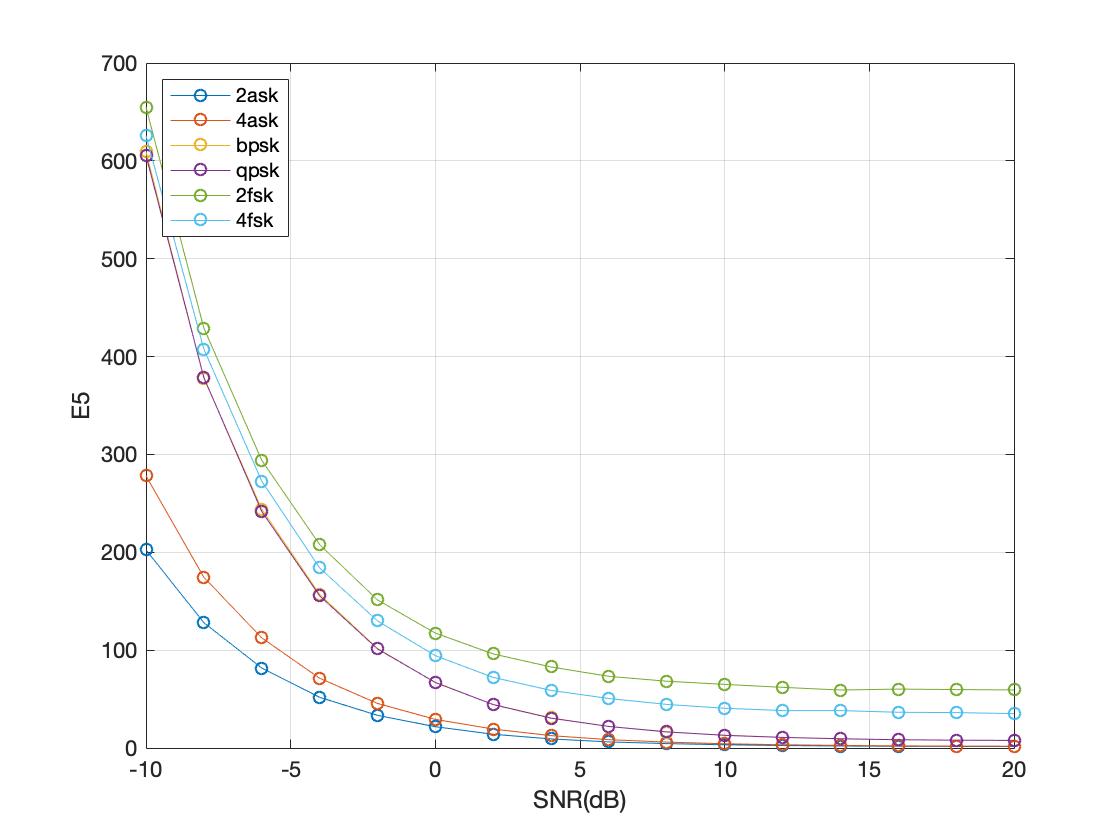}
\caption{Simulation values of wavelet transform of single signal, feature5}
\label{feature5}
\end{figure}

In this section, $\{2\mathrm{ASK}, 4\mathrm{ASK}, \mathrm{BPSK}, \mathrm{QPSK}, 2\mathrm{FSK}, 4\mathrm{FSK}\}$ are used for the simulation. 
Computer simulations are used to simulate the feature parameters.
The baseband signal is obtained after down-conversion, and then modulation recognition is performed. With carrier frequency of symbol modulation equal to 70Hz, we set the sampling rate equal to 400Hz, symbol rate equal to 2 bps, number of symbols equal to 1000, noise chosen as the Gaussian white noise and signal-to-noise ratio ranging from -10 to 20 dB, and take the average of 500 simulations for each signal. The simulation results of five feature parameters are shown in Figs. \ref{feature1} - \ref{feature5}. 


\paragraph{Mixed signal}\mbox{}

\begin{figure}[htb]
\centering  
\includegraphics[width=0.48\textwidth]{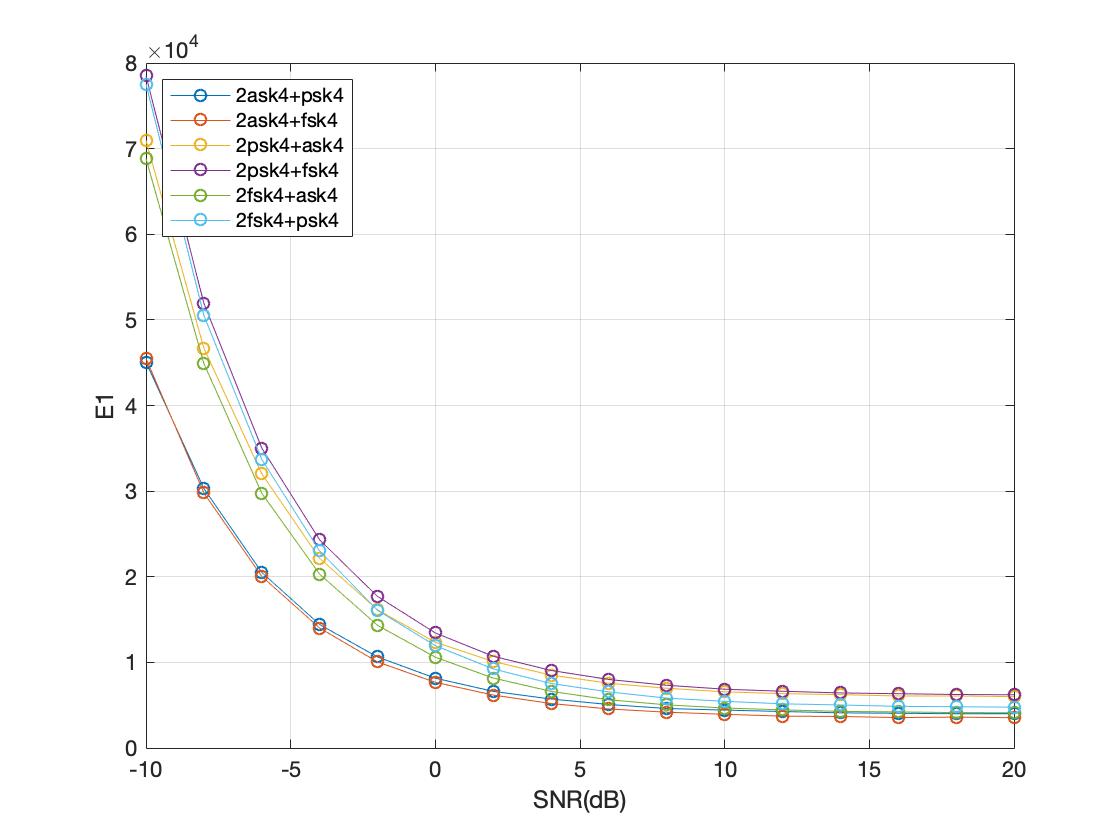}
\caption{Simulation values of wavelet transform of mixed signal, feature1}
\label{MixedSimulationfeature1}
\end{figure}

\begin{figure}[htb]
\includegraphics[width=0.48\textwidth]{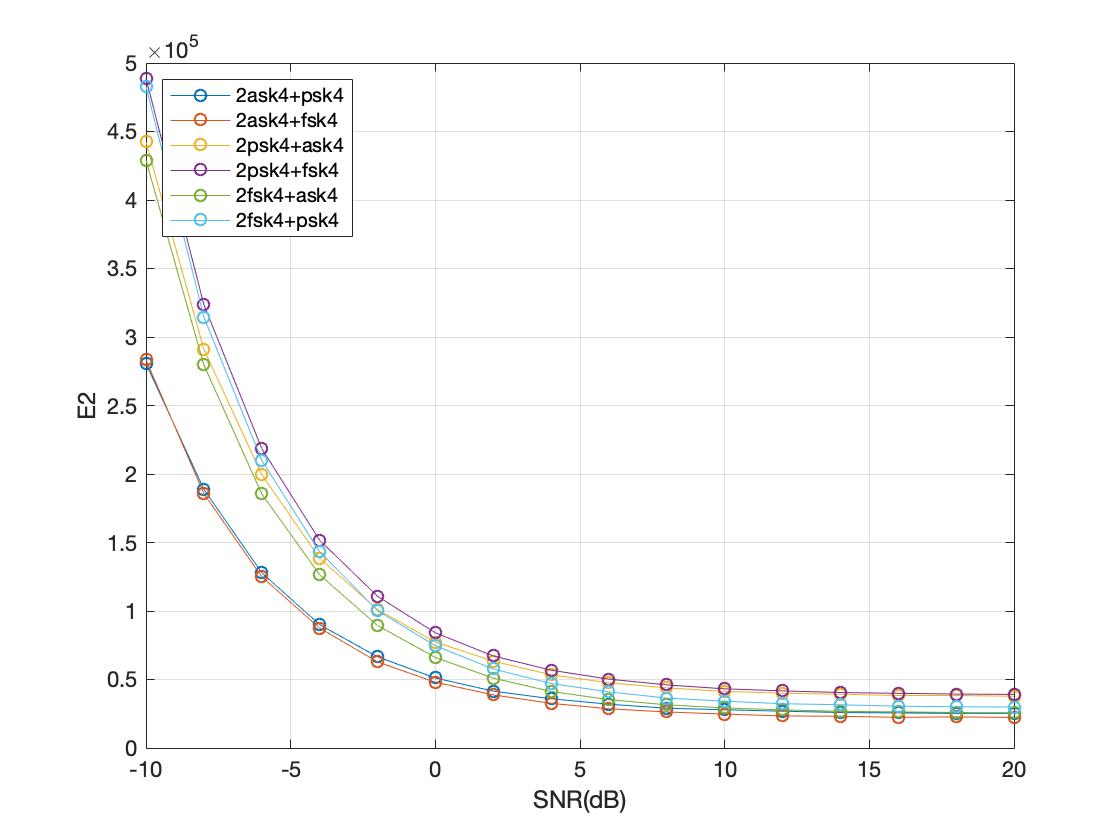}
\caption{Simulation values of wavelet transform of mixed signal, feature2}
\label{MixedSimulationfeature2}
\end{figure}

\begin{figure}[htb]

\includegraphics[width=0.48\textwidth]{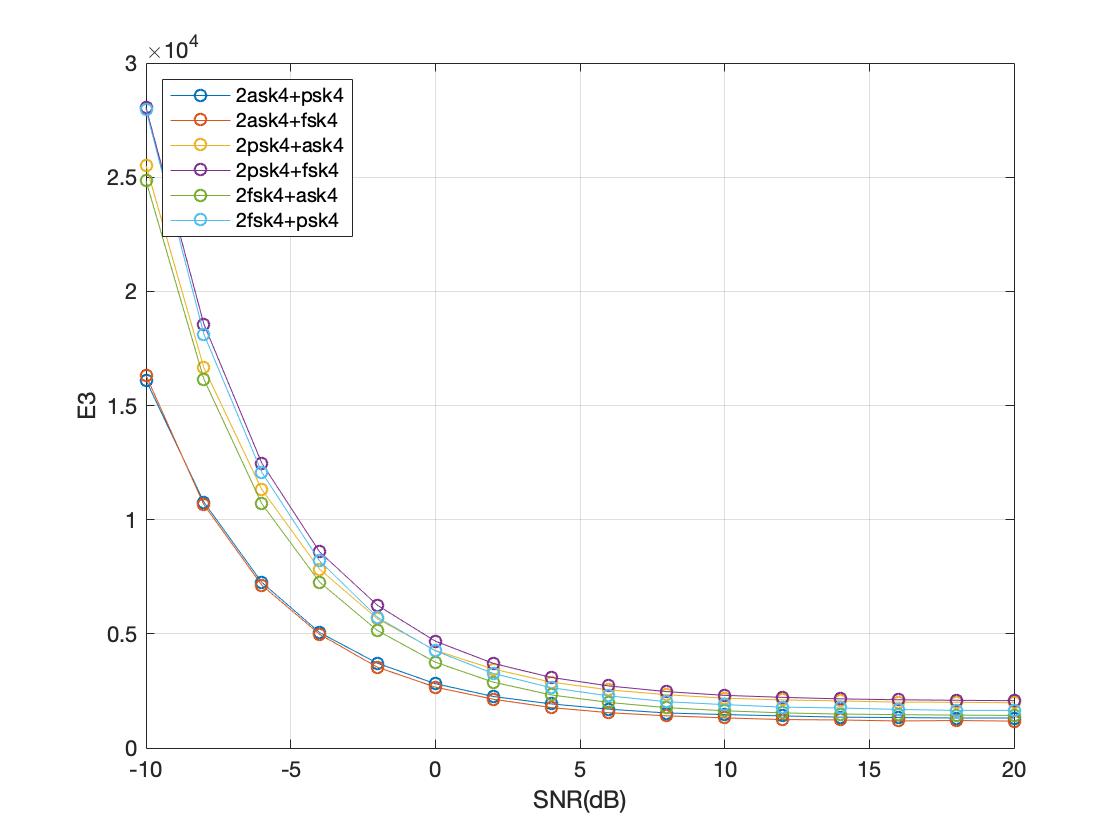}
\caption{Simulation values of wavelet transform of mixed signal, feature3}
\label{MixedSimulationfeature3}
\end{figure}

\begin{figure}[htb]
\includegraphics[width=0.48\textwidth]{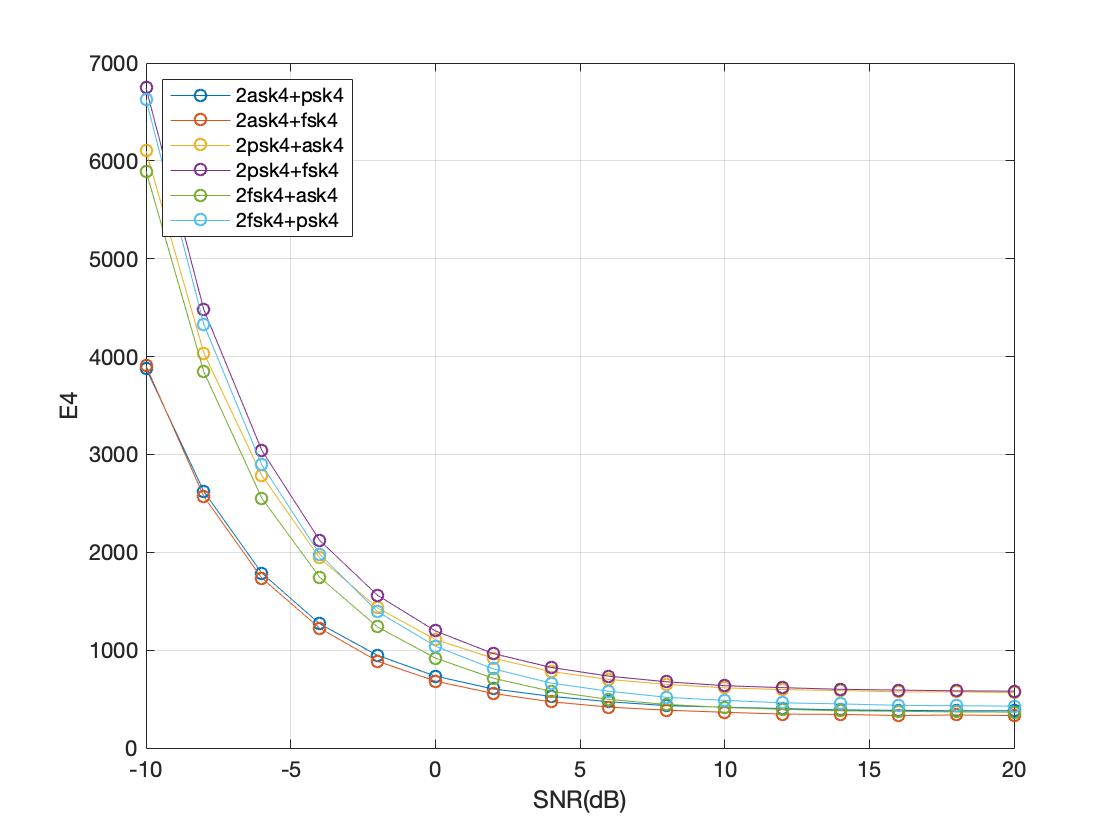}
\caption{Simulation values of wavelet transform of mixed signal, feature4}
\label{MixedSimulationfeature4}
\end{figure}

\begin{figure}[htb]
\subfigure[Simulation values of feature5]{

\includegraphics[width=0.48\textwidth]{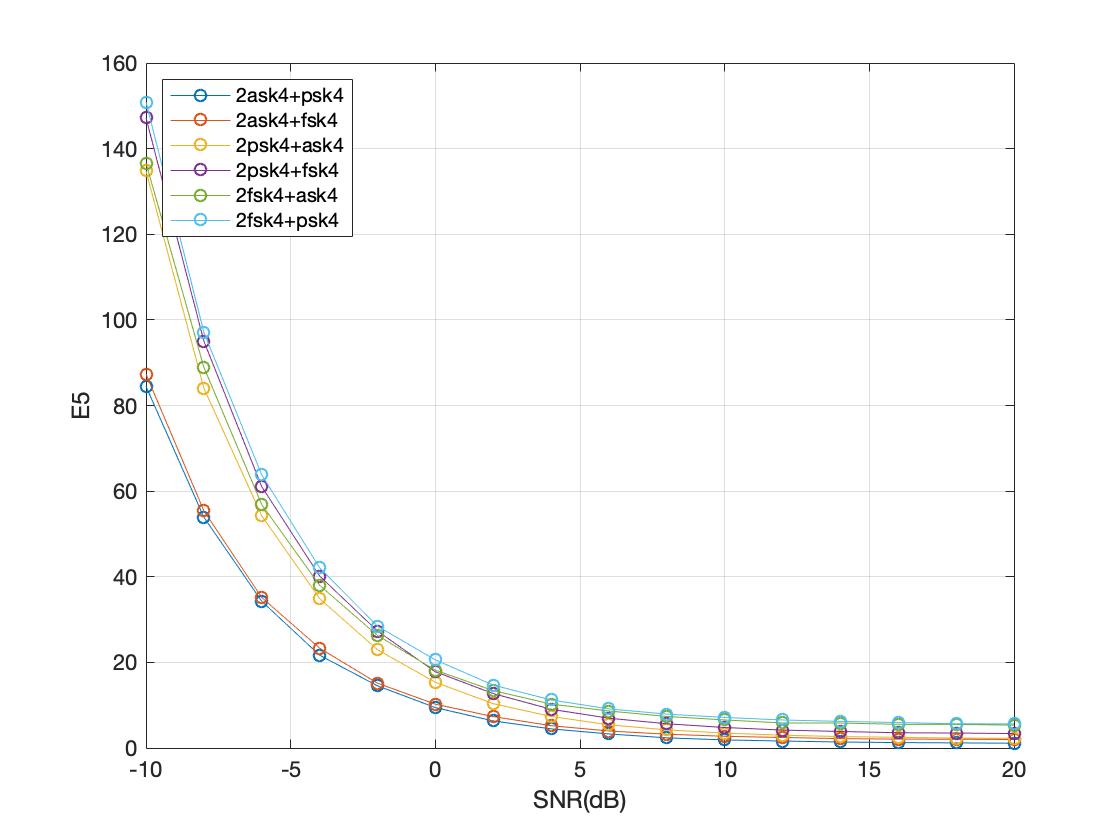}
}
\caption{Simulation values of wavelet transform of mixed signal, feature5}
\label{MixedSimulationfeature5}
\end{figure}

In this section, three digital signals $\{4\mathrm{ASK}, \mathrm{QPSK}, 4\mathrm{FSK}\}$ are divided into six mixed signals according to the signal power ratio, which is equal to 2:1, namely $\{2\mathrm{ASK}4+\mathrm{PSK}4,2\mathrm{ASK}4+\mathrm{FSK}4,2\mathrm{PSK}4+\mathrm{ASK}4,2\mathrm{PSK}4+\mathrm{FSK}4,2\mathrm{FSK}4+\mathrm{ASK}4 , 2\mathrm{FSK}4+\mathrm{PSK}4\}$. The signals would be used to do the simulation and get the results of each value of different features. The parameters sets are the same as those in the single signal section.The simulation results of five feature parameters are shown in Figs. \ref{MixedSimulationfeature1} - \ref{MixedSimulationfeature5}.

\section{SVM model} \label{sec:SVM}
\subsection{Support vector machine classifier overview}
Support vector machine is a machine learning method based on statistical learning theory. It is practical and feasible to solve practical problems such as classification and regression. It can balance the requirements of model complexity and learning ability, and shows better performance than traditional methods. For linearly separable data, a linearly separable support vector machine can be directly learned. When the training data is linearly inseparable, the non-linear data is mapped to the high-dimensional feature space through the kernel function to transform the complexity of the data distribution in a new high-dimensional space to find the linear segmentation surface, and finally get the classification method with the highest correct rate\cite{NandiA.K1997Mrua}.

Given training data:
\begin{equation}
\left(\boldsymbol{x}_{1}, y_{1}\right), \cdots,\left(\boldsymbol{x}_{\ell}, y_{\ell}\right), \boldsymbol{x}_{i} \in \boldsymbol{R}^{n}, y_{i} \in\{+1,-1\} 
\label{eq:svm1}
\end{equation}
In (\ref{eq:svm1}), $i\in\{1,2, \cdots, \lambda\}$ and $x_{i}$ is the input mode set by the two kinds of dots; $y_{i}$ is a category index. If $x_{i}$ belongs to the first class, then $y_{i}=1$ is labeled, otherwise, $y_{i}=-1$ is labeled. The goal of learning is to construct a discriminant function to separate the two types of patterns as accurate as possible. The construction of the decision function can finally be transformed into a typical quadratic programming (QP) problem, that is, under the constraints:
\begin{equation}
y_{i}(\boldsymbol{w} \cdot \boldsymbol{x}-b)+\xi \geqslant 1, \quad \xi \geqslant 0, \quad i=1,2, \cdots, \lambda
\label{eq:svm2}
\end{equation}
Find the minimum value of the following function:
\begin{equation}
\phi(\boldsymbol{w})=\frac{1}{2}\|\boldsymbol{w}\|^{2}+c \sum_{i=1}^{\lambda} \xi
\label{eq:svm3}
\end{equation}
In (\ref{eq:svm2}) and (\ref{eq:svm3}), $w$ is the weight coefficient vector of the classification plane; $b$ is the classification domain value; $c>0$ is the custom penalty coefficient; $\lambda$ can be regarded as the deviation of the training sample separating from the hyperplane. When training is linearly separable, $\lambda=0$ and when the training data is linearly inseparable or if it is not known in advance whether it is linearly separable, $\lambda>0$. The above optimization problem can be solved using standard Lagrange multiplier method, and the final classification function can be obtained as:
\begin{equation}
\phi(\boldsymbol{w})=\frac{1}{2}\|\boldsymbol{w}\|^{2}+c \sum_{i=1}^{\lambda} \xi
\label{eq:svm4}
\end{equation}
In (\ref{eq:svm4}), $a_{i}$ is the Lagrange multiplier; $b$ is the classification domain value. For the nonlinear case, the separation surface may be introduced by non-linear mapping $\varphi(x): \boldsymbol{R}^{n} \rightarrow \boldsymbol{F}$ to reflect input space $\boldsymbol{R}^{n} $ into high dimensional inner product space $\boldsymbol{F}$, then the optimal hyperplane is constructed in $\boldsymbol{F}$ and use linear classifier to complete the classification. According to the related theory of functionals, under the condition of $Mercer$, this kind of nonlinear mapping can be realized by defining an appropriate kernel function. Then the formula above becomes:
\begin{equation}
f(x)=\operatorname{sgn}\left[\sum_{i=1}^{\lambda} \alpha_{i} y_{i} K\left(\boldsymbol{x}_{i}, \boldsymbol{x}\right)+b\right]
\end{equation}
Where $K\left(\boldsymbol{x}_{i}, \boldsymbol{x}\right)=\phi\left(\boldsymbol{x}_{i}\right) \cdot \phi(\boldsymbol{x})$ is a kernel function that satisfies $Mercer$ condition. The above facts point out that the inner product operation in the high-dimensional feature space can be transformed into a simple function operation on the low-dimensional input space. Support vector machine can be completely characterized by training set and kernel function.
\subsection{Key parameter selection}

Using support vector machine to realize the modulation of the digital signal, in order to achieve the best classification effect, it is necessary to select the appropriate kernel function and find the optimal parameter C. The selection of these key parameters will be explained below \cite{ChanY.T1989Iotm}.
\begin{enumerate}
\item Choice of kernel function:
In the SVM algorithm, the kernel function assumes the role of mapping the original data to the high-dimensional space through nonlinear operations. A kernel function represents a nonlinear processing method. Therefore, only by selecting the appropriate kernel function can the SVM perform optimally classification ability. In current research on kernel functions, there are four commonly used kernel functions:
\begin{itemize}
\item  Linear kernel function: $K\left(x, x_{i}\right)=x^{\mathrm{T}} x_{i}$
\item Polynomial kernel function: $K\left(x, x_{i}\right)=\left(x^{\mathrm{T}} x_{i}+1\right)^{q}$
\item Radial basis kernel function: $K\left(x, x_{i}\right)=\exp \left(-g\left\|x-x_{i}\right\|^{2}\right)$
\item Multilayer perceptron kernel function: $K\left(x, x_{i}\right)=\tanh \left(\gamma x^{\mathrm{T}} x_{i}+r\right)$
\end{itemize}

In this work, we choose radial basis function (also called Gaussian kernel function) to do inner product operation, because a large number of research results show that when there is no certain prior knowledge, the processing effect of using radial basis kernel function is often better than other kernel functions. And there is only one parameter that needs to be adjusted, which is quite flexible.

\item Selection of penalty parameter $\mathrm{C}$ and kernel function parameter $\mathrm {g}$:
After selecting the radial basis kernel function, we need to consider the optimal value of the penalty parameter $\mathrm{C}$ and the kernel function parameter  $\mathrm {g}$. Improper selection of key parameters will seriously affect the non-linear processing capability of SVM, and the recognition effect will be greatly reduced. At present, the commonly used methods for parameter optimization of support vector machines are cross-validation, genetic algorithm and particle swarm optimization algorithm. Among them, cross-validation algorithm requires shorter optimization time when achieving similar performance, so it is widely used in radial basis Kernel function parameter optimization is in progress.
In this paper, the optimal value of parameter $\mathrm{C}$ is set to 0.5, and the optimal value of parameter $\mathrm {g}$ is set to 0.05.

\item Data normalization processing:
Before inputting the data into the support vector machine, first normalize the data, limit the data to be processed within the range of [0,1], weaken the influence of singular values in the sample, and speed up the convergence of the program. The normalization method is as follows:
\begin{equation}
y=\frac{x-x_{\min }}{x_{\max }-x_{\min }}
\end{equation}
Where $x_{\min }=\min (x), x_{\max }=\max (x)$. The map min max function in MATLAB can realize the above normalization function. The original data $\mathrm{x}$ can be mapped to the interval [0,1] through the $\mathrm{ps}$ structure by using the statement $[y, p s]=\operatorname{map} \min \max (x, 0,1)$ to complete the data normalization process.
\end{enumerate}

\begin{figure}[htb]  
\centering
\includegraphics[width=0.48\textwidth]{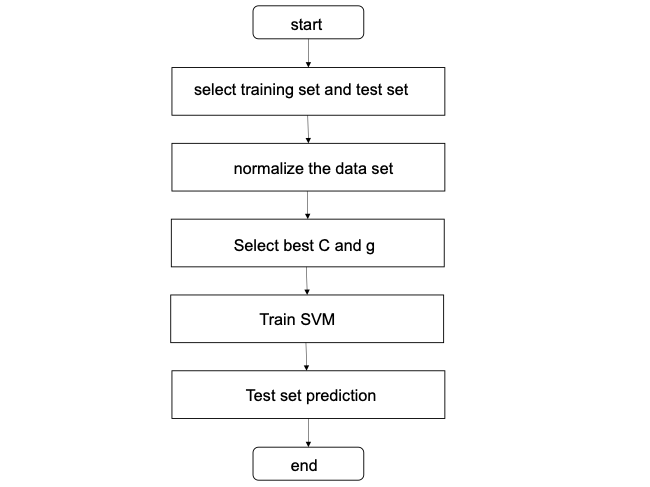}
\caption{Flow chart of the SVM model}
\label{Flow chart of SVM model}
\end{figure}

The flow chart of the SVM model is shown in Fig. \ref{Flow chart of SVM model}. First, the data is divided into a certain proportion of training set and test set, then the training set and test set data are normalized. The third step is to set the best parameters $\mathrm{C}$ and $\mathrm {g}$, followed by the classification model training. Finally, the prediction results of the test set are obtained.

\section{Conclusion\label{cha:conclusion}}

 The paper focused on the feature extraction, modulation and recognition of single and mixed signals. 
We investigated modulation recognition based on the likelihood ratio test algorithm, feature extractions using decision tree and support vector machine. For single signal modulation, the modulation recognition of four and eight digital signals is completed respectively. When SNR is greater than or equal to 8dB, the recognition rate of SVM has reached 80\% or more. For mixed signal modulation, the modulation recognition of six digital signals is completed. When SNR is greater than or equal to 6dB, the recognition rate based on decision tree algorithm has reached over 70\%. When SNR is greater than or equal to 6dB, the recognition rate based on SVM has reached 80\% or more. 

So far, the signal modulation classification problem based on machine/deep learning is still in its infancy. This paper only conducted a preliminary study on this problem, there are still a lot of problems to be solved. The research directions in the future are as follows:

 1. This paper studies both single and mixed signals modulation classification based on some machine learning algorithms, and there is still a certain gap from the theoretical upper limit. Deep learning frameworks such as capsule networks and those models developed in our existing works, e.g., \cite{IREALCARE1,IREALCARE2,IREALCARE3,IREALCARE4,IREALCARE5} can be used in this problem for follow-up research.

2. The number of signal categories included in the classified signal set studied in this paper is small, and only linear non-memory digital modulation systems are considered. In the future, coded modulation schemes, such as Continuous Phase Modulation (CPM) \cite{codedcpm2,codedcpm3} and trellis coded CPM \cite{codedcpm1,codedcpm4,codedcpm5,codedcpm6}, and modulation combined with different coding schemes used in some practical systems \cite{b3a,b3b,b3c,distributedRaptor,Raptor_ML,JNCC,RCRC,NC0,NC1,NC2,NC3,NC4,WRN,network_capacity,UAVdownlink,UAV_THz,UAV_2,RF_energy1,RF_energy2,RF_energy3,RF_energy4,privacy2,pos1,pos2,GI3}, can also be considered. For a richer set of classified signals, how the modulation recognition performance of various modulation recognition technologies based on deep learning will perform is also a direction worthy of further research.



\begin{thebibliography}{99}  
\bibitem{dobre2007survey}O. A. Dobre, A. Abdi, Y. Bar-Ness, and W. Su, “Survey of automatic modulation classification techniques: Classical approaches and new trends,” IET Commun., vol. 1, no. 2, pp. 137–156, Apr. 2007.

\bibitem{Jiahao2022} J. Xu, Z. Lin,  "Modulation and Classification of Mixed Signals Based on Deep Learning", arXiv preprint arXiv:2205.09916. 2022 May 20.

\bibitem{DorieL.2006Aslm}L. Dorie, S. Le Nours, O. Pasquier and J. Diouris, ‘A system level model for software
defined radio design,’ in Proceedings - 2006 IEEE Radio and Wireless Symposium,
vol. 2006, 2006, pp. 463–466, ISBN: 0780394127.

\bibitem{alma991031525652105106} Z. Zhu, Automatic modulation classification : principles, algorithms, and applications,
eng. Chichester, England: Wiley, ISBN: 1-118-90650-0.

\bibitem{SapianoP.C1996MlPc}P. Sapiano and J. Martin, ‘Maximum likelihood psk classifier,’ in Proceedings of
MILCOM ’96 IEEE Military Communications Conference, vol. 3, IEEE, 1996, 1010–
1014 vol.3.
\bibitem{WenWei2000Mcfd} W. Wei and J. Mendel, ‘Maximum-likelihood classification for digital amplitude-phase
modulations,’ IEEE Transactions on Communications, vol. 48, no. 2, pp. 189–193,
2000, ISSN: 0090-6778.
\bibitem{SillsJ.A1999Mmcf} J. Sills, ‘Maximum-likelihood modulation classification for psk/qam,’ eng, in MILCOM
1999. IEEE Military Communications. Conference Proceedings (Cat. No.99CH36341),
vol. 1, IEEE, 1999, 217–220 vol.1, ISBN: 0780355385.

\bibitem{HuangCy1995LMFM} C. Huang and A. Polydoros, ‘Likelihood methods for mpsk modulation classification,’
English, Ieee Transactions On Communications, vol. 43, no. 2-4, pp. 1493–1504, 1995,
ISSN: 0090-6778.
\bibitem{PolydorosA1990Otda} A. Polydoros and K. Kim, ‘On the detection and classification of quadrature digital
modulations in broad-band noise,’ eng, IEEE Transactions on Communications, vol. 38,
no. 8, pp. 1199–1211, 1990, ISSN: 0090-6778.
74

\bibitem{LiangHong2002Aalm} L. Hong and K. Ho, ‘Antenna array likelihood modulation classifier for bpsk and qpsk
signals,’ eng, in MILCOM 2002. Proceedings, vol. 1, IEEE, 2002, 647–651 vol.1, ISBN:
0780376250.
\bibitem{ChuggK.M1995Clpe} K. Chugg, C.-S. Long and A. Polydoros, ‘Combined likelihood power estimation
and multiple hypothesis modulation classification,’ eng, in Conference Record of The
Twenty-Ninth Asilomar Conference on Signals, Systems and Computers, vol. 2, IEEE,
1995, 1137–1141 vol.2, ISBN: 0818673702.
\bibitem{DobreO.A2006LAfL} O. Dobre and F. Hameed, ‘Likelihood-based algorithms for linear digital modulation
classification in fading channels,’ eng, in 2006 Canadian Conference on Electrical and
Computer Engineering, IEEE, 2006, pp. 1347–1350, ISBN: 1424400384.

\bibitem{ZhuZ.2015MciM} Z. Zhu and A. Nandi, ‘Modulation classification in mimo fading channels via expecta-
tion maximization with non-data-aided initialization,’ vol. 2015-, Institute of Electrical
and Electronics Engineers Inc., 2015, pp. 3014–3018, ISBN: 9781467369978.
\bibitem{JingwenZhang2017CMCf} J. Zhang, D. Cabric, F. Wang and Z. Zhong, ‘Cooperative modulation classification for
multipath fading channels via expectation-maximization,’ eng, IEEE Transactions on
Wireless Communications, vol. 16, no. 10, pp. 6698–6711, 2017, ISSN: 1536-1276.

\bibitem{FengXiang2012HMCA} X. Feng and H.-B. Yuan, ‘Hierarchical modulation classification algorithm based on
higher-order cyclic cumulants and support vector machines,’ chi, Dianxun Jishu /Telecommunications Engineering, vol. 52, no. 6, pp. 878–882, 2012, ISSN: 1001-893X.

\bibitem{ZhangQ2009Otsc} Q. Zhang, O. Dobre, S. Rajan and R. Inkol, ‘On the second-order cyclostationarity
for joint signal detection and classification in cognitive radio systems,’ eng, in 2009
Canadian Conference on Electrical and Computer Engineering, IEEE, 2009, pp. 204–
208, ISBN: 9781424435098.
\bibitem{YanPeng-Zhan2010ARoD} P. -Z. Yan and Z. -Y. Wang, ‘Automatic recognition of digital modulation signals by
applying high order cumulants and support vector machines,’ chi, Dianxun Jishu /
Telecommunications Engineering, vol. 50, no. 10, pp. 36–40, 2010, ISSN: 1001-893X.

\bibitem{AzzouzE.E1995Aiod}E. Azzouz and A. Nandi, ‘Automatic identification of digital modulation types,’ eng,
Signal processing, vol. 47, no. 1, pp. 55–69, 1995, ISSN: 0165-1684.
\bibitem{AzzouzE1996Pfar} E. Azzouz and A. Nandi, ‘Procedure for automatic recognition of analogue and digital
modulations,’ eng, IEE Proceedings: Communications, vol. 143, no. 5, pp. 259–266,
1996, ISSN: 1350-2425.
\bibitem{alma991031688625105106} E. Azzouz, Automatic Modulation Recognition of Communication Signals, eng, 1st ed.
1996. New York, NY: Springer US, ISBN: 9781475724691.

\bibitem{GardnerWilliamA1986Tsct} W. A. Gardner, ‘The spectral correlation theory of cyclostationary time-series,’ eng,
Signal processing, vol. 11, no. 1, pp. 13–36, 1986, ISSN: 0165-1684.
\bibitem{XiaoYan2017IRMC} X. Yan, G. Feng, H.-C. Wu, W. Xiang and Q. Wang, ‘Innovative robust modulation
classification using graph-based cyclic-spectrum analysis,’ eng, IEEE Communications
Letters, vol. 21, no. 1, pp. 16–19, 2017, ISSN: 1089-7798.
\bibitem{LiangYe2018RoDS} Y. Liang, J. Hao and R. Shi, ‘Recognition of digital signal modulation mode based on
wavelet neural network,’ Journal of Jilin University.
Science Edition, no. 2, 2018, ISSN: 1671-5489.
\bibitem{ChenJian2006Miod} J. Chen, Y.-H. Kuo, J.-D. Li and Y.-B. Ma, ‘Modulation identification of digital signals
with wavelet transform.,’ chi, Dianzi Yu Xinxi Xuebao/Journal of Electronics and
Information Technology, vol. 28, no. 11, pp. 2026–2029, 2006, ISSN: 1009-5896.
\bibitem{SunJian-Cheng2006Mamr} J.-C. Sun, T.-Y. Zhang and H. -Y. Liu, ‘Multi-class analogue modulation recognition
algorithms based on support vector machines.,’ chi, Journal of University of Electronic
Science and Technology of China, vol. 35, no. 2, pp. 149–152, 2006, ISSN: 1001-0548.

\bibitem{cellular1}Z. Lin, P. Xiao and B. Vucetic, “Analysis of Receiver Algorithms for LTE SC-FDMA Based Uplink MIMO Systems”, IEEE Transactions on Wireless Communications, Vol. 9, No. 1, Nov. 2010, pp. 60-65. 

\bibitem{cellular2}Y. Chen, M. Ding, D. Lopez-Perez, J. Li, Z. Lin, B. Vucetic, "Dynamic reuse of unlicensed spectrum: An inter-working of LTE and WiFi", IEEE Wireless Communications 24 (5), 52-59

\bibitem{cellular3}Y Chen, J Li, Z Lin, G Mao, B Vucetic, "User association with unequal user priorities in heterogeneous cellular networks", IEEE Transactions on Vehicular Technology 65 (9), 7374-7388

\bibitem{cellular4}Y. Chen, M. Ding, D. López-Pérez, Z. Lin and G. Mao, "A Space-Time Analysis of LTE and Wi-Fi Inter-Working," in IEEE Journal on Selected Areas in Communications, vol. 34, no. 11, pp. 2981-2998, Nov. 2016, doi: 10.1109/JSAC.2016.2614922.

\bibitem{cellular5} K. Wei, G. Mao, W. Zhang, Y. Yang, Z. Lin and C. S. Chen, "Optimal microcell deployment for effective mobile device energy saving in heterogeneous networks," 2014 IEEE International Conference on Communications (ICC), 2014, pp. 4048-4053, doi: 10.1109/ICC.2014.6883954.

\bibitem{cellular6}J. Qiu, Z. Lin, W. Hardjawana, B. Vucetic, C. Tao and Z. Tan, "Resource allocation for OFDMA system under high-speed railway condition," 2014 IEEE Wireless Communications and Networking Conference (WCNC), 2014, pp. 2683-2687, doi: 10.1109/WCNC.2014.6952832.

\bibitem{cellular7}Z. Lin, P. Xiao, T. B. Sørensen, and B. Vucetic, (2010), Spatial frequency scheduling for uplink SC-FDMA based linearly precoded multiuser MIMO systems. Eur. Trans. Telecomm., 21: 213-223. https://doi.org/10.1002/ett.1372

\bibitem{cellular8}Z. Lin, P. Xiao and B. Vucetic, "SINR distribution for LTE downlink multiuser MIMO systems," 2009 IEEE International Conference on Acoustics, Speech and Signal Processing, 2009, pp. 2833-2836, doi: 10.1109/ICASSP.2009.4960213.

\bibitem{cellular9}Z. Lin, T. B. Sorensen and P. E. Mogensen, "Downlink SINR Distribution of Linearly Precoded Multiuser MIMO Systems," in IEEE Communications Letters, vol. 11, no. 11, pp. 850-852, November 2007, doi: 10.1109/LCOMM.2007.071082.

\bibitem{cellular10}M. Ding, D. López-Pérez, G. Mao and Z. Lin, "Microscopic Analysis of the Uplink Interference in FDMA Small Cell Networks," in IEEE Transactions on Wireless Communications, vol. 15, no. 6, pp. 4277-4291, June 2016, doi: 10.1109/TWC.2016.2538261.

\bibitem{cellular11}J. Yang, M. Ding, G. Mao and Z. Lin, "Interference Management in In-Band D2D Underlaid Cellular Networks," in IEEE Transactions on Cognitive Communications and Networking, vol. 5, no. 4, pp. 873-885, Dec. 2019, doi: 10.1109/TCCN.2019.2927568.

\bibitem{cellular12} M. Ding, P. Wang, D. López-Pérez, G. Mao and Z. Lin, "Performance Impact of LoS and NLoS Transmissions in Dense Cellular Networks," in IEEE Transactions on Wireless Communications, vol. 15, no. 3, pp. 2365-2380, March 2016, doi: 10.1109/TWC.2015.2503391.

\bibitem{cellular13}M. Ding, D. Lopez-Perez, G. Mao, P. Wang and Z. Lin, "Will the Area Spectral Efficiency Monotonically Grow as Small Cells Go Dense?," 2015 IEEE Global Communications Conference (GLOBECOM), 2015, pp. 1-7, doi: 10.1109/GLOCOM.2015.7416981.

\bibitem{cellular14}J. Yang, M. Ding, G. Mao, Z. Lin, D. -G. Zhang and T. H. Luan, "Optimal Base Station Antenna Downtilt in Downlink Cellular Networks," in IEEE Transactions on Wireless Communications, vol. 18, no. 3, pp. 1779-1791, March 2019, doi: 10.1109/TWC.2019.2897296.

\bibitem{MIMO_capacity}Z. Lin, B. Vucetic, J. Mao, "Ergodic capacity of LTE downlink multiuser MIMO systems", 2008 IEEE International Conference on Communications, 3345-3349.

\bibitem{MallatS.G1989Atfm} S. Mallat, ‘A theory for multiresolution signal decomposition: The wavelet representa-
tion,’ eng, IEEE Transactions on Pattern Analysis and Machine Intelligence, vol. 11,
no. 7, pp. 674–693, 1989, ISSN: 0162-8828.
\bibitem{NandiA.K1997Mrua} A. Nandi and E. Azzouz, ‘Modulation recognition using artificial neural networks,’ eng,
Signal processing, vol. 56, no. 2, pp. 165–175, 1997, ISSN: 0165-1684.
\bibitem{ChanY.T1989Iotm} Y. Chan and L. Gadbois, ‘Identification of the modulation type of a signal,’ eng, Signal
processing, vol. 16, no. 2, pp. 149–154, 1989, ISSN: 0165-1684.

\bibitem{IREALCARE1}L. Meng, K. Ge, Y. Song, D. Yang, and Z. Lin, "Long-term wearable electrocardiogram signal monitoring and analysis based on convolutional neural network", the IEEE Transactions on Instrumentation \& Measurement, Vol. 70, April 2021, DOI:10.1109/TIM.2021.3072144

\bibitem{IREALCARE2} Z. Chen, Z. Lin, P. Wang, and M. Ding, Negative-ResNet: noisy ambulatory electrocardiogram signal classification scheme, Neural Computing and Applications, Vol. 33, Issue 14, July 2021, pp.  8857-8869

\bibitem{IREALCARE3} P, Wang, Z. Lin, Z. Chen, X. Yan, M. Ding, A Wearable ECG Monitor for Deep Learning-Based Real-Time Cardiovascular Disease Detection, https://arxiv.org/abs/2201.10083

\bibitem{IREALCARE4} X. Yan, Z. Lin, P. Wang, Wireless Electrocardiograph Monitoring Based on Wavelet Convolutional Neural Network, Proceedings of the IEEE WCNC 2020.

\bibitem{IREALCARE5} M. Liu, Z. Lin, P. Xiao, and W. Xiang. "Human Biometric Signals Monitoring based on WiFi Channel State Information using Deep Learning." arXiv preprint arXiv:2203.03980 (2022).


\bibitem{codedcpm2}Z. Lin and T. Aulin, “On Combined Ring Convolutional Coded Quantization and CPM for Joint Source and Channel Coding”, Transactions on Emerging Telecommunications Technologies, Special Issue on ’New Directions in Information Theory’, Vol.19, No.4. June 2008, pp. 443-453. 

\bibitem{codedcpm3} Z. Lin and T. Aulin, “Joint Source-Channel Coding using Combined TCQ/CPM: Iterative Decoding”, IEEE Transactions on Communications, VOL.53, NO. 12, Dec. 2005, pp. 1991-1995.


\bibitem{codedcpm1}Z. Lin and T. Aulin, “Joint Source and Channel Coding using Punctured Ring Convolutional Coded CPM”, IEEE Transactions on Communications, Vol. 56, No. 5, May, 2007, pp. 712-723. 

\bibitem{codedcpm4}	Z. Lin and B. Vucetic, “Performance Analysis on Ring Convolutional Coded CPM”, IEEE Transactions on Wireless Communications, Vol. 8, No. 9, Sept. 2009, pp. 4848-4854. The latest impact factor is 4.951
\bibitem{codedcpm5}	Z. Lin and B. Vucetic, “Spatial Frequency Scheduling for SC-FDMA Based Uplink Multi-user MIMO Systems”, IET Communications, V. 3, No. 7, July, 2009, pp. 163-165.

\bibitem{codedcpm6}	Z. Lin and T. Aulin, “On Joint Source and Channel Coding using trellis coded CPM: Analytical Bounds on the Channel Distortion”, IEEE Transactions on Information Theory, Vol. 53, No. 13, Sept. 2007. pp. 3081-3094. The latest impact factor is 2.679

\bibitem{b3a}	Y. Hu, P. Wang, Z. Lin, M. Ding, YC. Liang, “Performance Analysis of Ambient Backscatter Systems with LDPC-coded Source Signals” in IEEE Transactions on Vehicular Technology, vol. 70, no. 8, pp. 7870-7884, Aug. 2021, doi: 10.1109/TVT.2021.3093912.

\bibitem{b3b}	Y. Hu, P. Wang, Z. Lin, M. Ding, YC. Liang, “Machine Learning Based Signal Detection for Ambient Backscatter Communications”, 2019 IEEE International Conference on Communications (ICC): SAC Internet of Things Track. 

\bibitem{b3c} Z. Xing, Z. Lin, M. Ding,  “Outage Capacity Analysis for Ambient Backscatter Communication Systems”, 2018 28th International Telecommunication Networks and Applications Conference (ITNAC).

\bibitem{distributedRaptor}J. Yue, Z. Lin, B. Vucetic, G. Mao, T. Aulin, "Performance analysis of distributed raptor codes in wireless sensor networks", IEEE Transactions on Communications 61 (10), 2013, 4357-4368

\bibitem{Raptor_ML}P. Wang, G. Mao, Z. Lin, M Ding, W. Liang, X. Ge, Z. Lin, "Performance analysis of raptor codes under maximum likelihood decoding", IEEE Transactions on Communications 64 (3), 2016, 906-917

\bibitem{JNCC}K. Pang, Z. Lin, Y. Li, B. Vucetic, "Joint network-channel code design for real wireless relay networks", the 6th International Symposium on Turbo Codes \& Iterative Information, 2010, 429-433.

\bibitem{RCRC}Z. Lin, A. Svensson, "New rate-compatible repetition convolutional codes",
IEEE Transactions on Information Theory 46 (7),  2651-2659

\bibitem{NC0} Z. Lin, “Design of Network Coding Schemes in Wireless Network”, CRC Press, Taylor \& Francis. Books, ISBN: 9781032067766, June 2022.

\bibitem{NC1}Z Lin, B Vucetic, "Power and rate adaptation for wireless network coding with opportunistic scheduling", 2008 IEEE International Symposium on Information Theory, 21-25

\bibitem{NC2} T Ding, M Ding, G Mao, Z Lin, AY Zomaya, D López-Pérez, "Performance analysis of dense small cell networks with dynamic TDD", IEEE Transactions on Vehicular Technology 67 (10), 9816-9830, 2018.

\bibitem{NC3}P Wang, G Mao, Z Lin, X Ge, BDO Anderson, "Network coding based wireless broadcast with performance guarantee", IEEE Transactions on Wireless Communications 14 (1), 532-544, 2014.

\bibitem{NC4}K Pang, Z Lin, Y Li, B Vucetic,"Distributed network-channel codes design with short cycles removal", IEEE Wireless Communications Letters 2 (1), 62-65, 2012.

\bibitem{WRN}J. Yue; Z. Lin; B. Vucetic; G. Mao; M. Xiao; B. Bai; K. Pang, "Network Code Division Multiplexing for Wireless Relay Networks,"  IEEE Transactions on Wireless Communications, vol.14, no.10, pp.5736-5749, Oct. 2015.


\bibitem{network_capacity}G. Mao, Z. Lin, X. Ge, Y. Yang, "Towards a simple relationship to estimate the capacity of static and mobile wireless networks", IEEE transactions on wireless communications 12 (8), 2014, 3883-3895	

\bibitem{UAVdownlink}D López-Pérez, M Ding, H Li, LG Giordano, G Geraci, A Garcia-Rodriguez, Z. Lin, M. Hassan, "On the downlink performance of UAV communications in dense cellular networks", 2018 IEEE global communications conference (GLOBECOM), 1-7

\bibitem{UAV_THz}	X. Wang, P. Wang, M. Ding, Z. Lin, L. Hanzo and B. Vucetic, “Performance Analysis of TeraHertz Unmanned Aerial Vehicular Networks”, in IEEE Transactions on Vehicular Technology, vol. 69, no. 12, pp. 16330-16335, Dec. 2020, doi: 10.1109/TVT.2020.3035831. 

\bibitem{UAV_2}C Liu, M Ding, C Ma, Q Li, Z Lin, YC Liang, "Performance analysis for practical unmanned aerial vehicle networks with LoS/NLoS transmissions", IEEE International Conference on Communications Workshops (ICC Workshops), 2018, 1-6.

\bibitem{RF_energy1}J. Wang, B. Li, G. Wang, Z. Lin, H. Wang, and G. Chen, Optimal Power Splitting for MIMO SWIPT Relaying Systems with Direct Link in IoT Networks, Physical Communication, Volume 43, December 2020. 

\bibitem{RF_energy2}D. Zhai, H. Chen, Z. Lin. Y. Li and B. Vucetic, “Accumulate Then Transmit: Multi-user Scheduling in Full-Duplex Wireless-Powered IoT Systems”, IEEE Internet of Things Journal, Volume: 5 , Issue: 4 , Aug. 2018. 

\bibitem{RF_energy3} H. Chen, Y. Ma. Z. Lin, Y. Li and B. Vucetic, “Distributed Power Control in Interference Channels with QoS Constraints and RF Energy Harvesting: A Game-Theoretic Approach”, IEEE Transactions on Vehicular Technology, Volume: 65, Issue: 12, Dec. 2016. pp.10063 – 10069. 

\bibitem{RF_energy4} Y. Ma, H. Chen, Z. Lin, Y. Li and B. Vucetic, "Distributed and Optimal Resource Allocation for Power Beacon-Assisted Wireless-Powered Communications,", IEEE Transactions on  Communications, vol.63, no.10, pp.3569-3583, Oct. 2015.


 \bibitem{privacy2}	S. Shaham, M. Ding, B. Liu, S. Dang Z. Lin, J. Li, “Privacy Preservation in Location-Based Services: A Novel Metric and Attack Model”, IEEE Transactions on Mobile Computing, vol. 20, no. 10, pp. 3006-3019, 1 Oct. 2021. doi: 10.1109/TMC.2020.2993599.  





\bibitem{pos1}X. Shi, B. Anderson, G. Mao, Z. Yang, J. Chen, and Z. Lin, “Robust Localization Using Time Difference of Arrivals”, IEEE Signal processing letters, Vol. 23, Issue 10, Oct. 2016, pp. 1320 –1324. 

\bibitem{pos2}D. Zhai and Z. Lin, “RSS-based Indoor Positioning with Biased Estimator and Local Geographical Factor”, Proceedings of the 22nd International Conference on Telecommunications (ICT 2015), Sydney, Australia 




\bibitem{GI3} Z. Zhang, R. Luo, X. Wang, and Z. Lin, “Microwave ghost imaging via
LTE-DL signals,” in 2018 International Conference on Radar (RADAR),
2018, pp. 1–5



\end{thebibliography}
\end{document}